\documentclass[10pt]{article}
\usepackage{amsmath,amsthm,amssymb, bm}
\DeclareMathOperator*{\argmax}{argmax}

\usepackage{fullpage}
\usepackage{xcolor}
\usepackage{graphicx}
\usepackage{comment}
\usepackage[ruled,vlined]{algorithm2e}
\usepackage{algorithmic}
\usepackage{caption}
\usepackage{subcaption}
\usepackage{hyperref}
\usepackage{enumitem}

\usepackage[sorting=none,citestyle=numeric-comp,giveninits=true,doi=false,isbn=false,url=false,eprint=false,maxbibnames=99]{biblatex}

\newtheorem{theorem}{Theorem}[section]

\newtheorem{proposition}[theorem]{Proposition}

\theoremstyle{definition}
\newtheorem{definition}[theorem]{Definition}

\theoremstyle{remark}

\numberwithin{equation}{section}

\newcommand{\quality}{\theta^*}
\newcommand{\qualityest}{\widehat{\theta}}

\newcommand{\score}{s}

\newcommand{\scorematrix}{S}

\newcommand{\reviewerfunction}{\beta}
\newcommand{\reviewerfunctionpdf}{f}

\newcommand{\assignmentrv}{\mathcal{A}}
\newcommand{\assignment}{A}

\newcommand{\jointscorepdf}{f'}

\newcommand{\decisionrv}{\bm{D}}

\newcommand{\decision}{D}

\newcommand{\calibrationrv}{\bm{C}}

\newcommand{\probfunction}{g}
\newcommand{\calibratefunction}{h}

\newcommand{\noise}{\epsilon}

\newcommand{\Estrategy}{\zeta}

\newcommand{\calibrateprob}{q}

\newcommand{\normalcdf}{\Phi}

\newcommand{\Econference}{\mathcal{E}_C}

\newcommand{\Eadversary}{\mathcal{E}_A}

\title{Calibration with Privacy in Peer Review}
\author{Wenxin Ding \qquad Gautam Kamath \textcircled{r} Weina Wang \textcircled{r} Nihar B. Shah\thanks{Wenxin Ding is at the University of Chicago (wenxind@uchicago.edu), Gautam Kamath is at the University of Waterloo (g@csail.mit.edu), and Weina Wang and Nihar B. Shah are at Carnegie Mellon University (\{weinaw, nihars\}@cs.cmu.edu).}}
\date{}

\begin{document}

\maketitle

\begin{abstract}
Reviewers in peer review are often miscalibrated: they may be strict, lenient, extreme, moderate, etc. A number of algorithms have previously been proposed to calibrate reviews. Such attempts of calibration can however leak sensitive information about which reviewer reviewed which paper. In this paper, we identify this problem of calibration with privacy, and provide a foundational building block to address it. Specifically, we present a theoretical study of this problem under a simplified-yet-challenging model involving two reviewers, two papers, and an MAP-computing adversary. Our main results establish the Pareto frontier of the tradeoff between privacy (preventing the adversary from inferring reviewer identity) and utility (accepting better papers), and design explicit computationally-efficient algorithms that we prove are Pareto optimal.

%~\\{\bf Old abstract with two challenges:}  Reviewers in peer review are often miscalibrated: they may be strict, lenient, extreme, moderate, etc. A number of algorithms have previously been proposed to calibrate reviews. These algorithms however suffer from two key challenges: (1) the calibration algorithms may leak information about which reviewer reviewed which paper, and (2) they suffer from the bottleneck of a small number of samples (reviews) per reviewer. In this paper, to mitigate challenge \#2, we consider using exogenously obtained information about reviewers' calibration, such as data from past conferences. Our focus is challenge \#1, towards which we provide a foundational building block for calibration with privacy. Specifically, we present a theoretical study of this problem under a simplified-yet-challenging model involving two reviewers, two papers, and a MAP-computing adversary. Our main results establish the Pareto frontier of the tradeoff between privacy and utility (accepting better papers), and design explicit computationally-efficient algorithms that we prove are Pareto optimal.
\end{abstract}

\section{Introduction}

It is well known that scores provided by people are frequently miscalibrated. In the application of peer review, reviewers may be strict, lenient, extreme, moderate, etc. This leads to unfairness in peer review, for instance, disadvantaging papers that happen to go to strict reviewers~\cite{siegelman1991assassins}: \emph{``the
existence of disparate categories of
reviewers creates the potential for
unfair treatment of authors. Those
whose papers are sent by chance to
assassins/demoters are at an unfair
disadvantage, while zealots/pushovers give authors an unfair advantage.''} 

A number of algorithms~\cite{flach2010kdd,ge13bias,roos2011calibrate,roos2012statistical, paul1981calibration,baba2013quality,mackay2017calibration} are proposed in the literature to address the problem of miscalibration. There are two key challenges, however, towards any attempts of calibration using such algorithms:

{\bf Challenge \#1: The calibration algorithms may leak information about which reviewer reviewed which paper.} Here is an example showing how a na\"ive attempt at calibration can compromise privacy. Consider an adversary trying to guess the reviewer of a paper between two possibilities -- reviewer X or reviewer Y. The review for the paper is lukewarm, and for simplicity suppose this is the only review. We consider the ``open review'' model where all submitted papers, reviews, and final decisions are public (but reviewer identities are not). Also suppose it is known that reviewer X is strict but reviewer Y is not. Then the conference will not accept the paper unless the conference performs a calibration using this information \emph{and} the reviewer is X. The acceptance of the paper will provide the adversary with the necessary information to infer the reviewer as X. 

{\bf Challenge \#2: The bottleneck of a small number of samples (reviews) per reviewer.} Many conferences have each reviewer reviewing only a handful papers (typically 1 to 6 papers), as well as have each paper reviewed by a handful of reviewers. As a consequence, it is often hard to decipher the miscalibration of any reviewer, particularly since human miscalibration can be quite complex~\cite{brenner2005modeling}. Indeed, program chairs of conferences have tried to use some algorithms to calibrate reviewers' scores, but have found the outcomes to be unsatisfactory. For instance, John Langford, the program chair of the ICML 2012 conference says that~\emph{``We experimented with reviewer normalization and generally found it significantly harmful''}~\cite{langford2012icml}. 

Our focus on this paper is challenge \#1 of privacy. Towards challenge \# 2, we assume that the conference has exogenous information about the miscalibration of reviewers, such as reviewers' calibration information from other conferences where they have reviewed. (Appendix~\ref{AppSimulations} presents simulations illustrating benefits of calibration with exogeneous information.) 
Tackling the problem of privacy in calibration that we identify is quite challenging in full generality. In this paper, our goal is to initiate research towards this grander goal by providing a foundational building block for it. We consider a simplified--yet highly challenging--model with two reviewers, two papers, and (exogenously) known miscalibration functions where an adversary attempts to guess the reviewer assignment based on maximum a posteriori (MAP) computation. We provide a comprehensive analysis under this model. Our contributions are summarized as follows:
\begin{itemize}
    \item We identify the problem of privacy in calibration, and we initiate a theoretical study with the formulation of a problem that incorporates various key challenges of the more general setting.
    
    \item We provide explicit computationally-efficient algorithms for calibration with privacy that optimally trades off the error of the conference (in terms of accepting the better paper) and the error of the adversary (in terms of guessing the reviewer).
    \item We establish the structure of the Pareto optimal curve between the two aforementioned desiderata. We observe that interestingly, there is a linear tradeoff between the two errors up to a certain point, after which the error of the adversary does not decrease even if the conference adds more randomness in its protocols.
\end{itemize}
%Our envision building on this work to develop principled and practical methods of addressing the important problem of miscalibration without compromising privacy of reviewers. 

%%%%%%%%%%%%%%%%%%%%%%%%%%%%

\section{Related Work}
\label{SecRelated}
Peer review is extensively used for evaluating scientific papers and grant proposals. However, conference peer review also incurs various challenges such as miscalibration~\cite{flach2010kdd,ge13bias,roos2011calibrate,roos2012statistical, paul1981calibration,baba2013quality,mackay2017calibration,wang2018your,shah2017design}, biases~\cite{tomkins2017reviewer,manzoor2020uncovering,stelmakh2019testing,}, subjectivity~\cite{lee2015commensuration,noothigattu2018choosing,mahoney1977publication}, dishonesty~\cite{balietti2016peer,stelmakh2020catch,xu2018strategyproof,littman2021collusion,jecmen2020manipulation,wu2021making}, and others. See~\cite{shah2021survey} for a survey. 

The problem of miscalibration is well recognized in the literature. A common approach to design calibration algorithms is to assume a certain model of miscalibration, and under the assumed model, estimate the calibrated scores (or the model parameters) from the scores given by reviewers. This line of literature~\cite{flach2010kdd,ge13bias,roos2011calibrate,roos2012statistical, paul1981calibration,baba2013quality,mackay2017calibration} assumes affine models for miscalibration: they assume that each paper has some ``true'' real-valued quality and that the score provided by any reviewer is some affine transform (plus noise) of this true quality. In our formulation (detailed subsequently in Section~\ref{SecFormulation}) we also assume papers have true qualities, and a part of our work also assumes affine miscalibrations.

A second line of literature~\cite{mitliagkas2011,ammar2012aggregation,freund2003boosting} recognizes the problem of miscalibration, and takes the approach of using only the ranking of papers induced by the scores given by any individual reviewer, or alternatively, asking each reviewer to only provide a ranking of the papers they are reviewing. Using rankings alone thus gets rid of any miscalibrations, but on the downside, can lose some information contained in scores. Moreover, a recent work~\cite{wang2018your} showed that under certain settings, scores can yield more information than rankings even if the miscalibration is adversarial. 

Notably, these works consider addressing miscalibration using data from within the conference at hand, and moreover do not consider the issue of compromise of privacy.

We assume an ``open review'' model where all submitted papers and all reviews are available publicly, but where information of who reviews which paper is not. Such an open review model is gaining increasing popularity: see, for instance, \url{openreview.net} and \url{scipost.org}. This model is followed in the ICLR conference as well as other venues. In a survey~\cite{soergel2013open} at the ICLR 2013 conference, researchers felt that this open review model leads to benefits of more accountability of authors (in terms of not submitting below-par papers) as well as reviewers (in terms of giving high-quality reviews). The publicly available data has resulted in another benefit: it has yielded a rich dataset for research on peer review~\cite{xu2018strategyproof,kang2018dataset,manzoor2020uncovering,tran2020open,bharadhwaj2020anonymization,yuan2021can}. A downside of the open review approach is that if a rejected paper is resubmitted elsewhere, the (publicly available) knowledge of previous rejection may bias the reviewer~\cite{stelmakh2020resubmissions}. 

Our work considers explicitly randomized assignments and decisions. In practice, the assignments and decision protocols are typically deterministic (although some variations naturally arise due to human involvement in various parts of the peer-review process). The assignment of reviewers to papers is done by solving a certain optimization problem~\cite{goldsmith2007ai,taylor2008optimal,charlin13tpms,Garg2010papers,stelmakh2018forall,kobren19localfairness} involving similarities computed between each reviewer-paper pair~\cite{mimno07topicbased,charlin13tpms,fiez2019super,meir2020market}. Decisions are arrived at after discussions between the reviewers. That said, there are notable instances where randomization has been explicitly used in practice in peer review: randomization can help mitigate dishonest behavior~\cite{jecmen2020manipulation} and can help make more fair decisions for borderline papers or grants~\cite{liu2020acceptability,chawla2021swiss}. A recent survey of researchers finds support for randomized decisions~\cite{philipps2021research}. Finally, the algorithms in the theoretical work~\cite{wang2018your} comparing scores and rankings in the context of miscalibration also employ randomization.

Issues of privacy in peer review also arise when releasing data to researchers. The program chairs of the WSDM 2017 conference performed a remarkable controlled experiment to test for biases in peer review, and in their paper~\cite{tomkins2017reviewer} they point out privacy-related concerns in releasing data: 
\emph{``We would prefer to make available the raw data used in our study, but after some effort we have not been able to devise an anonymization scheme that will simultaneously protect the identities of the parties involved and allow accurate aggregate statistical analysis. We are familiar with the literature around privacy preserving dissemination of data for statistical analysis and feel that releasing our data is not possible using current state-of-the-art techniques.''} We are aware of two past works which deal with privacy in peer review~\cite{ding2020privacy,jecmen2020manipulation}. In particular, both papers consider privacy-preserving release of peer-review data. The paper~\cite{ding2020privacy} provides an algorithm to optimize utility when releasing histograms of certain functions of the review scores. The paper~\cite{jecmen2020manipulation} uses randomized assignments to guarantee privacy of the reviewer-paper assignment when data pertaining to similarities between reviewer-paper pairs is released. 

Differential privacy~\cite{dwork2016calibrating} is a popular rigorous notion of data privacy.
Roughly speaking, an algorithm is differentially private if its distribution over outputs is similar when provided with ``neighboring'' inputs. 
In our problem with two papers and two reviewers, one can consider neighboring inputs to be those that differ only in the assignment. 
We provide a tight characterization of the adversary's ability to determine which of the two possible assignments is the true one.
Thus, it may be a useful building block towards more complex private calibration schemes.
We note that our calibration algorithms are related to a form of randomized response~\cite{Warner65}, the canonical algorithm for local differential privacy~\cite{Warner65, EvfimievskiGS03, KasiviswanathanLNRS11}.
Though differential privacy is not the focus of our work, we further elaborate on this connection in Appendix~\ref{sec:ldp}.

\section{Problem Formulation and Preliminaries}
\label{SecFormulation}

In this section, we present the formal problem specification. We will introduce some notation in this section, and this notation is also summarized in Table~\ref{TabNotation}. 
\begin{table}[]
    \centering
    \begin{tabular}{|p{.15\textwidth}|p{.7\textwidth}|}
    \hline
        Notation & Meaning\\\hline
         $i \in \{1,2\}$ & Index for paper \\
         $j \in \{1,2\}$ & Index for reviewer\\
         $\quality_i \in \mathbb{R}$ & True quality of paper $i$ \\
         $\qualityest_i \in \mathbb{R}$ & Estimated quality of paper $i$ \\
         $\assignment_1$ and $\assignment_2$ & The two possible assignments\\
         $\score_i \in \mathbb{R}$ & Score received by paper $i$; $\scorematrix = [\score_1, \score_2]$\\
         $\reviewerfunction_j: \mathbb{R} \to \mathbb{R}$ & Calibration function of reviewer $j$ \\
         $\noise_j \in \mathbb{R}$ & Noise of reviewer $j$ \\
         $\reviewerfunctionpdf_j: \mathbb{R} \to \mathbb{R}$ & Marginal probability density function of score given by reviewer $j$, that is, distribution of $\reviewerfunction_j(\quality)$ where $\quality \sim N(0,1)$ \\
         $\Econference \in [0,1]$ & Error of the conference \\
         $\Eadversary \in [0,1]$ & Error of the adversary \\
         \hline
    \end{tabular}
    \caption{Summary of the main notation used in the paper.}
    \label{TabNotation}
\end{table}
\paragraph{Papers and reviewers.} 
We consider a setting with two reviewers and two papers. Each paper $i \in \{1,2\}$ has some latent true quality $\quality_i \in \mathbb{R}$. We assume that the qualities $\quality_1$ and $\quality_2$ are drawn i.i.d.\ according to the standard normal distribution (and hence we have $\quality_1 \neq \quality_2$ with probability $1$). 

\paragraph{Reviewer assignment.} Each reviewer reviews one paper and each paper is reviewed by one reviewer. There are thus two possible assignments: we let $\assignment_1$ denote the assignment of reviewer 1 to paper 1 and reviewer 2 to paper 2, and $\assignment_2$ denote the assignment of reviewer 1 to paper 2 and reviewer 2 to paper 1. We assume that the assignment is chosen uniformly at random from these two possibilities. We assume that the true assignment is known (only) to the conference. We let $\assignmentrv$ denote a random variable representing the assignment. Finally, in our exposition we will refer to the realization of $\assignmentrv$ as the ``true'' assignment (and the unrealized assignment as the ``wrong'' assignment). 

\paragraph{Miscalibration and reviewer scores.} For each paper $i \in \{1,2\}$, we let $\score_i$ denote the score received by by paper $i$. Note that this notation is not indexed by the reviewer for brevity since each paper receives exactly one review. For convenience, we define the vector $\scorematrix = [\score_1, \score_2]$. Following the popular ``open review'' model (\href{https://openreview.net}{OpenReview.net}, \href{https://scipost.org}{scipost.org}), we assume that the scores $\score_1$ and $\score_2$ are known publicly.\footnote{Even if the conference operates in a non-open-review setting where the scores are not public, our guarantees on privacy and conference's error continue to hold. However, our algorithm may not be optimal and the suboptimality may depend on assumptions about the adversary's knowledge of the scores.}

Following~\cite{wang2018your}, we assume that each reviewer $j \in \{1,2\}$ has a function $\reviewerfunction_j: \mathbb{R} \to \mathbb{R}$ which captures their miscalibration. If reviewer $j \in \{1,2\}$ reviews paper $i \in \{1,2\}$, we assume that the reviewer provides a score $\score_i \in \mathbb{R}$ given as:
\begin{align*}
  \score_i = \reviewerfunction_j(\quality_i) + \noise_j,
\end{align*}
where $\noise_j$ is a zero-mean Gaussian random variable independent of everything else. We assume that $\noise_1$ and $\noise_2$ are identically distributed. The value of the noise is unknown but its distribution is publicly known. We call $\reviewerfunction_j$ the \emph{reviewer's miscalibration function} %\ns{call it  ``reviewer's miscalibration function''} 
for reviewer $j$. We assume that the functions $\reviewerfunction_1$ and $\reviewerfunction_2$ are increasing and invertible. In one part of our work, we further make an assumption that the miscalibration functions are affine, and we detail this subsequently in the associated section. As discussed previously, our aim is to use exogenous information about the reviewer miscalibrations in order to mitigate the miscalibration, and to this end, we assume that the functions $\reviewerfunction_1$ and $\reviewerfunction_2$ are known publicly.

For any reviewer $j$, we let $\reviewerfunctionpdf_j$ denote the marginal probability density function of the final score given by that reviewer, that is, $\reviewerfunctionpdf_j$ is the distribution of $\reviewerfunction_j(\quality)$ where $\quality \sim N(0,1)$.

\paragraph{Conference's error.} 
The goal of the conference is to accept the paper with the higher true quality $\argmax_{i \in \{1,2\}} \quality_i$. Note that even if the noise terms were zero, simply choosing the paper with higher score (i.e., $\argmax_{i \in \{1,2\}} \score_i$) may be erroneous due to the miscalibration of the reviewers. The conference can however calibrate the scores, that is, use the information about the miscalibration functions of the reviewers and the knowledge of the assignment to potentially make a better decision. In our analysis, we will measure the conference's performance towards its goal in terms of two types of errors: 
\begin{enumerate}[label=(\alph*)]
    \item {\it Per-instance error}: For any given $\scorematrix=[\score_1, \score_2]$, the per-instance error of the conference is defined as $\Econference([\score_1, \score_2]) := \Pr(\text{conference accepts lower-quality paper} \mid \scorematrix=[\score_1, \score_2])$. 
    \item {\it Average-case error}: The average-case error of the conference is the per-instance error averaged over the distribution of the scores: $\int_{\score_1} \int_{\score_2} \Econference([\score_1, \score_2]) \jointscorepdf_{\scorematrix}([\score_1, \score_2])$ where $\jointscorepdf_{\scorematrix}$ is the p.d.f.\ of the joint distribution of $\scorematrix=[\score_1, \score_2]$. %\ns{the distrbution of the revewer score is also denoted by $f$.Is it the same $f$ as this? If not, this can be confusing}
\end{enumerate}

In conjunction with the goal of minimizing the error, the conference must also ensure that information about which reviewer reviewed which paper is not leaked.

\paragraph{Privacy.} 
We assume that the protocols followed by the conference are public. 
A challenge for the conference is that performing calibration may leak information about the assignment. As a simple example, suppose that reviewer $1$ is known to be strict and reviewer $2$ is known to be lenient. Suppose that paper $1$ is reviewed by reviewer $1$ and paper $2$ by reviewer $2$. Suppose paper $2$ receives a higher score than paper $1$, but the conference decides to accept paper $1$ after performing calibration. This decision leaks information that paper $2$ was reviewed by the lenient reviewer, that is, by reviewer $2$. Note that this issue of compromise of privacy arises whether or not the reviewer miscalibration functions are known to the conference.

To formalize the notion of privacy, we assume an adversary in the process. The goal of the adversary is to guess the assignment. In addition to knowing the scores received by both papers, the miscalibration functions of both reviewers, the noise distributions, and the final decision of the conference, the adversary also knows the calibration strategy used by the conference to make the decision. 

The adversary does not know the assignment, and aims to guess the assignment. We consider an adversary with no additional information, in which case, we assume it predicts the assignment via maximum a posteriori (MAP) estimation. Formally, if the conference decides to accept paper $P \in \{1,2\}$, then the adversary computes: 
\begin{align*}
\argmax_{\assignment \in \{ \assignment_1, \assignment_2\}} \Pr(\assignmentrv = \assignment\ |\ \scorematrix = [\score_1, \score_2], \text{ paper $P$ accepted by the conference}),
\end{align*}
where $\assignmentrv$ is the random variable representing the assignment. We make no assumptions on the computational power of the adversary and aim to guarantee privacy assuming they can compute the aforementioned argmax. 

As in the case of the conference's error, we also measure the error of the adversary in two ways:
\begin{enumerate}[label=(\alph*)]
\item {\it Per-instance error}: For any given $\scorematrix=[\score_1, \score_2]$, the per-instance error of the adversary is defined as $\Eadversary([\score_1, \score_2]) := \Pr(\text{adversary guesses wrong assignment} \mid \scorematrix=[\score_1, \score_2])$. 
\item {\it Average-case error:} The average-case error of the adversary is the per-instance error averaged over the distribution of the scores: $\int_{\score_1} \int_{\score_2} \Eadversary([\score_1, \score_2])  \jointscorepdf_{\scorematrix}([\score_1, \score_2])$ where $\jointscorepdf_{\scorematrix}$ is the p.d.f.\ of the joint distribution of $\scorematrix=[\score_1, \score_2]$. 
\end{enumerate}

\paragraph{Goal.} 
Our goal is to design methods to decide which paper to accept in a manner that simultaneously minimizes the conference's error and maximizes the adversary's error. The methods will inherently rely on calibrating reviewer decisions to accept the better paper, and hence we sometimes refer to them as the calibration strategy. 

The two aforementioned objectives may conflict with one another: a decision that reduces the chances of accepting the lower quality paper via calibration can also leak more information about the assignment. In this work, we thus establish the Pareto frontier of this tradeoff. We define the Pareto frontier as the set of all points of the (conference's error, adversary's error) tradeoff such that the adversary's error cannot be increased without increasing the conference's error. We call a calibration strategy Pareto optimal if for any given threshold on conference's error, it maximizes the adversary's error while ensuring that the conference's error does not exceed the given threshold.

%%%%%%%%%%%%%%%%%%%%%%%%%%%%%%%%%%%%

\section{Main Results}

In what follows, we present results for two settings: (1) a noiseless setting, where the noise in the reviewer-provided scores is zero; and (2) a noisy setting, where the noise in a reviewer score has a positive variance. We begin by a few preliminaries which we subsequently use to derive and present our main results. 

\subsection{Preliminaries}
We now formalize the calibration strategies that a conference can follow in a general form, and then derive a specific form that can be used without loss of optimality. Our subsequent results will then use this form of the calibration strategies. 

At a high level, the calibration strategies introduce a certain amount of randomness in the acceptance decisions. In the example in the `privacy' paragraph earlier in this section, suppose the conference does the calibration, and then tosses a coin. With probability 0.9, it accepts the paper it thinks is better and otherwise it accepts the other paper. This randomness ensures that an adversary who observes that paper 1 is accepted cannot be certain that paper 1 was reviewed by reviewer 1, due to the possibility that paper 1 was reviewed by the lenient reviewer 2 but was still accepted due to the randomness. However, due to the randomness introduced, the conference incurs an error in terms of accepting the paper which it thought was actually better. There is thus a tradeoff between the conference's error and the adversary's error, and our goal is to design calibration strategies that are optimal with respect to this tradeoff.

Let us now formalize the notion of a calibration strategy. The conference observes the scores $\scorematrix = [\score_1, \score_2]$ and the assignment $\assignment$. Given these values, a generic calibration strategy is specified by a function $\probfunction: \scorematrix \times \assignment \to [0,1]$ --- the conference accepts accept paper $1$ with probability $\probfunction(\scorematrix, \assignment)$ and accepts paper $2$ otherwise. Note that the function $\probfunction$ is publicly known but its realization is known only to the conference. For any function $\probfunction$ used by the conference, the conference's error is then given by
\begin{align*}
\Econference(\scorematrix, \assignment) &= 
\Big((1-\probfunction(\scorematrix, \assignment_1))\Pr(\assignmentrv = \assignment_1 | \quality_1 > \quality_2, \scorematrix) + (1-\probfunction(\scorematrix, \assignment_2))\Pr(\assignmentrv = \assignment_2 | \quality_1 > \quality_2, \scorematrix)\Big) \Pr(\quality_1 > \quality_2 | \scorematrix ) \\ 
& \qquad + \Big(\probfunction(\scorematrix, \assignment_1) \Pr(\assignmentrv = \assignment_1 | \quality_1 < \quality_2, \scorematrix) + \probfunction(\scorematrix, \assignment_2) \Pr(\assignmentrv = \assignment_2 | \quality_1 < \quality_2, \scorematrix)\Big) \Pr(\quality_1 < \quality_2 | \scorematrix).
\end{align*}

Having specified this general form of calibration strategy, we now discuss a specific variant. If one did not care about the privacy, then the conference's error can be minimized via maximum a posteriori (MAP) estimation: given scores $\scorematrix$ and the assignment $\assignment$, the conference accepts paper $1$ if $\Pr(\quality_1 > \quality_2 | \scorematrix, \assignment) > 0.5$ and accepts paper $2$ otherwise (breaking ties uniformly at random). Now under our scenario also involving privacy, consider the following class of calibration strategies. The strategy is governed by a function $\calibratefunction: \scorematrix \times \assignment \to [0,1]$. Given $\scorematrix = [\score_1, \score_2]$ and the assignment $\assignment$:
\begin{itemize}
    \item With probability $\calibratefunction(\scorematrix, \assignment)$, the conference executes MAP estimation under scores $\scorematrix$ and the (true) assignment $\assignment$, 
    \item otherwise (with probability $1 - \calibratefunction(\scorematrix, \assignment)$), the conference executes MAP estimation under scores $\scorematrix$ and the {\bf wrong} assignment $\{\assignment_1,\assignment_2\}\backslash \assignment$.
\end{itemize}
As before, we assume that function $\calibratefunction$ is known publicly but its realization or the random bits are not.

A calibration strategy is {\bf Pareto optimal} if any other strategy that incurs a lower conference error must also induce a lower error of the adversary, and any other strategy that induces a higher error of the adversary must also incur a higher conference error. The {\bf Pareto frontier} is the set of all (conference error, adversary error) pairs achieved by Pareto optimal strategies.
The following proposition states that without loss of optimality, one can restrict attention to the class of strategies specified by functions $\calibratefunction$. 
\begin{proposition}
\label{proposition:calibrationfunction}
For any values of error of the conference and error of the adversary $(\Econference, \Eadversary)$ achieved by a calibration strategy $\probfunction$, there exists a function $\calibratefunction$ such that under $\calibratefunction$, the error of the conference is no larger than $\Econference$ and the error of the adversary is no smaller than $\Eadversary$.
\end{proposition}
The proof of this proposition is available in Appendix~\ref{sec:proof:proposition:calibrationfunction}. Hence, without loss of Pareto optimality, any generic calibration strategy $\probfunction$ can be replaced with a strategy involving the calibration function $\calibratefunction$. 
Thus, in the sequel we restrict attention to calibration strategies using function $\calibratefunction$.

\subsection{Noiseless Setting}\label{sec:results-noiseless}
We first study the noiseless setting where the noise in the reviewer-provided scores is zero, that is, where $\noise_1=\noise_2=0$. Observe that in this setting the conference can obtain the true qualities of the papers from the scores by inverting the reviewer functions. We first explicitly characterize the Pareto frontier for per-instance errors of the conference and the adversary. Based on this characterization, we then design Pareto optimal strategies for conference calibration with respect to the per-instance error and the average-case error.

\subsubsection{Pareto Frontier for Per-Instance Errors}

In the following theorem, we present the main result of this section establishing the Pareto frontier for per-instance errors in the noiseless setting.

\begin{theorem}
\label{thm:noiselessPareto}
Consider the peer-review system in the noiseless setting.
The Pareto frontier of (per-instance error of the conference, per-instance error of the adversary) with scores $\scorematrix = [\score_1, \score_2]$ is given as follows.
\begin{enumerate}[label=(\arabic*),leftmargin=*]
\item If $\score_1 \ge \max \{\reviewerfunction_2(\reviewerfunction_1^{-1}(\score_2)), \reviewerfunction_1(\reviewerfunction_2^{-1}(\score_2))\}$ or $\score_1 \le \min \{\reviewerfunction_2(\reviewerfunction_1^{-1}(\score_2)), \reviewerfunction_1(\reviewerfunction_2^{-1}(\score_2))\}$, then the Pareto frontier consists of a single point $(0,\frac{\min\{\reviewerfunctionpdf_1(\score_1)\reviewerfunctionpdf_2(\score_2), \reviewerfunctionpdf_2(\score_1)\reviewerfunctionpdf_1(\score_2)\}}{\reviewerfunctionpdf_1(\score_1)\reviewerfunctionpdf_2(\score_2) + \reviewerfunctionpdf_2(\score_1)\reviewerfunctionpdf_1(\score_2)})$.
\item Otherwise, if $\min \{\reviewerfunction_2(\reviewerfunction_1^{-1}(\score_2)), \reviewerfunction_1(\reviewerfunction_2^{-1}(\score_2))\} < \score_1 < \max \{\reviewerfunction_2(\reviewerfunction_1^{-1}(\score_2)), \reviewerfunction_1(\reviewerfunction_2^{-1}(\score_2))\}$, then the Pareto frontier of conference error and adversary error is a line segment of slope $1$ starting from the origin $(0,0)$ to $\left(\frac{\min\{\reviewerfunctionpdf_1(\score_1)\reviewerfunctionpdf_2(\score_2), \reviewerfunctionpdf_2(\score_1)\reviewerfunctionpdf_1(\score_2)\}}{\reviewerfunctionpdf_1(\score_1)\reviewerfunctionpdf_2(\score_2) + \reviewerfunctionpdf_2(\score_1)\reviewerfunctionpdf_1(\score_2)},\frac{\min\{\reviewerfunctionpdf_1(\score_1)\reviewerfunctionpdf_2(\score_2), \reviewerfunctionpdf_2(\score_1)\reviewerfunctionpdf_1(\score_2)\}}{\reviewerfunctionpdf_1(\score_1)\reviewerfunctionpdf_2(\score_2) + \reviewerfunctionpdf_2(\score_1)\reviewerfunctionpdf_1(\score_2)}\right)$.
\end{enumerate}
\end{theorem}

\begin{figure}[t]
\centering
  \includegraphics[scale=0.3]{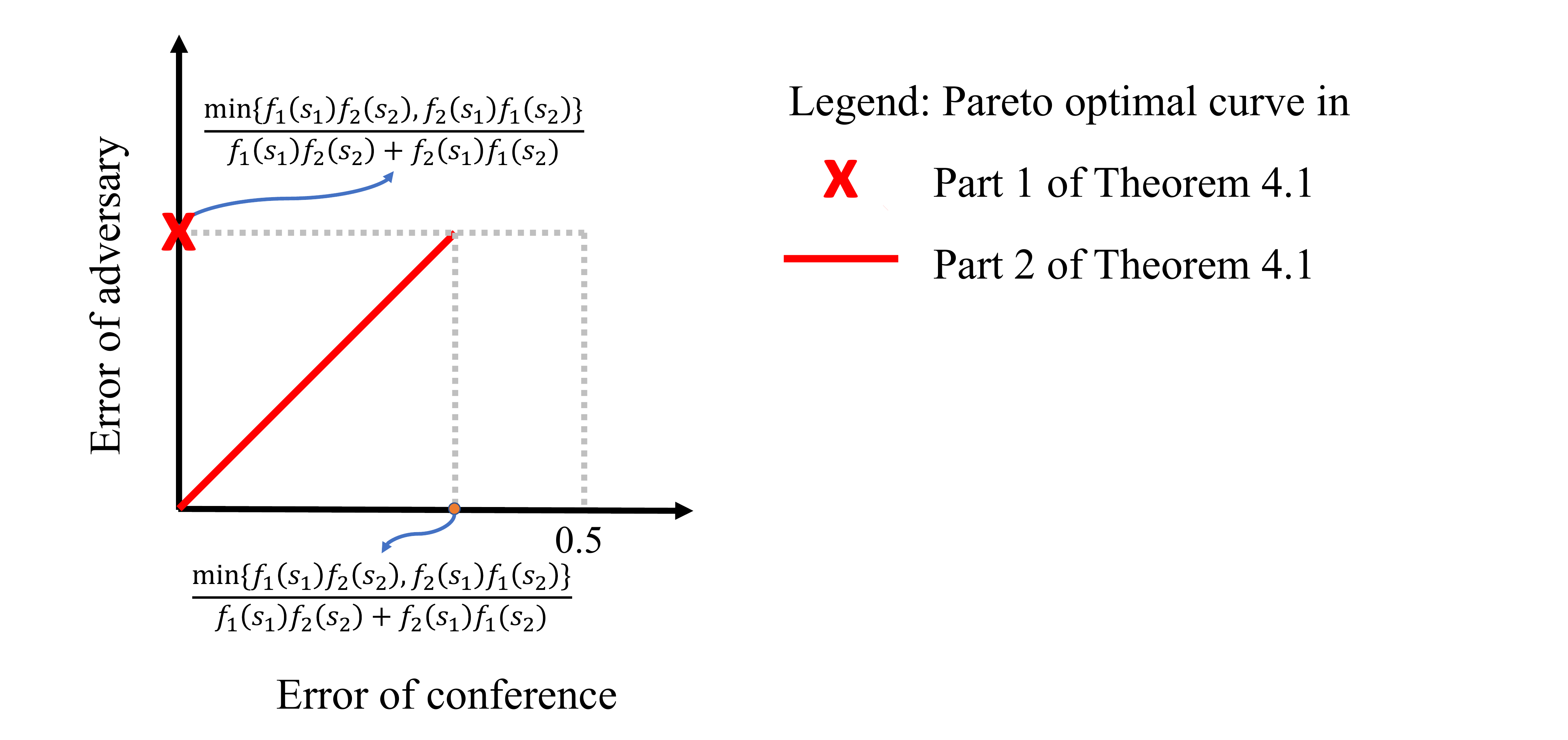}
  \caption{\label{fig:noiselessPareto}Pareto frontier for per-instance errors in the noiseless setting. }
\end{figure}

The proof of Theorem~\ref{thm:noiselessPareto} is provided in Appendix~\ref{sec:proof:thm:noiselessPareto}. The Pareto frontier established in Theorem~\ref{thm:noiselessPareto} is illustrated in Figure~\ref{fig:noiselessPareto}. 

We now unpack the result of Theorem~\ref{thm:noiselessPareto}, beginning with part $(1)$. Recall that in this noiseless setting, given scores $\scorematrix = [\score_1, \score_2]$ and knowing the reviewers' miscalibration functions, the conference can estimate the qualities of papers under each assignment. We use $\qualityest_i \in \mathbb{R}$ to denote the estimated quality of paper $i$. If the conference estimates the qualities assuming that $\assignment_1$ was the actual assignment, we get $\qualityest_1 = \reviewerfunction_1^{-1}(\score_1)$ and $\qualityest_2 = \reviewerfunction_2^{-1}(\score_2)$. If the conference estimates the qualities assuming that $\assignment_2$ was the actual assignment, we get $\qualityest_1 = \reviewerfunction_2^{-1}(\score_1)$ and $\qualityest_2 = \reviewerfunction_1^{-1}(\score_2)$. If $\score_1 \ge \max \{\reviewerfunction_2(\reviewerfunction_1^{-1}(\score_2)), \reviewerfunction_1(\reviewerfunction_2^{-1}(\score_2))\}$, then $\qualityest_1 \ge \qualityest_2$ under both assignments (and hence paper 1 should be accepted). Similarly, if $\score_1 \le \min \{\reviewerfunction_2(\reviewerfunction_1^{-1}(\score_2)), \reviewerfunction_1(\reviewerfunction_2^{-1}(\score_2))\}$, then $\qualityest_1 \le \qualityest_2$ under both assignments (and hence paper 2 should be accepted). Therefore, under the condition of part $(1)$ of the theorem, the same paper has higher estimated quality under both assignments, and hence that paper will be accepted irrespective of the function $\calibratefunction$. Thus, under this condition, the Pareto optimal curve comprises just a single point where the conference has zero error, and the adversary obtains no additional information from the acceptance decision as compared to the scores $\scorematrix = [\score_1, \score_2]$. The error of the adversary is $\frac{\min\{\reviewerfunctionpdf_1(\score_1)\reviewerfunctionpdf_2(\score_2), \reviewerfunctionpdf_2(\score_1)\reviewerfunctionpdf_1(\score_2)\}}{\reviewerfunctionpdf_1(\score_1)\reviewerfunctionpdf_2(\score_2) + \reviewerfunctionpdf_2(\score_1)\reviewerfunctionpdf_1(\score_2)}$ when it guesses the assignment using only the scores and not the decision.

Let us now discuss part $(2)$ of Theorem~\ref{thm:noiselessPareto}. For scores $\scorematrix = [\score_1, \score_2]$ that do not satisfy the condition in part $(1)$, the conference would accept different papers when performing MAP calibration under the two possible assignments. In this case, the function $\calibratefunction$ does influence the outcomes. The Pareto frontier includes the origin since the conference can ensure zero error in this noiseless setting, but this zero-error acceptance decision will also perfectly reveal the assignment to the adversary since the zero-error decisions would be different under the two assignments. Then in the proof, we find the \emph{maximum} per-instance error of the adversary given per-instance error of the conference. We find that the adversary's error no longer increases if the conference is allowed an error greater than $\frac{\min\{\reviewerfunctionpdf_1(\score_1)\reviewerfunctionpdf_2(\score_2), \reviewerfunctionpdf_2(\score_1)\reviewerfunctionpdf_1(\score_2)\}}{\reviewerfunctionpdf_1(\score_1)\reviewerfunctionpdf_2(\score_2) + \reviewerfunctionpdf_2(\score_1)\reviewerfunctionpdf_1(\score_2)}$. At this value of the conference's error, the maximum per-instance error of the adversary is also $\frac{\min\{\reviewerfunctionpdf_1(\score_1)\reviewerfunctionpdf_2(\score_2), \reviewerfunctionpdf_2(\score_1)\reviewerfunctionpdf_1(\score_2)\}}{\reviewerfunctionpdf_1(\score_1)\reviewerfunctionpdf_2(\score_2) + \reviewerfunctionpdf_2(\score_1)\reviewerfunctionpdf_1(\score_2)}$. We further show in the proof of the theorem that the Pareto frontier is precisely the line segment joining these two points. 
Therefore, the Pareto frontier for scores satisfy the condition is a line segment from the origin to the point $\left(\frac{\min\{\reviewerfunctionpdf_1(\score_1)\reviewerfunctionpdf_2(\score_2), \reviewerfunctionpdf_2(\score_1)\reviewerfunctionpdf_1(\score_2)\}}{\reviewerfunctionpdf_1(\score_1)\reviewerfunctionpdf_2(\score_2) + \reviewerfunctionpdf_2(\score_1)\reviewerfunctionpdf_1(\score_2)},\frac{\min\{\reviewerfunctionpdf_1(\score_1)\reviewerfunctionpdf_2(\score_2), \reviewerfunctionpdf_2(\score_1)\reviewerfunctionpdf_1(\score_2)\}}{\reviewerfunctionpdf_1(\score_1)\reviewerfunctionpdf_2(\score_2) + \reviewerfunctionpdf_2(\score_1)\reviewerfunctionpdf_1(\score_2)}\right)$ as shown in Figure~\ref{fig:noiselessPareto}.

\subsubsection{Optimal Calibration Strategy under Per-Instance Errors}\label{SecAlgoPerNoiseless}

In the previous section, we characterized the fundamental tradeoff between the conference's per-instance error and the adversary's per-instance error through the Pareto frontier. In this section, we design an explicit calibration strategy that achieves per-instance errors on the Pareto frontier, and is thus optimal for per-instance errors.

Since $\scorematrix$ is a fixed realization in the analysis of per-instance errors, to simplify the notation we define
\begin{align*}
    \calibrateprob_1 = \calibratefunction(\scorematrix, \assignment_1) \qquad \text{and}\qquad 
    \calibrateprob_2 = \calibratefunction(\scorematrix, \assignment_2).
\end{align*}
Under this notation, $\calibrateprob_1$ is the probability with which the conference calibrates under the true assignment when the true assignment is $\assignment_1$, and $\calibrateprob_2$ is the probability with which the conference calibrates under the true assignment when the true assignment is $\assignment_2$. Therefore, from Proposition~\ref{proposition:calibrationfunction}, given the maximum allowable error of the conference $\Econference$, our goal is to find values of $\calibrateprob_1$ and $\calibrateprob_2$ that are Pareto optimal. We present our proposed algorithm for this setting as Algorithm~\ref{alg:noiselessworste}.

\begin{theorem}
\label{thm:noiselessInstance}
The calibration algorithm described in Algorithm~\ref{alg:noiselessworste} ensures the maximum per-instance error of the adversary for any given value of the maximum allowable per-instance error $\Econference([\score_1, \score_2])$ for the conference, and is hence Pareto optimal. 
\end{theorem}

\begin{algorithm}[tb]
   \caption{Conference calibration with per-instance error in the noiseless setting}
   \label{alg:noiselessworste}
\begin{algorithmic}
\Indm
   \STATE {\bfseries Input:} scores $\scorematrix = [\score_1, \score_2]$, maximum allowable per-instance error of the conference $\Econference([\score_1, \score_2])$
   \IF {$\score_1 \ge \max \{\reviewerfunction_1(\reviewerfunction_2^{-1}(\score_2)), \reviewerfunction_2(\reviewerfunction_1^{-1}(\score_2))\}$}
   \STATE {accept paper 1}
   \ELSIF {$\score_1 \le \min \{\reviewerfunction_1(\reviewerfunction_2^{-1}(\score_2)), \reviewerfunction_2(\reviewerfunction_1^{-1}(\score_2))\}$}
   \STATE {accept paper 2}
    \ELSIF{$\Econference([\score_1, \score_2]) \ge \frac{\min \left \{ \reviewerfunctionpdf_1(\score_1)\reviewerfunctionpdf_2(\score_2), \reviewerfunctionpdf_2(\score_1)\reviewerfunctionpdf_1(\score_2) \right\}}{\reviewerfunctionpdf_1(\score_1)\reviewerfunctionpdf_2(\score_2) + \reviewerfunctionpdf_2(\score_1)\reviewerfunctionpdf_1(\score_2)}$}
   \STATE {choose $\calibrateprob_1, \calibrateprob_2 \in [0,1]$ such that $ \reviewerfunctionpdf_1(\score_1)\reviewerfunctionpdf_2(\score_2) \calibrateprob_1 + \reviewerfunctionpdf_2(\score_1)\reviewerfunctionpdf_1(\score_2) \calibrateprob_2 = \max \left \{ \reviewerfunctionpdf_1(\score_1)\reviewerfunctionpdf_2(\score_2), \reviewerfunctionpdf_2(\score_1)\reviewerfunctionpdf_1(\score_2) \right\}$ }
   \ELSE
   \STATE {choose $\calibrateprob_1, \calibrateprob_2 \in [0,1]$ such that $\Econference([\score_1, \score_2]) = 1 - \frac{\reviewerfunctionpdf_1(\score_1)\reviewerfunctionpdf_2(\score_2)\calibrateprob_1 + \reviewerfunctionpdf_2(\score_1)\reviewerfunctionpdf_1(\score_2)\calibrateprob_2}{\reviewerfunctionpdf_1(\score_1)\reviewerfunctionpdf_2(\score_2) + \reviewerfunctionpdf_2(\score_1)\reviewerfunctionpdf_1(\score_2)}$ }
   \ENDIF
\end{algorithmic}
\end{algorithm}

The proof of Theorem~\ref{thm:noiselessInstance} is presented in Appendix~\ref{sec:proof:thm:noiselessInstance}. If $\score_1 \ge \max \{\reviewerfunction_1(\reviewerfunction_2^{-1}(\score_2)), \reviewerfunction_2(\reviewerfunction_1^{-1}(\score_2))\}$ or $\score_1 \le \min \{\reviewerfunction_1(\reviewerfunction_2^{-1}(\score_2)), \reviewerfunction_2(\reviewerfunction_1^{-1}(\score_2))\}$, we are in part $(1)$ of Theorem~\ref{thm:noiselessPareto}. Under scores that satisfy this condition, the conference is guaranteed to accept the higher-quality paper and thus has zero error. The error of the adversary is also fixed because the adversary makes its guess based on the scores only.

Otherwise, for a Pareto optimal calibration strategy, the errors of the conference and the adversary should stay on the Pareto frontier as in Figure~\ref{fig:noiselessPareto}. When $\Econference([\score_1, \score_2]) \ge \frac{\min \left \{ \reviewerfunctionpdf_1(\score_1)\reviewerfunctionpdf_2(\score_2), \reviewerfunctionpdf_2(\score_1)\reviewerfunctionpdf_1(\score_2) \right\}}{\reviewerfunctionpdf_1(\score_1)\reviewerfunctionpdf_2(\score_2) + \reviewerfunctionpdf_2(\score_1)\reviewerfunctionpdf_1(\score_2)}$, the conference should choose $\calibrateprob_1$ and $\calibrateprob_2$ such that its per-instance error is $\frac{\min \left \{ \reviewerfunctionpdf_1(\score_1)\reviewerfunctionpdf_2(\score_2), \reviewerfunctionpdf_2(\score_1)\reviewerfunctionpdf_1(\score_2) \right\}}{\reviewerfunctionpdf_1(\score_1)\reviewerfunctionpdf_2(\score_2) + \reviewerfunctionpdf_2(\score_1)\reviewerfunctionpdf_1(\score_2)}$ since further sacrifice of accuracy cannot increase the per-instance error of the adversary as indicated by the Pareto frontier. On the other hand, if $\Econference([\score_1, \score_2]) < \frac{\min \left \{ \reviewerfunctionpdf_1(\score_1)\reviewerfunctionpdf_2(\score_2), \reviewerfunctionpdf_2(\score_1)\reviewerfunctionpdf_1(\score_2) \right\}}{\reviewerfunctionpdf_1(\score_1)\reviewerfunctionpdf_2(\score_2) + \reviewerfunctionpdf_2(\score_1)\reviewerfunctionpdf_1(\score_2)}$, the conference can choose $\calibrateprob_1$ and $\calibrateprob_2$ that yields the maximum allowable per-instance error. Since $\Econference([\score_1, \score_2]) < \frac{\min \left \{ \reviewerfunctionpdf_1(\score_1)\reviewerfunctionpdf_2(\score_2), \reviewerfunctionpdf_2(\score_1)\reviewerfunctionpdf_1(\score_2) \right\}}{\reviewerfunctionpdf_1(\score_1)\reviewerfunctionpdf_2(\score_2) + \reviewerfunctionpdf_2(\score_1)\reviewerfunctionpdf_1(\score_2)}$ and $\Econference([\score_1, \score_2]) = 1 - \frac{\reviewerfunctionpdf_1(\score_1)\reviewerfunctionpdf_2(\score_2)\calibrateprob_1 + \reviewerfunctionpdf_2(\score_1)\reviewerfunctionpdf_1(\score_2)\calibrateprob_2}{\reviewerfunctionpdf_1(\score_1)\reviewerfunctionpdf_2(\score_2) + \reviewerfunctionpdf_2(\score_1)\reviewerfunctionpdf_1(\score_2)}$, we can conclude that $\reviewerfunctionpdf_1(\score_1)\reviewerfunctionpdf_2(\score_2) \calibrateprob_1 + \reviewerfunctionpdf_2(\score_1)\reviewerfunctionpdf_1(\score_2) \calibrateprob_2 > \max \left \{ \reviewerfunctionpdf_1(\score_1)\reviewerfunctionpdf_2(\score_2), \reviewerfunctionpdf_2(\score_1)\reviewerfunctionpdf_1(\score_2) \right\}$ and the adversary has the same per-instance error under this condition. Therefore, in Algorithm~\ref{alg:noiselessworste}, the error of the adversary is the same as the error of the conference and the errors are always on the Pareto frontier.

\subsubsection{Optimal Calibration Strategy under Average-case Error}

In the previous section, we designed an optimal strategy under per-instance errors. In this section, we design a calibration strategy that achieves optimal average-case errors for the conference with respect to the average-case error of the adversary. Unlike for per-instance error, we do not have a closed form expression for average-case error. We present our proposed algorithm as Algorithm~\ref{alg:noiselessaveragee}. We now present our main result of this subsection, following which we discuss more details of the algorithm this result.

\begin{theorem}
\label{thm:noiselessAverage}
The calibration algorithm described in Algorithm~\ref{alg:noiselessaveragee} ensures the maximum average-case error of the adversary for any given value of the maximum allowable average-case error $\Econference$ for the conference, and is hence Pareto optimal. 
\end{theorem}

\begin{algorithm}[tb]
   \caption{Conference calibration with average-case error in the noiseless setting}
   \label{alg:noiselessaveragee}
\begin{algorithmic}
\Indm
   \STATE {\bfseries Input:} maximum allowable average-case error of the conference $\Econference$
   \STATE{Let $\Estrategy$ = error of the conference for adopting Algorithm~\ref{alg:noiselessworste} with $\Econference([\score_1, \score_2]) = 1$ for all $[\score_1, \score_2]$}
   %Algorithm~\ref{alg:noiselessstrategy}}
   \IF {$\Econference > \Estrategy$}
   \STATE {the desired conference error is Pareto inefficient and operate at $\Econference = \Estrategy$}
   \ELSIF {$\Econference = \Estrategy$}
   \STATE {run Algorithm~\ref{alg:noiselessworste} with $\Econference([\score_1, \score_2]) = 1$}
   %Algorithm~\ref{alg:noiselessstrategy}}
   \ELSIF {$\Econference < \Estrategy$}
   \STATE {toss a coin that has probability $\frac{\Econference}{\Estrategy}$ of head}
   \IF {coin toss outcome is head}
   \STATE {run Algorithm~\ref{alg:noiselessworste} with $\Econference([\score_1, \score_2]) = 1$}
   %Algorithm~\ref{alg:noiselessstrategy}}
   \ELSE
   \STATE {calibrate under true assignment}
   \ENDIF
   \ENDIF
\end{algorithmic}
\end{algorithm}

The proof of Theorem~\ref{thm:noiselessAverage} is provided in Appendix~\ref{sec:proof:thm:noiselessAverage}. 

In Algorithm~\ref{alg:noiselessaveragee}, running Algorithm~\ref{alg:noiselessworste} with $\Econference([\score_1, \score_2]) = 1$ is a strategy that yields no error when the same paper has higher estimated quality under both assignments and otherwise, error of the conference equals error of the adversary. Moreover, both per-instance error of the conference and per-instance error of the adversary is $\frac{\min \left \{ \reviewerfunctionpdf_1(\score_1)\reviewerfunctionpdf_2(\score_2), \reviewerfunctionpdf_2(\score_1)\reviewerfunctionpdf_1(\score_2) \right\}}{\reviewerfunctionpdf_1(\score_1)\reviewerfunctionpdf_2(\score_2) + \reviewerfunctionpdf_2(\score_1)\reviewerfunctionpdf_1(\score_2)}$. That is, the maximum per-instance error for the adversary. Thus, this strategy is Pareto optimal for any score pair and is also Pareto optimal under its average-error since the error of the adversary is maximized under such average-error of the conference.

In proof for optimality of Algorithm~\ref{alg:noiselessaveragee}, we take advantage of the fact that the Pareto frontier is either a point where the conference has no error or an increasing line with slope 1. Under this fact, the optimal average-case error of the conference is where the conference has zero error when the adversary guesses the assignment based on the scores only and has the same error as the adversary otherwise. Therefore, Algorithm~\ref{alg:noiselessaveragee} makes use of Algorithm~\ref{alg:noiselessworste} with $\Econference([\score_1, \score_2]) = 1$ and on average, the error of the conference and the adversary matches the Pareto optimality for the conference.

\subsection{Noisy Setting}
%\ww{Wenxin, can you make a pass to make sure this section follows a structure that's similar to that of Section~\ref{sec:results-noiseless}?}
We now study the noisy setting. We consider both reviewers' miscalibration functions $\reviewerfunction_1$ and $\reviewerfunction_2$ to be affine and both reviewers' noises $\noise_1$ and $\noise_2$ to be Gaussian. Furthermore, the distributions of the noise are the same for both reviewers with mean zero and some known variance $\sigma^2$. Formally, we assume: 
\begin{align*}
    &\reviewerfunction_1(\quality) = a_1  \quality + b_1,\quad
    \reviewerfunction_2(\quality) = a_2  \quality + b_2, \quad
    \noise_1 \sim N(0, \sigma^2),\quad \text{and}\quad
    \noise_2 \sim N(0, \sigma^2).
\end{align*}
As we will see below, the presence of noise makes the analysis much more complex, even when we assume affine miscalibration, as compared to the noiseless setting.

\subsubsection{Pareto Frontier for Per-Instance Errors}
\label{sec:noisyPareto}

We begin by establishing the Pareto frontier for per-instance errors in the noisy case. Let $\normalcdf$ denote the cumulative distribution function of the standard normal distribution. Also define notation $\Phi_1$ and $\Phi_2$ as:
\begin{subequations}
\label{align:phi}
\begin{align}
\label{align:phi1}
    &\Phi_1 = \normalcdf\left(\frac{a_2(a_1^2+\sigma^2)(\score_2-b_2)-a_1(a_2^2+\sigma^2)(\score_1-b_1)}{\sqrt{\sigma^2(a_1^2+a_2^2+2\sigma^2)(a_1^2+\sigma^2)(a_2^2+\sigma^2)}}\right)\\
\label{align:phi2}
    &\Phi_2 = \normalcdf\left(\frac{a_1(a_2^2 + \sigma^2)(\score_2-b_1)-a_2(a_1^2+\sigma^2)(\score_1-b_2)}{\sqrt{\sigma^2(a_1^2+a_2^2+2\sigma^2)(a_1^2+\sigma^2)(a_2^2+\sigma^2)}}\right).
    \end{align}
\end{subequations}

\begin{theorem}
\label{thm:noisyPareto}
Consider the peer-review system in the noisy setting.
The Pareto frontier of (per-instance error of the conference, per-instance error of the adversary) with scores $\scorematrix = [\score_1, \score_2]$ is as follows.
\begin{enumerate}[label=(\arabic*),leftmargin=*]
\item If $\score_1 \ge \max \left\{\frac{a_2(a_1^2+\sigma^2)(\score_2-b_2)}{a_1(a_2^2+\sigma^2)} + b_1, \frac{a_1(a_2^2 + \sigma^2)(\score_2-b_1)}{a_2(a_1^2+\sigma^2)}+b_2\right\}$ or $\score_1 \le \min \left\{\frac{a_2(a_1^2+\sigma^2)(\score_2-b_2)}{a_1(a_2^2+\sigma^2)} + b_1, \frac{a_1(a_2^2 + \sigma^2)(\score_2-b_1)}{a_2(a_1^2+\sigma^2)}+b_2\right\}$, then the Pareto frontier consists of a single point. 

Specifically, when $\score_1 \ge \max \left\{\frac{a_2(a_1^2+\sigma^2)(\score_2-b_2)}{a_1(a_2^2+\sigma^2)} + b_1, \frac{a_1(a_2^2 + \sigma^2)(\score_2-b_1)}{a_2(a_1^2+\sigma^2)}+b_2\right\}$, the Pareto frontier of conference error and adversary error is the point $\left (\frac{\reviewerfunctionpdf_1(\score_1)\reviewerfunctionpdf_2(\score_2) \Phi_1 + \reviewerfunctionpdf_2(\score_1)\reviewerfunctionpdf_1(\score_2) \Phi_2}{\reviewerfunctionpdf_1(\score_1)\reviewerfunctionpdf_2(\score_2) + \reviewerfunctionpdf_2(\score_1)\reviewerfunctionpdf_1(\score_2)}, \frac{\min\{\reviewerfunctionpdf_1(\score_1)\reviewerfunctionpdf_2(\score_2), \reviewerfunctionpdf_2(\score_1)\reviewerfunctionpdf_1(\score_2)\}}{\reviewerfunctionpdf_1(\score_1)\reviewerfunctionpdf_2(\score_2) + \reviewerfunctionpdf_2(\score_1)\reviewerfunctionpdf_1(\score_2)} \right)$. And similarly, when $\score_1 \le \min \left\{\frac{a_2(a_1^2+\sigma^2)(\score_2-b_2)}{a_1(a_2^2+\sigma^2)} + b_1, \frac{a_1(a_2^2 + \sigma^2)(\score_2-b_1)}{a_2(a_1^2+\sigma^2)}+b_2\right\}$, the Pareto frontier is the point $\left (\frac{\reviewerfunctionpdf_1(\score_1)\reviewerfunctionpdf_2(\score_2) (1-\Phi_1) + \reviewerfunctionpdf_2(\score_1)\reviewerfunctionpdf_1(\score_2) (1-\Phi_2)}{\reviewerfunctionpdf_1(\score_1)\reviewerfunctionpdf_2(\score_2) + \reviewerfunctionpdf_2(\score_1)\reviewerfunctionpdf_1(\score_2)}, \frac{\min\{\reviewerfunctionpdf_1(\score_1)\reviewerfunctionpdf_2(\score_2), \reviewerfunctionpdf_2(\score_1)\reviewerfunctionpdf_1(\score_2)\}}{\reviewerfunctionpdf_1(\score_1)\reviewerfunctionpdf_2(\score_2) + \reviewerfunctionpdf_2(\score_1)\reviewerfunctionpdf_1(\score_2)} \right)$.

\item If $\min \left\{\frac{a_2(a_1^2+\sigma^2)(\score_2-b_2)}{a_1(a_2^2+\sigma^2)} + b_1, \frac{a_1(a_2^2 + \sigma^2)(\score_2-b_1)}{a_2(a_1^2+\sigma^2)}+b_2\right\} < \score_1 < \max \left\{\frac{a_2(a_1^2+\sigma^2)(\score_2-b_2)}{a_1(a_2^2+\sigma^2)} + b_1, \frac{a_1(a_2^2 + \sigma^2)(\score_2-b_1)}{a_2(a_1^2+\sigma^2)}+b_2\right\}$, then the Pareto frontier is an increasing line.

\end{enumerate}
\end{theorem}

\begin{figure}[htp]
\centering
\begin{subfigure}[b]{0.5\textwidth}
  \centering
  \includegraphics[scale=0.3]{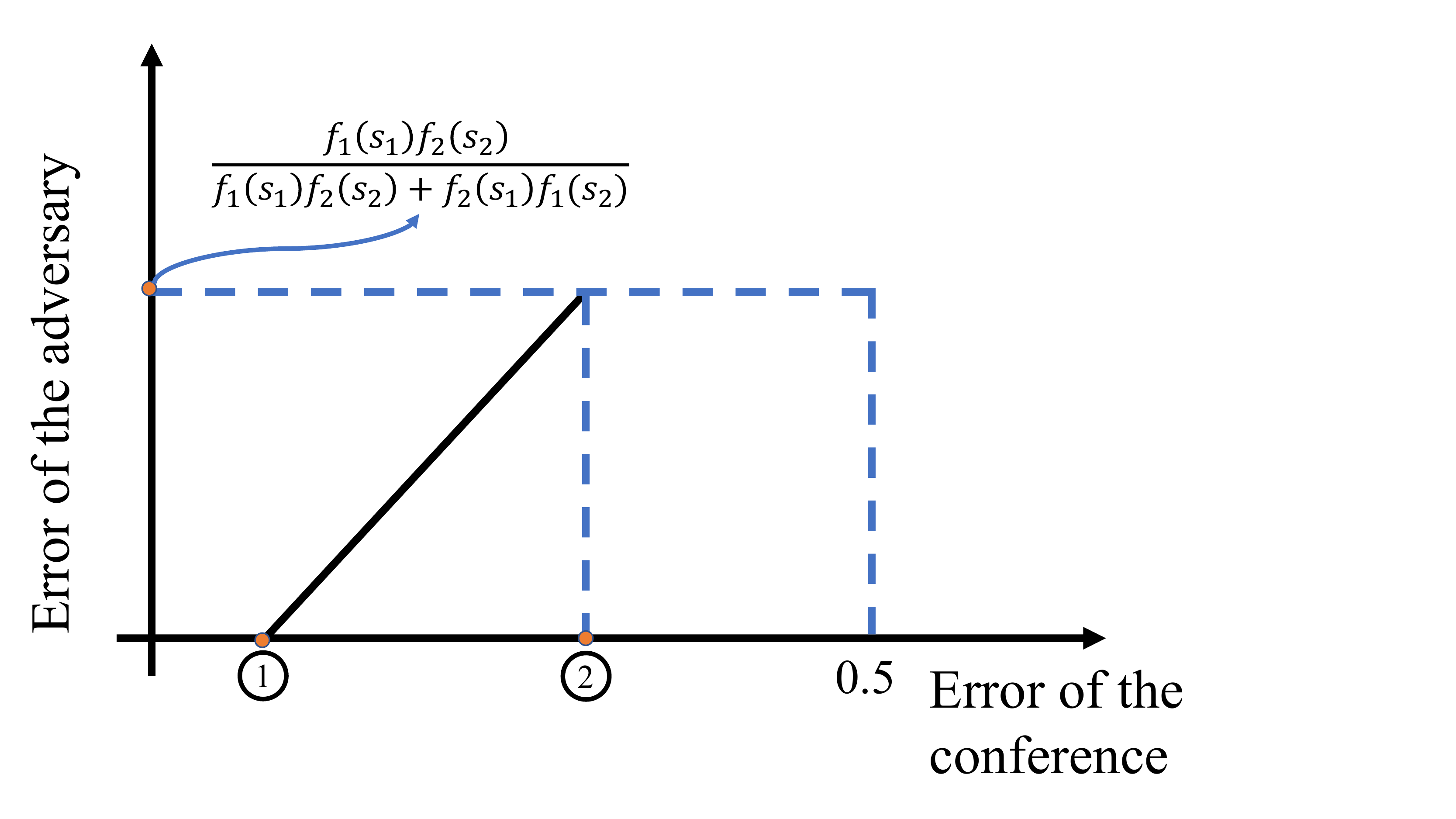}
  \end{subfigure}%\hspace{.09\textwidth}
\begin{subfigure}[b]{0.5\textwidth}
 \centering
 \includegraphics[scale=0.3]{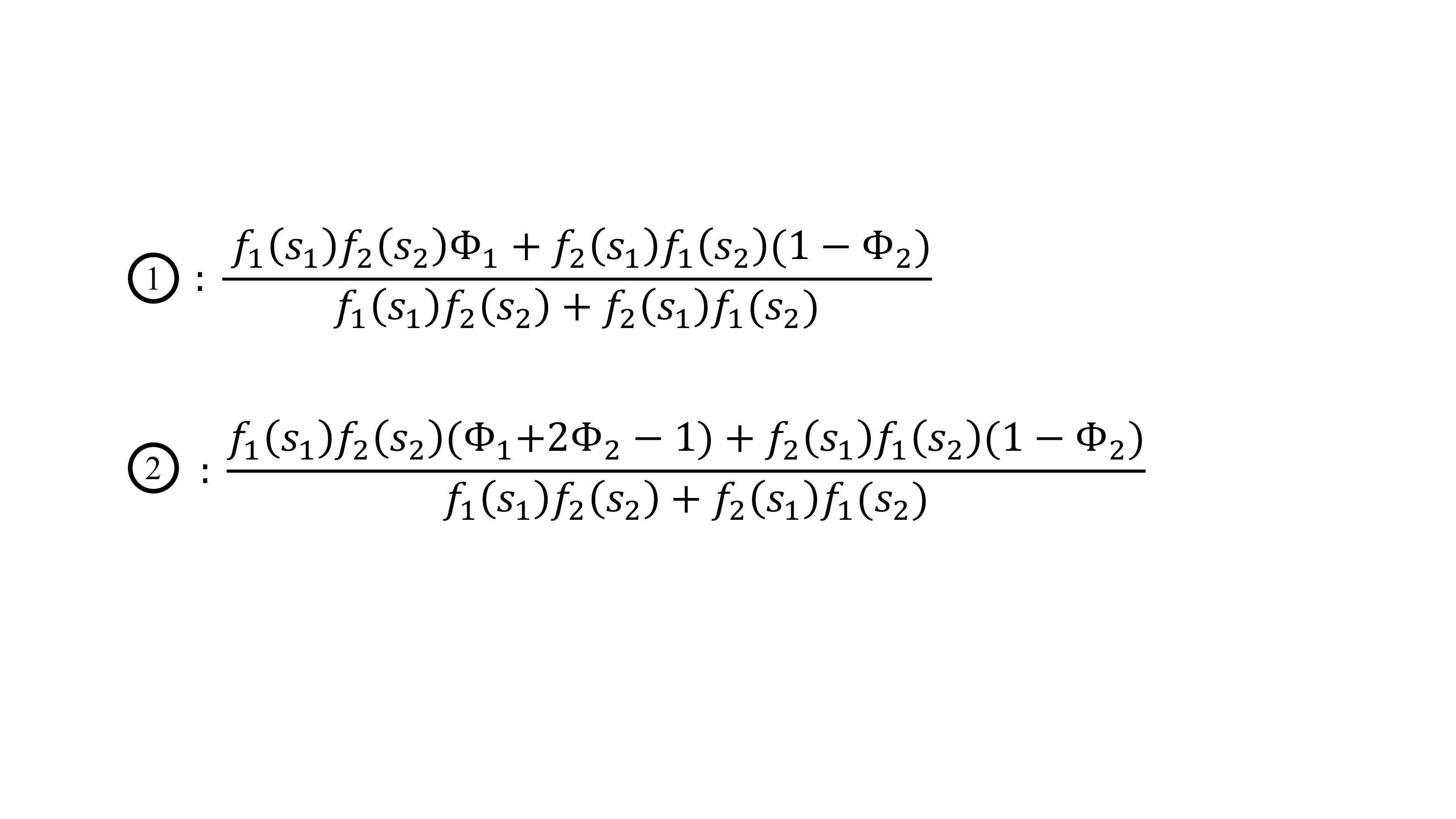}
\end{subfigure}
\caption{\label{fig:noisyPareto}A Pareto frontier in the noisy setting of part $(2)$ of Theorem~\ref{thm:noisyPareto} in the case that $\reviewerfunctionpdf_1(\score_1)\reviewerfunctionpdf_2(\score_2) < \reviewerfunctionpdf_2(\score_1)\reviewerfunctionpdf_1(\score_2)$ and $\Phi_1 < \frac{1}{2} < \Phi_2$ with $0 < \Phi_2 - \frac{1}{2} < \frac{1}{2} - \Phi_1$. The notations $\Phi_1$ and $\Phi_2$ are defined in~\eqref{align:phi}.}
\end{figure}

The proof of Theorem~\ref{thm:noisyPareto} is provided in Appendix~\ref{sec:proof:thm:noisyPareto}. We now unpack this result and specify precisely the Pareto frontier in both parts of the theorem.

Given scores $\scorematrix = [\score_1, \score_2]$ and knowing the reviewers' miscalibration functions, the conference can estimate the qualities of papers under each assignment. Under assignment $\assignment_1$, we have $\Pr(\quality_1 > \quality_2 | \assignmentrv = \assignment_1, \scorematrix = [\score_1, \score_2]) = 1-\Phi_1$. And under assignment $\assignment_2$, we have $\Pr(\quality_1 > \quality_2 | \assignmentrv = \assignment_2, \scorematrix = [\score_1, \score_2]) = 1-\Phi_2$.

Let us now consider part $(1)$ of Theorem~\ref{thm:noisyPareto}. If the condition specified in the statement of the theorem is satisfied, then we know that either $(\Phi_1 \le \frac{1}{2}, \Phi_2 \le \frac{1}{2})$ in which case paper 1 has a higher estimated quality under either assignment, or $(\Phi_1 \ge \frac{1}{2}, \Phi_2 \ge \frac{1}{2})$ in which paper 2 has a higher estimated quality under either assignment. Thus, if the condition in part $(1)$ is met, the same paper has higher estimated quality under both assignments and hence the decision does not depend on $\calibratefunction$. Thus, for such pair of scores, the Pareto optimal situation is where the conference has minimum error and the adversary guesses the assignment based on the scores alone.

Let us now move to part $(2)$ of Theorem~\ref{thm:noisyPareto}, and consider parameters that satisfy the condition stated therein. Under this condition, the conference would accept different papers by calibrating under different assignments, and the function $\calibratefunction$ needs to be carefully designed. 
We study the Pareto frontier for scores in the range.

We consider a specific case where $\reviewerfunctionpdf_1(\score_1)\reviewerfunctionpdf_2(\score_2) < \reviewerfunctionpdf_2(\score_1)\reviewerfunctionpdf_1(\score_2)$, $\Phi_1 < \frac{1}{2}$ and $\Phi_2 > \frac{1}{2}$ with $0 < \Phi_2 - \frac{1}{2} < \frac{1}{2} - \Phi_1$. All other cases can be derived in a similar fashion to the proof in Appendix~\ref{sec:proof:thm:noisyPareto}. The Pareto frontier with these assumptions are shown in Figure~\ref{fig:noisyPareto}. We first find the maximum per-instance error of the adversary given per-instance error of the conference. We find that the adversary's error no longer increases if the conference increase its error larger than $\frac{\reviewerfunctionpdf_1(\score_1)\reviewerfunctionpdf_2(\score_2)(\Phi_1+2\Phi_2-1)+\reviewerfunctionpdf_2(\score_1)\reviewerfunctionpdf_1(\score_2)(1-\Phi_2)}{\reviewerfunctionpdf_1(\score_1)\reviewerfunctionpdf_2(\score_2) + \reviewerfunctionpdf_2(\score_1)\reviewerfunctionpdf_1(\score_2)}$ in this case. The maximum per-instance error of the adversary is $\frac{\reviewerfunctionpdf_1(\score_1)\reviewerfunctionpdf_2(\score_2)}{\reviewerfunctionpdf_1(\score_1)\reviewerfunctionpdf_2(\score_2) + \reviewerfunctionpdf_2(\score_1)\reviewerfunctionpdf_1(\score_2)}$. Therefore, the Pareto frontier for scores satisfy the condition is an increasing line from $\left(\frac{\reviewerfunctionpdf_1(\score_1)\reviewerfunctionpdf_2(\score_2)\Phi_1+\reviewerfunctionpdf_2(\score_1)\reviewerfunctionpdf_1(\score_2)(1-\Phi_2)}{\reviewerfunctionpdf_1(\score_1)\reviewerfunctionpdf_2(\score_2)+\reviewerfunctionpdf_2(\score_1)\reviewerfunctionpdf_1(\score_2)},0\right)$ to $\left(\frac{\reviewerfunctionpdf_1(\score_1)\reviewerfunctionpdf_2(\score_2)(\Phi_1+2\Phi_2-1)+\reviewerfunctionpdf_2(\score_1)\reviewerfunctionpdf_1(\score_2)(1-\Phi_2)}{\reviewerfunctionpdf_1(\score_1)\reviewerfunctionpdf_2(\score_2)+\reviewerfunctionpdf_2(\score_1)\reviewerfunctionpdf_1(\score_2)}, \frac{\reviewerfunctionpdf_1(\score_1)\reviewerfunctionpdf_2(\score_2)}{\reviewerfunctionpdf_1(\score_1)\reviewerfunctionpdf_2(\score_2)+\reviewerfunctionpdf_2(\score_1)\reviewerfunctionpdf_1(\score_2)}\right)$.

We show the Pareto frontier in the case described above in Figure~\ref{fig:noisyPareto}. 
In all other cases, the shape of the Pareto frontier is the same as Figure~\ref{fig:noisyPareto} but has different coordinates.
The relationship between $\reviewerfunctionpdf_1(\score_1)\reviewerfunctionpdf_2(\score_2)$ and $\reviewerfunctionpdf_2(\score_1)\reviewerfunctionpdf_1(\score_2)$ combining with the values of $\Phi_1$ and $\Phi_2$ and their distance to $\frac{1}{2}$, we have eight different combinations of these values. In all eight cases, the Pareto frontier contains either a single point or an increasing line depending on the scores. Moreover, the maximum error of the adversary is $\frac{\min\{\reviewerfunctionpdf_1(\score_1)\reviewerfunctionpdf_2(\score_2), \reviewerfunctionpdf_2(\score_1)\reviewerfunctionpdf_1(\score_2)\}}{\reviewerfunctionpdf_1(\score_1)\reviewerfunctionpdf_2(\score_2) + \reviewerfunctionpdf_2(\score_1)\reviewerfunctionpdf_1(\score_2)}$ in all cases.

\subsubsection{Optimal Calibration Strategy under Per-Instance Errors}
\label{noisywce}

In the previous section, we characterized the fundamental tradeoff between the conference's per-instance error and the adversary's per-instance error through the Pareto frontier. In this section, we design a calibration strategy that achieves per-instance errors on the Pareto frontier, meaning that the strategy is optimal under per-instance errors.

Since $\scorematrix$ is a fixed realization in the analysis of per-instance errors, to simplify the notation we define (similar to Section~\ref{SecAlgoPerNoiseless}): 
\begin{align*}
    \calibrateprob_1 = \calibratefunction(\scorematrix, \assignment_1) \qquad \text{and}\qquad 
    \calibrateprob_2 = \calibratefunction(\scorematrix, \assignment_2).
\end{align*}
Under this notation, given the maximum allowable error of the conference $\Econference$, our goal is to find values of $\calibrateprob_1$ and $\calibrateprob_2$ that maximize the error of the adversary $\Eadversary$. We continue to use the notations $\Phi_1$ and $\Phi_2$ introduced in~\ref{align:phi}.

We present our proposed algorithm as Algorithm~\ref{alg:noisyworste}.
\begin{theorem}
\label{thm:noisyInstance}
The calibration algorithm described in Algorithm~\ref{alg:noisyworste} ensures the maximum per-instance error of the adversary for any given value of the maximum allowable per-instance error $\Econference([\score_1, \score_2])$ for the conference, and is hence Pareto optimal. 
\end{theorem}

The proof of Theorem~\ref{thm:noisyInstance} is provided in Appendix~\ref{sec:proof:thm:noisyInstance}. For a moment, consider the case of $\reviewerfunctionpdf_1(\score_1)\reviewerfunctionpdf_2(\score_2) < \reviewerfunctionpdf_2(\score_1)\reviewerfunctionpdf_1(\score_2)$ and $\Phi_1 < \frac{1}{2} < \Phi_2$ with $0 < \Phi_2-\frac{1}{2} < \frac{1}{2}-\Phi_1$, for a Pareto optimal calibration strategy, the error of the conference and the adversary should stay on the Pareto frontier as in Figure~\ref{fig:noisyPareto}. If the required error of the conference is less than $\frac{\reviewerfunctionpdf_1(\score_1)\reviewerfunctionpdf_2(\score_2)\Phi_1 + \reviewerfunctionpdf_2(\score_1)\reviewerfunctionpdf_1(\score_2)(1-\Phi_2)}{\reviewerfunctionpdf_1(\score_1)\reviewerfunctionpdf_2(\score_2) + \reviewerfunctionpdf_2(\score_1)\reviewerfunctionpdf_1(\score_2)}$, then due to the noise, there is no feasible calibration strategy that satisfies this requirement. Otherwise, the error of the conference and the error of the adversary adhere to the Pareto frontier.

\begin{algorithm}[tb]
   \caption{Conference calibration with per-instance error in the noisy setting}
   \label{alg:noisyworste}
\begin{algorithmic}
\Indm
   \STATE {\bfseries Input:} scores $\scorematrix = [\score_1, \score_2]$, maximum allowable per-instance error of the conference $\Econference([\score_1, \score_2])$
   \IF {$\score_1 > \max \left\{\frac{a_2(a_1^2+\sigma^2)(\score_2-b_2)}{a_1(a_2^2+\sigma^2)} + b_1, \frac{a_1(a_2^2 + \sigma^2)(\score_2-b_1)}{a_2(a_1^2+\sigma^2)}+b_2\right\}$}
   \STATE {accept paper 1}
   \ELSIF {$\score_1 < \min \left\{\frac{a_2(a_1^2+\sigma^2)(\score_2-b_2)}{a_1(a_2^2+\sigma^2)} + b_1, \frac{a_1(a_2^2 + \sigma^2)(\score_2-b_1)}{a_2(a_1^2+\sigma^2)}+b_2\right\}$}
   \STATE {accept paper 2}
   \ELSIF{$\Econference([\score_1, \score_2]) < \frac{\reviewerfunctionpdf_1(\score_1)\reviewerfunctionpdf_2(\score_2)\Phi_1 + \reviewerfunctionpdf_2(\score_1)\reviewerfunctionpdf_1(\score_2)(1-\Phi_2)}{\reviewerfunctionpdf_1(\score_1)\reviewerfunctionpdf_2(\score_2) + \reviewerfunctionpdf_2(\score_1)\reviewerfunctionpdf_1(\score_2)}$}
   \STATE {error conference of cannot be achieved}
   \ELSIF {$\Econference([\score_1, \score_2]) \ge \frac{\reviewerfunctionpdf_1(\score_1)\reviewerfunctionpdf_2(\score_2)(\Phi_1+2\Phi_2-1)+\reviewerfunctionpdf_2(\score_1)\reviewerfunctionpdf_1(\score_2)(1-\Phi_2)}{\reviewerfunctionpdf_1(\score_1)\reviewerfunctionpdf_2(\score_2) + \reviewerfunctionpdf_2(\score_1)\reviewerfunctionpdf_1(\score_2)}$}
   \STATE {choose $\calibrateprob_1 = 1, \calibrateprob_2 = \frac{(\reviewerfunctionpdf_2(\score_1)\reviewerfunctionpdf_1(\score_2)-\reviewerfunctionpdf_1(\score_1)\reviewerfunctionpdf_2(\score_2))(1-2\Phi_2)}{\reviewerfunctionpdf_1(\score_1)\reviewerfunctionpdf_2(\score_2) + \reviewerfunctionpdf_2(\score_1)\reviewerfunctionpdf_1(\score_2)}$}
   \ELSE
   \STATE {choose $\calibrateprob_1 = 1, \calibrateprob_2 = \frac{\Econference([\score_1, \score_2]) \cdot (\reviewerfunctionpdf_1(\score_1)\reviewerfunctionpdf_2(\score_2) + \reviewerfunctionpdf_2(\score_1)\reviewerfunctionpdf_1(\score_2)) - \reviewerfunctionpdf_1(\score_1)\reviewerfunctionpdf_2(\score_2)  (1-\Phi_1) - \reviewerfunctionpdf_2(\score_1)\reviewerfunctionpdf_1(\score_2) \Phi_2 -(2\Phi_1-1)\reviewerfunctionpdf_1(\score_1)\reviewerfunctionpdf_2(\score_2)}{(1 - 2\Phi_2)\reviewerfunctionpdf_2(\score_1)\reviewerfunctionpdf_1(\score_2)}$}
   \ENDIF
\end{algorithmic}
\end{algorithm}

Algorithm~\ref{alg:noisyworste} follows directly from the Pareto frontier established in Theorem~\ref{thm:noisyPareto}. The calibration probabilities $\calibrateprob_1$ and $\calibrateprob_2$ are chosen such that the error of the conference and the error of the adversary lie on the Pareto frontier. The first two cases of Algorithm~\ref{alg:noisyworste} correspond to part $(1)$ of Theorem~\ref{thm:noisyPareto} where the same paper has higher estimated quality under both assignments. In the noisy case, there is a minimum value for the per-instance error of the conference. Therefore, in the third case of Algorithm~\ref{alg:noisyworste}, when the maximum allowable per-instance error of the conference is too small, the conference cannot achieve such error. If $\Econference([\score_1, \score_2]) \ge \frac{\reviewerfunctionpdf_1(\score_1)\reviewerfunctionpdf_2(\score_2)(\Phi_1+2\Phi_2-1)+\reviewerfunctionpdf_2(\score_1)\reviewerfunctionpdf_1(\score_2)(1-\Phi_2)}{\reviewerfunctionpdf_1(\score_1)\reviewerfunctionpdf_2(\score_2) + \reviewerfunctionpdf_2(\score_1)\reviewerfunctionpdf_1(\score_2)}$, the error of the conference should be $\frac{\reviewerfunctionpdf_1(\score_1)\reviewerfunctionpdf_2(\score_2)(\Phi_1+2\Phi_2-1)+\reviewerfunctionpdf_2(\score_1)\reviewerfunctionpdf_1(\score_2)(1-\Phi_2)}{\reviewerfunctionpdf_1(\score_1)\reviewerfunctionpdf_2(\score_2) + \reviewerfunctionpdf_2(\score_1)\reviewerfunctionpdf_1(\score_2)}$ to stay Pareto optimal since further sacrifice of accuracy cannot increase error of the adversary. And for the rest of the per-instance error of the conference, we choose $\calibrateprob_1$ and $\calibrateprob_2$ such that the errors of the conference and the adversary stay on the Pareto frontier in Figure~\ref{fig:noisyPareto}.

\section{Discussion}

Our work is only a starting point towards addressing the important problem of calibration with privacy in its full generality. Several challenges need to be addressed in future work in order to design practical algorithms with guarantees for calibration and privacy. There are open problems pertaining to relaxations of assumptions made in this paper such as that of two reviewers and papers, homogeneity and knowledge of the noise variance, etc. An important open problem pertains to challenge \#2 discussed in the introduction, in conjunction with challenge \#1. Instead of assuming precise exogenous knowledge of the reviewers' miscalibration functions,
consider having some access to data from other conferences. Then how can one obtain and use meaningful estimates of reviewer miscalibrations from past conferences while guaranteeing privacy of the current as well as past conferences (``federated learning for calibration'')? In any of these endeavors, one may aim to uncover precise fundamental limits and optimal algorithms, or perhaps design algorithms that are readily applicable in practice with some basic theoretical guarantees.

\section*{Acknowledgments}
This work was supported by NSF CAREER award 1942124, NSF grant CIF 1763734, an NSERC Discovery Grant, and a Google Research Scholar Award. Most of this research was done when Wenxin Ding was at Carnegie Mellon University.

\printbibliography

%\newpage
~\\~\\

\appendix
~\\\noindent{\LARGE \bf Appendices}

\section{Simulations: Correcting miscalibration with and without exogeneous information}
\label{AppSimulations}

In the introduction section in the main text, we discussed two ways of reducing miscalibration: one where only the current conferences' data is used and another where miscalibration parameters of reviewers are obtained exogenously (e.g., from previous conferences). In this section, we conduct a simulation-based study to understand the performance of these approaches: What is the reduction in error if correcting for miscalibration? What if the reviewer-calibration parameters are known?

Our main results in the main text was focused on privacy and considered a setting with two reviewers and two papers. In this section we consider a larger number of reviewers and papers. The methods we simulate in this more general setting do not consider privacy. 

We first describe the simulation setting, and then discuss the results. The code for the simulations is available here: \url{https://github.com/wenxind/calibration-with-privacy-in-peer-review}.

{\bf Conference review setup: }
We consider 100 reviewers and 100 papers. We assign reviewers to papers uniformly at random with 3 reviewers per paper and 3 papers per reviewer. 

{\bf Miscalibration model: } We assume each reviewer has a linear miscalibration function: the miscalibration function $f_j$ of reviewer $j$ is given by $f_j(\quality) = a_j \quality + b_j$ where $\quality$ is the true quality of the paper being reviewed. For every paper $i$, its true quality $\quality_i$ is drawn from a Gaussian distribution with mean 0 and variance 1 independent of all else. The scalars $a_j$ in the reviewers' miscalibration functions are i.i.d.\ exponential random variables with rate 1. The biases $b_j$ are i.i.d.\ Gaussian random variables with mean 0 and variance 0.5. The score given by any reviewer $j$ to any paper $i$ is then given as $\quality_i$ is $a_j \quality_i + b_j + \noise_{ij}$ where $\noise_{ij}$ is a Gaussian random variable with mean 0, whose variance is varied in the plots.

{\bf Calibration methods: }
We consider the following three methods to calibrate the decisions. 
\begin{itemize}
    \item No calibration: The score for each paper is the mean score of the 3 review scores.
    \item Within-conference calibration: For each reviewer, we compute the mean score and standard deviation of the 3 review scores given by the reviewer and normalize the 3 scores by subtracting mean and dividing standard deviation. Then the score for each paper is the mean score of the 3 normalized review scores.
    \item Calibration with known parameters: We assume that the miscalibration parameters of the reviewers are known (exogeneously). Then we estimate the quality of each paper via maximum likelihood estimation as follows. For any paper $i$, let ${\cal R}_i$ denote the set of reviewers for paper $i$. Then the estimate of the score for any paper $i$ is $\frac{\sum_{j \in {\cal R}_i}a_j (\score_{ji} - b_j)}{\sum_{{j \in {\cal R}_i}} a_j^2}$ where $\score_{ji}$ denotes the score given by reviewer $j$ to paper $i$.
\end{itemize}
For each paper, we then take a mean of the calibrated scores across all its reviewers. 
The papers are then ranked according to these mean scores; we call this the estimated ranking.

\begin{figure}[t]
    \centering
    \begin{subfigure}[b]{0.45\textwidth}
    \centering
    \includegraphics[width=\textwidth]{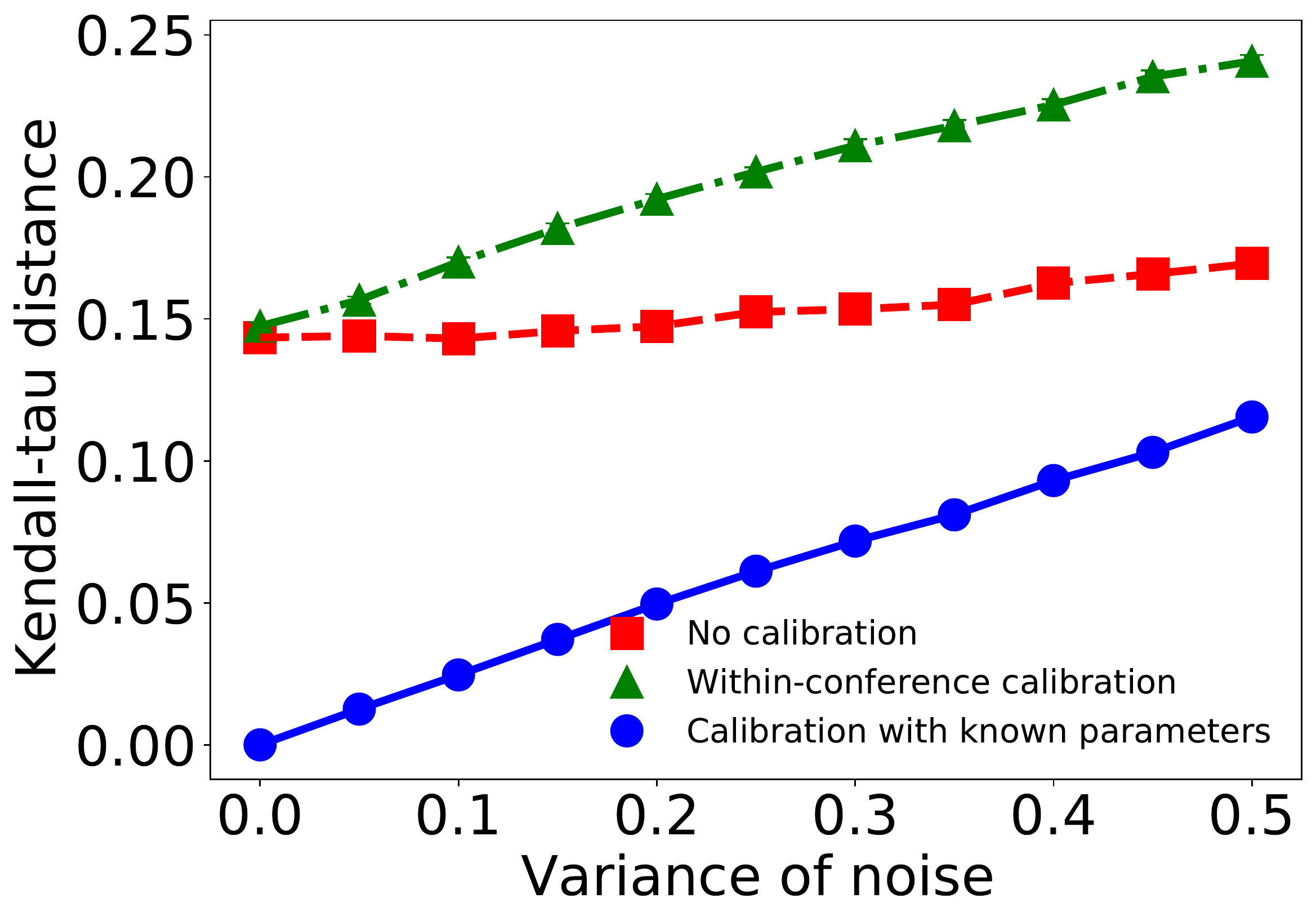}
    \caption{\label{fig:vary_as_kt}Kendall-$\tau$ distance.}
    \end{subfigure}~~~~~%\hspace{.09\textwidth}
    \begin{subfigure}[b]{0.45\textwidth}
    \centering
    \includegraphics[width=\textwidth]{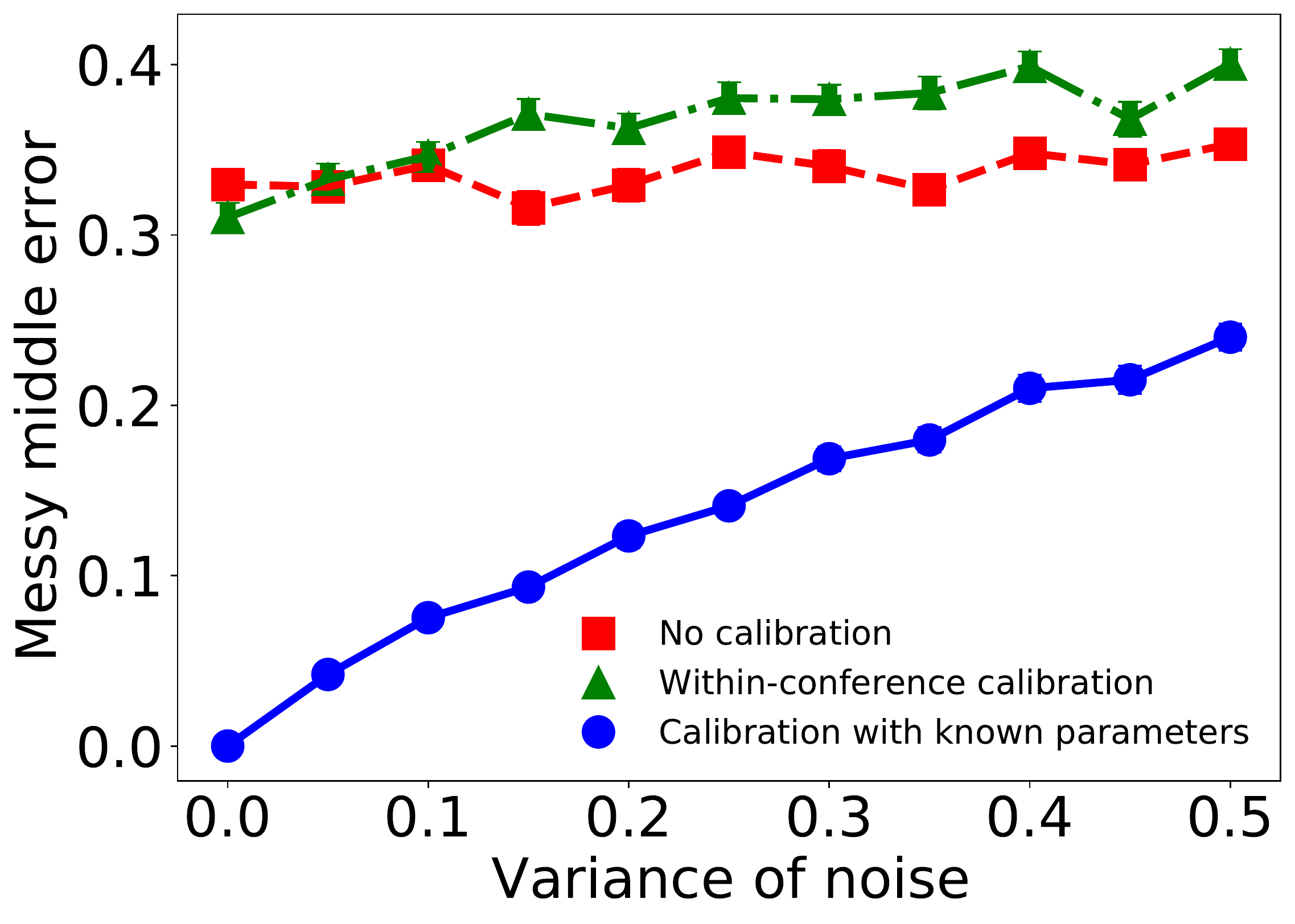}
    \caption{\label{fig:vary_as_mm}Messy middle error.}
    \end{subfigure}
    \caption{A simulation of the review process where the reviewers are miscalibrated. \label{FigCalibrationSimulation}}
\end{figure}

{\bf Error metrics: }
We consider two ways of measuring the error between the ranking of the papers in terms of their true scores and the ranking of the papers in terms of their estimated scores. 
\begin{itemize}
    \item Kendall tau distance: Given two rankings of the papers, the Kendall tau distance between the two ranking is $\frac{\text{number of discordant pairs}}{\text{total number of pairs}}$.
    
    \item Messy middle error: Given two rankings of the 100 papers, suppose that the conference wishes to accept the top 25 papers. Then we consider papers 11-40 as those that are marginal accepts, and we measure the error as the fraction of these papers which are (erroneously) rejected. In other words, the messy middle error equals $\frac{\text{number of papers whose true ranking is between 11--40 that are wrongly accepted/rejected}}{30}$. 
\end{itemize}

{\bf Results:} The results of the simulations are shown in Figure~\ref{FigCalibrationSimulation}. Each point depicts the mean from 100 iterations of these simulations. The error bars are too small to be visible. We see that correcting for miscalibration even without access to the parameters can lead to significant reduction in the error as compared to not correcting for the miscalibration. Furthermore, if the parameters were known (e.g., from other conferences) then it can lead to multi-fold further reductions in the error. 

\section{Connection to Local Differential Privacy}
\label{sec:ldp}
In this section, we discuss the connections between our algorithm and differential privacy (DP). 
We recall the definition of DP:
\begin{definition}[\cite{dwork2016calibrating}]
\label{def:dp}
An algorithm $M : \mathcal{X}^n \rightarrow \mathcal{Y}$ is $\epsilon$-differentially private (DP) if, for all $X, X' \in \mathcal{X}^n$ which differ in one entry (often called \emph{neighboring databases}) and $\mathcal{S} \subseteq{Y}$, we have that 
\[\Pr[M(X) \in \mathcal{S}] \leq e^\epsilon \Pr[M(X') \in \mathcal{S}]. \]
\end{definition}

Roughly speaking, a procedure involving $n$ users is \emph{$\epsilon$-locally differentially private} if each user applies an $\epsilon$-DP algorithm to their single datapoint and shares only the result with other users or a data curator.
The most familiar LDP algorithm is (binary) randomized response.
\begin{definition}\label{def:brr}
Binary randomized response with parameter $\gamma$ is an algorithm $M : \{0,1\} \rightarrow \{0, 1\}$, which, given input $x$, outputs $x$ with probability $\frac{1}{1+\gamma}$, and outputs $1-x$ with probability $\frac{\gamma}{1+\gamma}$.
\end{definition}

The following claim is immediate from the definition of differential privacy and randomized response.
\begin{proposition}
Binary randomized response with parameter $e^\epsilon$ is $\epsilon$-DP.
\end{proposition}

We now relate randomized response to the algorithms proposed in our setting. 
The private information for our calibration problem consists solely of the reviewer assignment $\assignment$, which takes one of two different values (i.e., reviewer 1 is assigned to paper 1 and reviewer 2 is assigned to paper 2, or vice versa).
These two reviewer assignments can be considered to be ``neighboring'' datasets, as mentioned in Definition~\ref{def:dp}.
All other information (paper scores $\scorematrix$ and reviewer calibration functions) are assumed to be public.

As argued in Proposition~\ref{proposition:calibrationfunction}, it is without loss of optimality to solely consider strategies of the form $\calibratefunction$, in which the conference calibrates according to the true assignment with probability $\calibratefunction(\scorematrix, \assignment)$ and according to the false assignment otherwise. 
This can be rephrased into the language of randomized response by considering the assignment $\assignment$ to be the input bit to binary randomized response, in which it is preserved with probability $\calibratefunction(\scorematrix, \assignment)$ (in the language of Definition~\ref{def:brr}, $\calibratefunction(\scorematrix, \assignment) = \frac{1}{1 + \gamma}$) and flipped otherwise, and then the conference calibrates with the resulting assignment.

We caution that this connection does \emph{not} directly imply that our algorithms are differentially private.
This is because the probability $\calibratefunction(\scorematrix, \assignment)$ is selected in a data-dependent way, whereas differential privacy requires it to be data independent. 
Nevertheless, our work provides tight guarantees on the probability that an MAP adversary can determine the true assignment.

\section{Appendix: Proofs}
In the appendix, we present complete proofs of the results claimed in the main text.

\subsection{Proof of Proposition~\ref{proposition:calibrationfunction}}\label{sec:proof:proposition:calibrationfunction}
The most generic calibration strategy can be represented using a function $\probfunction$ such that for any given score and assignment, $\probfunction$ outputs a probability for accepting paper 1. In other words, the conference accepts paper 1 with probability $\probfunction(\scorematrix, \assignment)$ and accepts paper 2 with probability $1 - \probfunction(\scorematrix, \assignment)$. We propose a calibration strategy using function $\calibratefunction$ instead of $\probfunction$, where $\calibratefunction$ outputs a probability that the conference calibrates under the true assignment by accepting the paper with higher estimated quality under the true assignment. 

Calibrating using the calibration strategy of function $\calibratefunction$ differs from calibrating using the calibration strategy of function $\probfunction$ only when the same paper has higher estimated quality under both assignments by the MAP. Since otherwise, when calibrating under the true assignment and calibrating assuming the wrong assignment lead to accepting different papers, either paper can have arbitrary non-zero probability of being accepted (their probabilities sum to 1) by adjusting the output of $\calibratefunction(\scorematrix, \assignment_1)$ and $\calibratefunction(\scorematrix, \assignment_2)$. Then it is the same calibration strategy as using function $\probfunction$.

Note that the adversary makes its guess using the MAP $\argmax_{\{\assignment = \assignment_1\text{or} \assignment = \assignment_2\}} \Pr(\assignmentrv = \assignment | \decisionrv = P, \scorematrix = [\score_1, \score_2])$ where $\decisionrv$ is the random variable for the decision made by the conference (acceptance of paper) and $P$ is the paper being accepted. By expanding the probability expression, we have that
\begin{align*}
     & \argmax_{\assignment = \assignment_1\text{or} \assignment = \assignment_2} \Pr(\assignmentrv = \assignment | \decisionrv = P, \scorematrix = [\score_1, \score_2]) \\
    = & \argmax_{\assignment = \assignment_1\text{or} \assignment = \assignment_2} \frac{\Pr(\assignmentrv = \assignment, \decisionrv = P | \scorematrix = [\score_1, \score_2])}{\Pr(\decisionrv = P | \scorematrix = [\score_1, \score_2])}\\
    = & \argmax_{\assignment = \assignment_1\text{or} \assignment = \assignment_2} \frac{\Pr(\decisionrv = P | \assignmentrv = \assignment, \scorematrix = [\score_1, \score_2]) \Pr(\assignmentrv = \assignment | \scorematrix = [\score_1, \score_2])}{\Pr(\decisionrv = P | \scorematrix = [\score_1, \score_2])}\\
    = & \argmax_{\assignment = \assignment_1\text{or} \assignment = \assignment_2} \Pr(\decisionrv = P | \assignmentrv = \assignment, \scorematrix = [\score_1, \score_2]) \Pr(\assignmentrv = \assignment | \scorematrix = [\score_1, \score_2]).
\end{align*}

If the same paper has higher estimated quality under both assignments, and the conference accepts the believed higher-quality paper, then the adversary guesses the assignment based on the scores only. Because the adversary knows the calibration strategy used by the conference, if $P$ is the paper that has higher quality under both assignments, then $\Pr(\decisionrv = P | \assignmentrv = \assignment, \scorematrix = [\score_1, \score_2])=1$ for both $\assignmentrv=\assignment_1$ and $\assignmentrv=\assignment_2$. Therefore, the MAP used by the adversary simplifies to $\argmax_{\assignment = \assignment_1\text{or} \assignment = \assignment_2} \Pr(\assignmentrv = \assignment | \scorematrix = [\score_1, \score_2])$. In this case, the conference does not have extra privacy leakage by accepting $P$ since the adversary is making its guess based on the information that is already public (the scores). In addition, if the conference has non-zero probability of accepting the other paper, its utility decreases by accepting the lower-quality paper. Even if the conference accepts the lower-quality paper, the error of the adversary remains unchanged as it can use the scores to guess the assignment without being affected by the conference decision. Thus, there is no need for the conference to have non-zero probability for accepting the paper that has lower-quality under both assignments.

In conclusion, calibrating using the calibration strategy of function $\calibratefunction$ instead of the calibration strategy of function $\probfunction$ does not reduce the optimally of the conference. Therefore, we consider the calibration strategy with function $\calibratefunction$ in our analysis.

\subsection{Proof of Theorem~\ref{thm:noiselessPareto}}\label{sec:proof:thm:noiselessPareto}
To find the Pareto frontier of per-instance error of the adversary against per-instance error of the conference, we first derive expressions for per-instance error of the conference and the adversary. We find a fixed expression for the error of the conference and calculate the maximum per-instance error of the adversary in different cases. We then analyze the relation between the errors and complete plots for maximum per-instance error of the adversary for any per-instance error of conference. Finally, we derive the Pareto frontier from the plots.

In the noiseless setting, the conference uses the inverse functions of reviewers' miscalibration functions and the scores to exactly compute the quality of the papers. If the conference estimates the qualities assuming that $\assignment_1$ was the actual assignment, we get $\qualityest_1 = \reviewerfunction_1^{-1}(\score_1)$ and $\qualityest_2 = \reviewerfunction_2^{-1}(\score_2)$. If the conference estimates the qualities assuming that $\assignment_2$ was the actual assignment, we get $\qualityest_1 = \reviewerfunction_2^{-1}(\score_1)$ and $\qualityest_2 = \reviewerfunction_1^{-1}(\score_2)$. If $\score_1 > \max \{\reviewerfunction_2(\reviewerfunction_1^{-1}(\score_2)), \reviewerfunction_1(\reviewerfunction_2^{-1}(\score_2))\}$, then $\qualityest_1 > \qualityest_2$ under both assignments (and hence paper 1 should be accepted). Similarly, if $\score_1 < \min \{\reviewerfunction_2(\reviewerfunction_1^{-1}(\score_2)), \reviewerfunction_1(\reviewerfunction_2^{-1}(\score_2))\}$, then $\qualityest_1 < \qualityest_2$ under both assignments and hence paper 2 should be accepted. Therefore, when $\score_1 > \max \{\reviewerfunction_2(\reviewerfunction_1^{-1}(\score_2)), \reviewerfunction_1(\reviewerfunction_2^{-1}(\score_2))\}$ or $\score_1 < \min \{\reviewerfunction_2(\reviewerfunction_1^{-1}(\score_2)), \reviewerfunction_1(\reviewerfunction_2^{-1}(\score_2))\}$, which is a subset of part $(1)$ of Theorem~\ref{thm:noiselessPareto}, the same paper has higher estimated quality under both assignments, and hence that paper will be accepted irrespective of the function $\calibratefunction$. Thus, under this condition, the Pareto optimal curve comprises just a single point where the conference has zero error, and the adversary obtains no additional information from the acceptance decision as compared to the scores $\scorematrix = [\score_1, \score_2]$. The error of the adversary is $\frac{\min\{\reviewerfunctionpdf_1(\score_1)\reviewerfunctionpdf_2(\score_2), \reviewerfunctionpdf_2(\score_1)\reviewerfunctionpdf_1(\score_2)\}}{\reviewerfunctionpdf_1(\score_1)\reviewerfunctionpdf_2(\score_2) + \reviewerfunctionpdf_2(\score_1)\reviewerfunctionpdf_1(\score_2)}$ when it guesses the assignment using only the scores and not the decision.

For the rest scores, the conference uses function $\calibratefunction$ to decide acceptance of paper. Since $\scorematrix$ is a fixed realization in the analysis, we simplify the calibration strategy for the conference as 
\begin{align*}
    \calibrateprob_1 &= \calibratefunction(\scorematrix, \assignment_1)\\
    \calibrateprob_2 &= \calibratefunction(\scorematrix, \assignment_2).
\end{align*}

We now consider the rest scores in part $(1)$ of the theorem. If $\score_1 = \reviewerfunction_2(\reviewerfunction_1^{-1}(\score_2))$, the conference accepts each paper uniform at random if calibrating under $\assignment_1$ and accepts paper 1 if calibrating under $\assignment_2$. Since paper 1 has higher or equal quality than paper 2, the conference only has error when paper 2 is accepted and $\assignmentrv = \assignment_2$.

\begin{align*}
    & \Pr(\text{conference accepts lower-quality paper} | \scorematrix = [\score_1, \score_2]) \\
    = & \Pr(\text{conference accepts lower-quality paper} | \scorematrix = [\score_1, \score_2], \decision = P_1) \Pr(\decision = P_1 | \scorematrix = [\score_1, \score_2]) \\
    & + \Pr(\text{conference accepts lower-quality paper} | \scorematrix = [\score_1, \score_2], \decision = P_2) \Pr(\decision = P_2 | \scorematrix = [\score_1, \score_2])\\
    = & \Pr(\text{conference accepts lower-quality paper} | \scorematrix = [\score_1, \score_2], \decision = P_2) \Pr(\decision = P_2 | \scorematrix = [\score_1, \score_2]).
\end{align*}

Given that the conference accepts $P_2$, the probability of the conference making error is the probability $\Pr(\assignmentrv = \assignment_2 | \scorematrix = [\score_1, \score_2])$.

\begin{align*}
    & \Pr(\decision = P_2 | \scorematrix = [\score_1, \score_2]) \\
    = & \Pr(\decision = P_2 | \scorematrix = [\score_1, \score_2], \assignmentrv = \assignment_1) \Pr(\assignmentrv = \assignment_1 | \scorematrix = [\score_1, \score_2]) \\
    & + \Pr(\decision = P_2 | \scorematrix = [\score_1, \score_2], \assignmentrv = \assignment_2) \Pr(\assignmentrv = \assignment_2 | \scorematrix = [\score_1, \score_2]) \\
    = & \frac{1}{2} \calibrateprob_1 \Pr(\assignmentrv = \assignment_1 | \scorematrix = [\score_1, \score_2]) + \frac{1}{2}(1-\calibrateprob_2)\Pr(\assignmentrv = \assignment_2 | \scorematrix = [\score_1, \score_2]).
\end{align*}

For the adversary, if paper 1 is accepted, it gains no information on the assignment other than the scores so its error is $\frac{\min \{\reviewerfunctionpdf_1(\score_1)\reviewerfunctionpdf_2(\score_2), \reviewerfunctionpdf_2(\score_1)\reviewerfunctionpdf_1(\score_2)\}}{\reviewerfunctionpdf_1(\score_1)\reviewerfunctionpdf_2(\score_2) + \reviewerfunctionpdf_2(\score_1)\reviewerfunctionpdf_1(\score_2)}$. Otherwise, it guesses $\assignmentrv = \assignment_1$ and its error is $\Pr(\assignmentrv = \assignment_2 | \scorematrix = [\score_1, \score_2])$. Note that error of the adversary does not exceed $\frac{\min \{\reviewerfunctionpdf_1(\score_1)\reviewerfunctionpdf_2(\score_2), \reviewerfunctionpdf_2(\score_1)\reviewerfunctionpdf_1(\score_2)\}}{\reviewerfunctionpdf_1(\score_1)\reviewerfunctionpdf_2(\score_2) + \reviewerfunctionpdf_2(\score_1)\reviewerfunctionpdf_1(\score_2)}$ since in the worst case for the adversary, it guesses the assignment solely based on the scores and ignore the conference decision.

\begin{align*}
    & \Pr(\text{adversary guesses assignment wrong} | \scorematrix = [\score_1, \score_2]) \\
    = & \Pr(\text{adversary guesses assignment wrong} | \scorematrix = [\score_1, \score_2], \decision = P_1) \Pr(\decision = P_1 | \scorematrix = [\score_1, \score_2]) \\
    & + \Pr(\text{adversary guesses assignment wrong} | \scorematrix = [\score_1, \score_2], \decision = P_2) \Pr(\decision = P_2 | \scorematrix = [\score_1, \score_2]) \\
    = & \frac{\min \{\reviewerfunctionpdf_1(\score_1)\reviewerfunctionpdf_2(\score_2), \reviewerfunctionpdf_2(\score_1)\reviewerfunctionpdf_1(\score_2)\}}{\reviewerfunctionpdf_1(\score_1)\reviewerfunctionpdf_2(\score_2) + \reviewerfunctionpdf_2(\score_1)\reviewerfunctionpdf_1(\score_2)} \left( (1-\frac{1}{2}\calibrateprob_1) \Pr(\assignmentrv = \assignment_1 | \scorematrix = [\score_1, \score_2]) + (\frac{1}{2} + \frac{1}{2} \calibrateprob_2) \Pr(\assignmentrv = \assignment_2 | \scorematrix = [\score_1, \score_2]) \right) \\
    & + \Pr(\assignmentrv = \assignment_2 | \scorematrix = [\score_1, \score_2]) \left( \frac{1}{2} \calibrateprob_1 \Pr(\assignmentrv = \assignment_1 | \scorematrix = [\score_1, \score_2]) + \frac{1}{2}(1-\calibrateprob_2)\Pr(\assignmentrv = \assignment_2 | \scorematrix = [\score_1, \score_2]) \right)
\end{align*}

Therefore, we can minimize the error of the conference to 0 by choosing $\calibrateprob_1 = 0$ and $\calibrateprob_2 = 1$, which results in the conference always accepts paper 1. Then error of the adversary is $\frac{\min \{\reviewerfunctionpdf_1(\score_1)\reviewerfunctionpdf_2(\score_2), \reviewerfunctionpdf_2(\score_1)\reviewerfunctionpdf_1(\score_2)\}}{\reviewerfunctionpdf_1(\score_1)\reviewerfunctionpdf_2(\score_2) + \reviewerfunctionpdf_2(\score_1)\reviewerfunctionpdf_1(\score_2)}$, which is maximized. Further increase of error of the conference cannot increase error of the adversary. So the Pareto optimal point is $(0, \frac{\min \{\reviewerfunctionpdf_1(\score_1)\reviewerfunctionpdf_2(\score_2), \reviewerfunctionpdf_2(\score_1)\reviewerfunctionpdf_1(\score_2)\}}{\reviewerfunctionpdf_1(\score_1)\reviewerfunctionpdf_2(\score_2) + \reviewerfunctionpdf_2(\score_1)\reviewerfunctionpdf_1(\score_2)})$. The same argument works when $\score_1 = \reviewerfunction_1(\reviewerfunction_2^{-1}(\score_2))$.

In the noiseless setting where $\min \{\reviewerfunction_2(\reviewerfunction_1^{-1}(\score_2)), \reviewerfunction_1(\reviewerfunction_2^{-1}(\score_2))\} < \score_1 < \max \{\reviewerfunction_2(\reviewerfunction_1^{-1}(\score_2)), \reviewerfunction_1(\reviewerfunction_2^{-1}(\score_2))\}$, which is part $(2)$ of Theorem~\ref{thm:noiselessPareto}, we first find the maximum per-instance error of the adversary given per-instance error of the conference in this range. We will show the proof with the assumptions that $\reviewerfunction_2(\reviewerfunction_1^{-1}(\score_2)) > \reviewerfunction_1(\reviewerfunction_2^{-1}(\score_2))$ and $\reviewerfunctionpdf_1(\score_1)\reviewerfunctionpdf_2(\score_2) > \reviewerfunctionpdf_2(\score_1)\reviewerfunctionpdf_1(\score_2)$. The proof follows the same procedure for other values of $\reviewerfunction_2(\reviewerfunction_1^{-1}(\score_2))$, $\reviewerfunction_1(\reviewerfunction_2^{-1}(\score_2))$, $\reviewerfunctionpdf_1(\score_1)\reviewerfunctionpdf_2(\score_2)$, and $\reviewerfunctionpdf_2(\score_1)\reviewerfunctionpdf_1(\score_2)$.

When the scores satisfy $\reviewerfunction_1(\reviewerfunction_2^{-1}(\score_2)) < \score_1 < \reviewerfunction_2(\reviewerfunction_1^{-1}(\score_2))$, the conference always accepts the higher-quality paper if it calibrates under the true assignment, and the conference always accepts the lower-quality paper if it calibrates assuming the wrong assignment. But the conference can calibrate assuming the wrong assignment for the purpose of misleading the adversary. We use $\assignmentrv$ to denote the random variable for the assignment, $\decisionrv$ to denote the random variable for the conference decision and $\scorematrix$ is the scores. In addition, we use $\calibrationrv$ to denote the calibration status. If the conference calibrates under the true assignment then $\calibrationrv = T$. Otherwise, $\calibrationrv = F$.

Therefore, the error of the conference is computed as
\begin{align*}
    & \Pr(\text{conference accepts lower-quality paper} | \scorematrix = [\score_1, \score_2]) \\
    = & \Pr(\calibrationrv = F | \scorematrix = [\score_1, \score_2]) \\
    = & \Pr(\calibrationrv = F, \assignmentrv = \assignment_1 | \scorematrix = [\score_1, \score_2]) + \Pr(\calibrationrv = F, \assignmentrv = \assignment_2 | \scorematrix = [\score_1, \score_2]) \\
    = & \Pr(\calibrationrv = F | \assignmentrv = \assignment_1, \scorematrix = [\score_1, \score_2]) \Pr(\assignmentrv = \assignment_1 | \scorematrix = [\score_1, \score_2]) \\
    & + \Pr(\calibrationrv = F | \assignmentrv = \assignment_2, \scorematrix = [\score_1, \score_2]) P(\assignmentrv = \assignment_2 | \scorematrix = [\score_1, \score_2]) \\
    = & (1-\calibrateprob_1) \cdot \frac{\reviewerfunctionpdf_1(\score_1)\reviewerfunctionpdf_2(\score_2)}{\reviewerfunctionpdf_1(\score_1)\reviewerfunctionpdf_2(\score_2) + \reviewerfunctionpdf_2(\score_1)\reviewerfunctionpdf_1(\score_2)} + (1-\calibrateprob_2) \cdot \frac{\reviewerfunctionpdf_2(\score_1)\reviewerfunctionpdf_1(\score_2)}{\reviewerfunctionpdf_1(\score_1)\reviewerfunctionpdf_2(\score_2) + \reviewerfunctionpdf_2(\score_1)\reviewerfunctionpdf_1(\score_2)}\\
    = & 1 -\frac{\reviewerfunctionpdf_1(\score_1)\reviewerfunctionpdf_2(\score_2)\calibrateprob_1 + \reviewerfunctionpdf_2(\score_1)\reviewerfunctionpdf_1(\score_2)\calibrateprob_2}{\reviewerfunctionpdf_1(\score_1)\reviewerfunctionpdf_2(\score_2) + \reviewerfunctionpdf_2(\score_1)\reviewerfunctionpdf_1(\score_2)}.
\end{align*}

The adversary uses MAP to guess the assignment. If the two assignments have the same a posteriori probability, then the adversary makes a random guess between the assignments where either assignment has probability $\frac{1}{2}$ of being guessed. When making a guess, the adversary observes the scores and the conference decision. So the adversary finds $\argmax_{\{\assignment = \assignment_1\text{or} \assignment = \assignment_2\}} \Pr(\assignmentrv = \assignment | \decisionrv = P, \scorematrix = [\score_1, \score_2])$ where $P$ is the paper being accepted. Following Section~\ref{sec:proof:proposition:calibrationfunction}, the adversary finds

\begin{align*}
     & \argmax_{\assignment = \assignment_1\text{or} \assignment = \assignment_2} \Pr(\assignmentrv = \assignment | \decisionrv = P, \scorematrix = [\score_1, \score_2]) \\
    = & \argmax_{\assignment = \assignment_1\text{or} \assignment = \assignment_2} \Pr(\decisionrv = P | \assignmentrv = \assignment, \scorematrix = [\score_1, \score_2]) \Pr(\assignmentrv = \assignment | \scorematrix = [\score_1, \score_2])\\
    = & \argmax_{\assignment = \assignment_1\text{or} \assignment = \assignment_2} (\Pr(\decisionrv = P | \assignmentrv = \assignment, \scorematrix = [\score_1, \score_2], \calibrationrv = T) \Pr(\calibrationrv = T | \assignmentrv = \assignment, \scorematrix = [\score_1, \score_2])\\
    & + \Pr(\decisionrv = P | \assignmentrv = \assignment, \scorematrix, \calibrationrv = F) \Pr(\calibrationrv = F | \assignmentrv = \assignment, \scorematrix = [\score_1, \score_2])) \cdot \Pr(\assignmentrv = \assignment | \scorematrix = [\score_1, \score_2])\\
    = & \argmax_{\assignment = \assignment_1\text{or} \assignment = \assignment_2} (\Pr(\decisionrv = P | \assignmentrv = \assignment, \scorematrix = [\score_1, \score_2], \calibrationrv = T) \cdot \calibratefunction(\scorematrix = [\score_1, \score_2], \assignmentrv = \assignment)\\
    & + \Pr(\decisionrv = P | \assignmentrv = \assignment, \scorematrix = [\score_1, \score_2], \calibrationrv = F) \cdot (1-\calibratefunction(\scorematrix = [\score_1, \score_2], \assignmentrv = \assignment))) \cdot \Pr(\assignmentrv = \assignment | \scorematrix = [\score_1, \score_2])
\end{align*}

Under our assumptions of $\reviewerfunction_2(\reviewerfunction_1^{-1}(\score_2)) > \reviewerfunction_1(\reviewerfunction_2^{-1}(\score_2))$ and $\reviewerfunctionpdf_1(\score_1)\reviewerfunctionpdf_2(\score_2) > \reviewerfunctionpdf_2(\score_1)\reviewerfunctionpdf_1(\score_2)$, paper 1 has higher estimated quality under $\assignment_1$ and paper 2 has higher estimated quality under $\assignment_2$. Suppose paper 1 is accepted, i.e., $\decisionrv = P_1$. The value of the above expression under $\assignmentrv = \assignment_1$ is
\begin{align*}
    & (\Pr(\decisionrv = P_1 | \assignmentrv = \assignment_1, \scorematrix = [\score_1, \score_2], \calibrationrv = T) \cdot \calibratefunction(\scorematrix = [\score_1, \score_2], \assignmentrv = \assignment_1) \\
    & + \Pr(\decisionrv = P_1 | \assignmentrv = \assignment_1, \scorematrix = [\score_1, \score_2], \calibrationrv = F) \cdot (1-\calibratefunction(\scorematrix = [\score_1, \score_2], \assignmentrv = \assignment_1))) \cdot \Pr(\assignmentrv = \assignment_1 | \scorematrix = [\score_1, \score_2]) \\
    = & (\calibrateprob_1 + 0) \cdot  \reviewerfunctionpdf_1(\score_1)\reviewerfunctionpdf_2(\score_2) \\
    = & \reviewerfunctionpdf_1(\score_1)\reviewerfunctionpdf_2(\score_2) \calibrateprob_1.
\end{align*}
On the other hand, suppose paper 1 is accepted, the value of the above expression under $\assignmentrv = \assignment_2$ is
\begin{align*}
    & (\Pr(\decisionrv = P_1 | \assignmentrv = \assignment_2, \scorematrix = [\score_1, \score_2], \calibrationrv = T) \cdot \calibratefunction(\scorematrix = [\score_1, \score_2], \assignmentrv = \assignment_2) \\
    & + \Pr(\decisionrv = P_1 | \assignmentrv = \assignment_2, \scorematrix = [\score_1, \score_2], \calibrationrv = F) \cdot (1-\calibratefunction(\scorematrix = [\score_1, \score_2], \assignmentrv = \assignment_2))) \cdot \Pr(\assignmentrv = \assignment_2 | \scorematrix = [\score_1, \score_2]) \\
    = & (0 + (1-\calibrateprob_2)) \cdot  \reviewerfunctionpdf_1(\score_1)\reviewerfunctionpdf_2(\score_2) \\
    = & \reviewerfunctionpdf_2(\score_1)\reviewerfunctionpdf_1(\score_2)(1-\calibrateprob_2).
\end{align*}
Therefore, when the conference accepts paper 1, if $\reviewerfunctionpdf_1(\score_1)\reviewerfunctionpdf_2(\score_2)\calibrateprob_1 > \reviewerfunctionpdf_2(\score_1)\reviewerfunctionpdf_1(\score_2)(1-\calibrateprob_2)$, then the adversary guesses $\assignmentrv = \assignment_1$. Otherwise, it guesses $\assignmentrv = \assignment_2$ except that when $\reviewerfunctionpdf_1(\score_1)\reviewerfunctionpdf_2(\score_2)\calibrateprob_1 = \reviewerfunctionpdf_2(\score_1)\reviewerfunctionpdf_1(\score_2)(1-\calibrateprob_2)$, it makes a random guess assigning probability $\frac{1}{2}$ to each assignment.
Similarly, if paper 2 is accepted, the adversary compares $\reviewerfunctionpdf_1(\score_1)\reviewerfunctionpdf_2(\score_2)(1-\calibrateprob_1)$ and $\reviewerfunctionpdf_2(\score_1)\reviewerfunctionpdf_1(\score_2)\calibrateprob_2$ where it guesses $\assignmentrv = \assignment_1$ when $\reviewerfunctionpdf_1(\score_1)\reviewerfunctionpdf_2(\score_2)(1-\calibrateprob_1) > \reviewerfunctionpdf_2(\score_1)\reviewerfunctionpdf_1(\score_2)\calibrateprob_2$. There are 2 papers and 2 possible assignments, so we have 4 scenarios combining decisions and assignments. 

\begin{enumerate}
    \item Scenario 1: $\assignmentrv = \assignment_1$ and $\decisionrv = P_1$
    
    This scenario happens with probability $\Pr(\assignmentrv = \assignment_1, \decisionrv = P_1 | \scorematrix = [\score_1, \score_2]) = \frac{\reviewerfunctionpdf_1(\score_1)\reviewerfunctionpdf_2(\score_2)\calibrateprob_1}{\reviewerfunctionpdf_1(\score_1)\reviewerfunctionpdf_2(\score_2) + \reviewerfunctionpdf_2(\score_1)\reviewerfunctionpdf_1(\score_2)}$.
    In this scenario, the adversary guesses wrong if $\reviewerfunctionpdf_1(\score_1)\reviewerfunctionpdf_2(\score_2)\calibrateprob_1 < \reviewerfunctionpdf_2(\score_1)\reviewerfunctionpdf_1(\score_2)(1-\calibrateprob_2)$. 
    
    \item Scenario 2: $\assignmentrv = \assignment_1$ and $\decisionrv = P_2$
    
    This scenario happens with probability $\Pr(\assignmentrv = \assignment_1, \decisionrv = P_2 | \scorematrix = [\score_1, \score_2]) = \frac{\reviewerfunctionpdf_1(\score_1)\reviewerfunctionpdf_2(\score_2)(1-\calibrateprob_1)}{\reviewerfunctionpdf_1(\score_1)\reviewerfunctionpdf_2(\score_2) + \reviewerfunctionpdf_2(\score_1)\reviewerfunctionpdf_1(\score_2)}$.
    In this scenario, the adversary guesses wrong if $\reviewerfunctionpdf_1(\score_1)\reviewerfunctionpdf_2(\score_2)(1-\calibrateprob_1) < \reviewerfunctionpdf_2(\score_1)\reviewerfunctionpdf_1(\score_2)\calibrateprob_2$. 
    
    \item Scenario 3: $\assignmentrv = \assignment_2$ and $\decisionrv = P_1$
    
    This scenario happens with probability $\Pr(\assignmentrv = \assignment_1, \decisionrv = P_1 | \scorematrix = [\score_1, \score_2]) = \frac{\reviewerfunctionpdf_2(\score_1)\reviewerfunctionpdf_1(\score_2)(1-\calibrateprob_2)}{\reviewerfunctionpdf_1(\score_1)\reviewerfunctionpdf_2(\score_2) + \reviewerfunctionpdf_2(\score_1)\reviewerfunctionpdf_1(\score_2)}$.
    In this scenario, the adversary guesses wrong if $\reviewerfunctionpdf_1(\score_1)\reviewerfunctionpdf_2(\score_2)\calibrateprob_1 > \reviewerfunctionpdf_2(\score_1)\reviewerfunctionpdf_1(\score_2)(1-\calibrateprob_2)$. 
    
    \item Scenario 4: $\assignmentrv = \assignment_2$ and $\decisionrv = P_2$
    
    This scenario happens with probability $\Pr(\assignmentrv = \assignment_1, \decisionrv = P_2 | \scorematrix = [\score_1, \score_2]) = \frac{\reviewerfunctionpdf_2(\score_1)\reviewerfunctionpdf_1(\score_2)\calibrateprob_2}{\reviewerfunctionpdf_1(\score_1)\reviewerfunctionpdf_2(\score_2) + \reviewerfunctionpdf_2(\score_1)\reviewerfunctionpdf_1(\score_2)}$.
    In this scenario, the adversary guesses wrong if $\reviewerfunctionpdf_1(\score_1)\reviewerfunctionpdf_2(\score_2)(1-\calibrateprob_1) > \reviewerfunctionpdf_2(\score_1)\reviewerfunctionpdf_1(\score_2)\calibrateprob_2$. 
\end{enumerate}

To compute the error of the adversary, we need to compare $\reviewerfunctionpdf_1(\score_1)\reviewerfunctionpdf_2(\score_2)$ and $\reviewerfunctionpdf_2(\score_1)\reviewerfunctionpdf_1(\score_2)$. So as in our assumptions, $\reviewerfunctionpdf_1(\score_1)\reviewerfunctionpdf_2(\score_2) > \reviewerfunctionpdf_2(\score_1)\reviewerfunctionpdf_1(\score_2)$. From the above 4 scenarios, 2 of them compare $\reviewerfunctionpdf_1(\score_1)\reviewerfunctionpdf_2(\score_2)\calibrateprob_1$ with $\reviewerfunctionpdf_2(\score_1)\reviewerfunctionpdf_1(\score_2)(1-\calibrateprob_2)$ and 2 of them compare $\reviewerfunctionpdf_1(\score_1)\reviewerfunctionpdf_2(\score_2)\calibrateprob_1$ with $\reviewerfunctionpdf_1(\score_1)\reviewerfunctionpdf_2(\score_2) - \reviewerfunctionpdf_2(\score_1)\reviewerfunctionpdf_1(\score_2)\calibrateprob_2$. To analyze the error of the adversary, we consider 5 cases of the value of $\reviewerfunctionpdf_1(\score_1)\reviewerfunctionpdf_2(\score_2)\calibrateprob_1$ separated by $\reviewerfunctionpdf_2(\score_1)\reviewerfunctionpdf_1(\score_2)(1-\calibrateprob_2)$ and $\reviewerfunctionpdf_1(\score_1)\reviewerfunctionpdf_2(\score_2) - \reviewerfunctionpdf_2(\score_1)\reviewerfunctionpdf_1(\score_2)\calibrateprob_2$. For each case, we refer to the 4 scenarios of $(\assignmentrv, \decisionrv)$ above. Also note that $\Econference([\score_1, \score_2]) = 1 -\frac{\reviewerfunctionpdf_1(\score_1)\reviewerfunctionpdf_2(\score_2)\calibrateprob_1 + \reviewerfunctionpdf_2(\score_1)\reviewerfunctionpdf_1(\score_2)\calibrateprob_2}{\reviewerfunctionpdf_1(\score_1)\reviewerfunctionpdf_2(\score_2) + \reviewerfunctionpdf_2(\score_1)\reviewerfunctionpdf_1(\score_2)}$ as computed above.

\begin{itemize}
    \item If $\reviewerfunctionpdf_1(\score_1)\reviewerfunctionpdf_2(\score_2)\calibrateprob_1 < \reviewerfunctionpdf_2(\score_1)\reviewerfunctionpdf_1(\score_2) - \reviewerfunctionpdf_2(\score_1)\reviewerfunctionpdf_1(\score_2)\calibrateprob_2$, the adversary guesses wrong in scenarios 1 and 4.
    
    Error of the adversary $\Eadversary([\score_1, \score_2])$ is $\frac{\reviewerfunctionpdf_1(\score_1)\reviewerfunctionpdf_2(\score_2)\calibrateprob_1 + \reviewerfunctionpdf_2(\score_1)\reviewerfunctionpdf_1(\score_2)\calibrateprob_2}{\reviewerfunctionpdf_1(\score_1)\reviewerfunctionpdf_2(\score_2) + \reviewerfunctionpdf_2(\score_1)\reviewerfunctionpdf_1(\score_2)}$. Since $\Econference([\score_1, \score_2]) = 1 -\frac{\reviewerfunctionpdf_1(\score_1)\reviewerfunctionpdf_2(\score_2)\calibrateprob_1 + \reviewerfunctionpdf_2(\score_1)\reviewerfunctionpdf_1(\score_2)\calibrateprob_2}{\reviewerfunctionpdf_1(\score_1)\reviewerfunctionpdf_2(\score_2) + \reviewerfunctionpdf_2(\score_1)\reviewerfunctionpdf_1(\score_2)}$, the relation between error of the adversary and error of the conference is $\Eadversary([\score_1, \score_2]) = 1 - \Econference([\score_1, \score_2])$.
    %which is the opposite of the error of the conference $\Econference([\score_1, \score_2])$. 
    %Error of the adversary ranges from 0 to $\frac{\reviewerfunctionpdf_2(\score_1)\reviewerfunctionpdf_1(\score_2)}{\reviewerfunctionpdf_1(\score_1)\reviewerfunctionpdf_2(\score_2) + \reviewerfunctionpdf_2(\score_1)\reviewerfunctionpdf_1(\score_2)}$. 
    For $0 \le \reviewerfunctionpdf_1(\score_1)\reviewerfunctionpdf_2(\score_2)\calibrateprob_1 < \reviewerfunctionpdf_2(\score_1)\reviewerfunctionpdf_1(\score_2) - \reviewerfunctionpdf_2(\score_1)\reviewerfunctionpdf_1(\score_2)\calibrateprob_2$,  $\Econference([\score_1, \score_2]) \in (\frac{\reviewerfunctionpdf_1(\score_1)\reviewerfunctionpdf_2(\score_2)}{\reviewerfunctionpdf_1(\score_1)\reviewerfunctionpdf_2(\score_2) + \reviewerfunctionpdf_2(\score_1)\reviewerfunctionpdf_1(\score_2)}, 1]$.

    \item If $\reviewerfunctionpdf_1(\score_1)\reviewerfunctionpdf_2(\score_2)\calibrateprob_1 = \reviewerfunctionpdf_2(\score_1)\reviewerfunctionpdf_1(\score_2) - \reviewerfunctionpdf_2(\score_1)\reviewerfunctionpdf_1(\score_2)\calibrateprob_2$, the adversary makes random guess in scenarios 1 and 3 and guesses wrong in scenario 4. 
     
    Error of the adversary $\Eadversary([\score_1, \score_2])$ is $\frac{\reviewerfunctionpdf_2(\score_1)\reviewerfunctionpdf_1(\score_2)}{\reviewerfunctionpdf_1(\score_1)\reviewerfunctionpdf_2(\score_2) + \reviewerfunctionpdf_2(\score_1)\reviewerfunctionpdf_1(\score_2)}$ and error of the conference $\Econference([\score_1, \score_2])$ is $\frac{\reviewerfunctionpdf_1(\score_1)\reviewerfunctionpdf_2(\score_2)}{\reviewerfunctionpdf_1(\score_1)\reviewerfunctionpdf_2(\score_2) + \reviewerfunctionpdf_2(\score_1)\reviewerfunctionpdf_1(\score_2)}$.

    \item If $\reviewerfunctionpdf_2(\score_1)\reviewerfunctionpdf_1(\score_2) - \reviewerfunctionpdf_2(\score_1)\reviewerfunctionpdf_1(\score_2)\calibrateprob_2 < \reviewerfunctionpdf_1(\score_1)\reviewerfunctionpdf_2(\score_2)\calibrateprob_1 < \reviewerfunctionpdf_1(\score_1)\reviewerfunctionpdf_2(\score_2) - \reviewerfunctionpdf_2(\score_1)\reviewerfunctionpdf_1(\score_2)\calibrateprob_2$, the adversary guesses wrong in scenarios 3 and 4. 
    
    Error of the the adversary $\Eadversary([\score_1, \score_2])$ is $\frac{\reviewerfunctionpdf_2(\score_1)\reviewerfunctionpdf_1(\score_2)}{\reviewerfunctionpdf_1(\score_1)\reviewerfunctionpdf_2(\score_2) + \reviewerfunctionpdf_2(\score_1)\reviewerfunctionpdf_1(\score_2)}$, which is constant. %The relation between error of the adversary and error of the conference is $\Eadversary([\score_1, \score_2]) = \frac{\reviewerfunctionpdf_2(\score_1)\reviewerfunctionpdf_1(\score_2)}{\reviewerfunctionpdf_1(\score_1)\reviewerfunctionpdf_2(\score_2) + \reviewerfunctionpdf_2(\score_1)\reviewerfunctionpdf_1(\score_2)}$ 
    In this case, since error of the conference $\Econference([\score_1, \score_2]) = 1 -\frac{\reviewerfunctionpdf_1(\score_1)\reviewerfunctionpdf_2(\score_2)\calibrateprob_1 + \reviewerfunctionpdf_2(\score_1)\reviewerfunctionpdf_1(\score_2)\calibrateprob_2}{\reviewerfunctionpdf_1(\score_1)\reviewerfunctionpdf_2(\score_2) + \reviewerfunctionpdf_2(\score_1)\reviewerfunctionpdf_1(\score_2)}$,  we can find that $\Econference([\score_1, \score_2])$ ranges from $(\frac{\reviewerfunctionpdf_2(\score_1)\reviewerfunctionpdf_1(\score_2)}{\reviewerfunctionpdf_1(\score_1)\reviewerfunctionpdf_2(\score_2) + \reviewerfunctionpdf_2(\score_1)\reviewerfunctionpdf_1(\score_2)}$ to $\frac{\reviewerfunctionpdf_1(\score_1)\reviewerfunctionpdf_2(\score_2)}{\reviewerfunctionpdf_1(\score_1)\reviewerfunctionpdf_2(\score_2) + \reviewerfunctionpdf_2(\score_1)\reviewerfunctionpdf_1(\score_2)})$.

    \item If $\reviewerfunctionpdf_1(\score_1)\reviewerfunctionpdf_2(\score_2)\calibrateprob_1 = \reviewerfunctionpdf_1(\score_1)\reviewerfunctionpdf_2(\score_2) - \reviewerfunctionpdf_2(\score_1)\reviewerfunctionpdf_1(\score_2)\calibrateprob_2$, the adversary makes random guess in scenarios 2 and 4 and guesses wrong in scenario 3. 
    
    Error of the adversary $\Eadversary([\score_1, \score_2])$ is $\frac{\reviewerfunctionpdf_2(\score_1)\reviewerfunctionpdf_1(\score_2)}{\reviewerfunctionpdf_1(\score_1)\reviewerfunctionpdf_2(\score_2) + \reviewerfunctionpdf_2(\score_1)\reviewerfunctionpdf_1(\score_2)}$ and error of the conference $\Econference([\score_1, \score_2])$ is also $\frac{\reviewerfunctionpdf_2(\score_1)\reviewerfunctionpdf_1(\score_2)}{\reviewerfunctionpdf_1(\score_1)\reviewerfunctionpdf_2(\score_2) + \reviewerfunctionpdf_2(\score_1)\reviewerfunctionpdf_1(\score_2)}$.

    \item If $\reviewerfunctionpdf_1(\score_1)\reviewerfunctionpdf_2(\score_2)\calibrateprob_1 > \reviewerfunctionpdf_1(\score_1)\reviewerfunctionpdf_2(\score_2) - \reviewerfunctionpdf_2(\score_1)\reviewerfunctionpdf_1(\score_2)\calibrateprob_2$, the adversary guesses wrong in scenarios 2 and 3. 
    
    Error of the adversary $\Eadversary([\score_1, \score_2])$ is $1 - \frac{\reviewerfunctionpdf_1(\score_1)\reviewerfunctionpdf_2(\score_2)\calibrateprob_1 + \reviewerfunctionpdf_2(\score_1)\reviewerfunctionpdf_1(\score_2)\calibrateprob_2}{\reviewerfunctionpdf_1(\score_1)\reviewerfunctionpdf_2(\score_2) + \reviewerfunctionpdf_2(\score_1)\reviewerfunctionpdf_1(\score_2)}$, which is the same as the error of the conference $\Econference([\score_1, \score_2])$. The relation between error of the adversary and error of the conference is $\Eadversary([\score_1, \score_2]) = \Econference([\score_1, \score_2])$. 
    %Error of the adversary ranges from 0 to $\frac{\reviewerfunctionpdf_2(\score_1)\reviewerfunctionpdf_1(\score_2)}{\reviewerfunctionpdf_1(\score_1)\reviewerfunctionpdf_2(\score_2) + \reviewerfunctionpdf_2(\score_1)\reviewerfunctionpdf_1(\score_2)}$. The relation between error of the adversary and error of the conference is $\Eadversary([\score_1, \score_2]) = \Econference([\score_1, \score_2])$ 
    For $1 \ge \reviewerfunctionpdf_1(\score_1)\reviewerfunctionpdf_2(\score_2)\calibrateprob_1 > \reviewerfunctionpdf_1(\score_1)\reviewerfunctionpdf_2(\score_2) - \reviewerfunctionpdf_2(\score_1)\reviewerfunctionpdf_1(\score_2)\calibrateprob_2$, $\Econference([\score_1, \score_2]) \in [0, \frac{\reviewerfunctionpdf_2(\score_1)\reviewerfunctionpdf_1(\score_2)}{\reviewerfunctionpdf_1(\score_1)\reviewerfunctionpdf_2(\score_2) + \reviewerfunctionpdf_2(\score_1)\reviewerfunctionpdf_1(\score_2)})$.
    
\end{itemize}

\begin{figure}[t!]
\centering
\begin{subfigure}[b]{0.5\textwidth}
  \centering
  \includegraphics[page=1, scale=0.22]{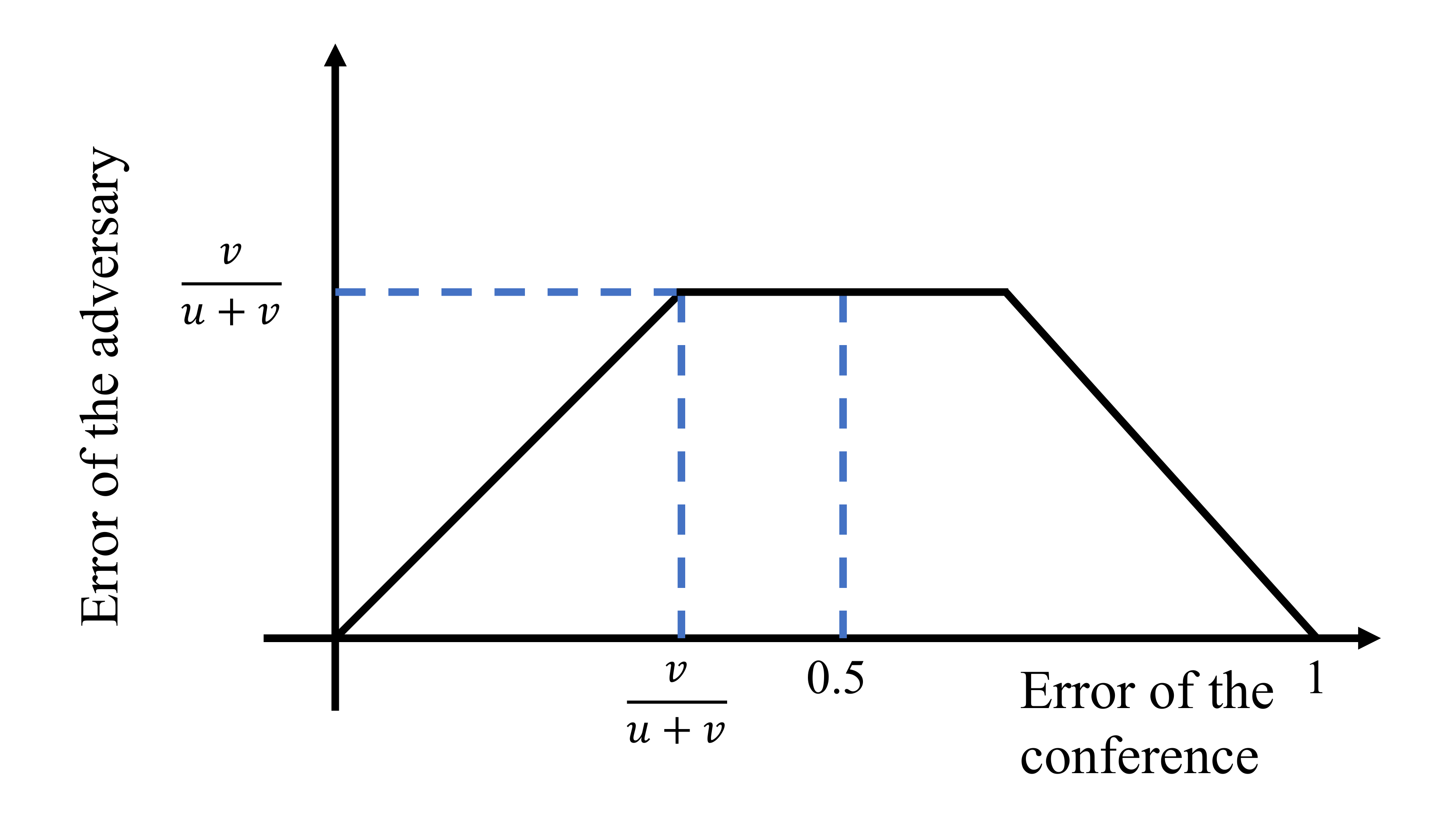}
  \caption{\label{fig:EAECnoiseless1}Maximum per-instance error of the adversary given per-instance error of the conference when $u>v$.}
\end{subfigure}\hspace{.09\textwidth}
\begin{subfigure}[b]{0.5\textwidth}
  \centering
  \includegraphics[page=2, scale=0.22]{figures/EAECnoiseless.pdf}
  \caption{\label{fig:EAECnoiseless2}Maximum per-instance error of the adversary given per-instance error of the conference when $u \le v$}
\end{subfigure}
\caption{\label{fig:EAECnoiseless}Relation between error of the adversary and error of the conference with $u = \reviewerfunctionpdf_1(\score_1)\reviewerfunctionpdf_2(\score_2)$ and $v = \reviewerfunctionpdf_2(\score_1)\reviewerfunctionpdf_1(\score_2)$.} 
\end{figure}

Therefore, the relation between error of the adversary and error of the conference when $\reviewerfunctionpdf_1(\score_1)\reviewerfunctionpdf_2(\score_2) > \reviewerfunctionpdf_2(\score_1)\reviewerfunctionpdf_1(\score_2)$ is of the shape of a trapezoid in $[0,1]$ with the three line segments of the slope +1, 0, and -1 as in Figure~\ref{fig:EAECnoiseless1}. Note that the relation between the per-instance errors does not change with the relation between values of $\reviewerfunctionpdf_1(\score_1)\reviewerfunctionpdf_2(\score_2)$ and $\reviewerfunctionpdf_2(\score_1)\reviewerfunctionpdf_1(\score_2)$. So Figure~\ref{fig:EAECnoiseless1} is the relation between the errors when $u>v$. Similarly, Figure~\ref{fig:EAECnoiseless2} is the relation between the errors when $u \le v$.

From Figure~\ref{fig:EAECnoiseless} we see that the conference should keep its per-instance error less than $\frac{\min\{u,v\}}{u+v}$ to stay optimal. Because if the error of the conference is greater than $\frac{\min\{u,v\}}{u+v}$, increasing its error does not increase the error of the adversary and thus is not optimal. Thus, the Pareto frontier of per-instance error of the adversary against error of the conference is the first line segment with slope 1 in both Figure~\ref{fig:EAECnoiseless1} and Figure~\ref{fig:EAECnoiseless2} when $\min \{\reviewerfunction_2(\reviewerfunction_1^{-1}(\score_2)), \reviewerfunction_1(\reviewerfunction_2^{-1}(\score_2))\} < \score_1 < \max \{\reviewerfunction_2(\reviewerfunction_1^{-1}(\score_2)), \reviewerfunction_1(\reviewerfunction_2^{-1}(\score_2))\}$. 

\subsection{Proof of Theorem~\ref{thm:noiselessInstance}}\label{sec:proof:thm:noiselessInstance}

We prove that Algorithm~\ref{alg:noiselessworste} is optimal for each instance of scores $\scorematrix = [\score_1, \score_2]$ with desired error of the conference $\Econference([\score_1, \score_2])$ in the noiseless setting.

From Theorem~\ref{thm:noiselessPareto} we know that if a paper has higher estimated quality under both assignments, the conference should accept the paper. This is the optimal calibration strategy for the conference.

Otherwise when $\min \{\reviewerfunction_2(\reviewerfunction_1^{-1}(\score_2)), \reviewerfunction_1(\reviewerfunction_2^{-1}(\score_2))\} < \score_1 < \max \{\reviewerfunction_2(\reviewerfunction_1^{-1}(\score_2)), \reviewerfunction_1(\reviewerfunction_2^{-1}(\score_2))\}$, we use the Pareto frontier in Theorem~\ref{thm:noiselessPareto} to explain the optimality of our algorithm. Suppose $\reviewerfunctionpdf_1(\score_1)\reviewerfunctionpdf_2(\score_2) \le \reviewerfunctionpdf_2(\score_1)\reviewerfunctionpdf_1(\score_2)$, then the endpoint on the Pareto frontier has both error of the conference and error of the adversary being $\frac{\reviewerfunctionpdf_1(\score_1)\reviewerfunctionpdf_2(\score_2)}{\reviewerfunctionpdf_1(\score_1)\reviewerfunctionpdf_2(\score_2) + \reviewerfunctionpdf_2(\score_1)\reviewerfunctionpdf_1(\score_2)}$. If $\Econference([\score_1, \score_2]) < \frac{\reviewerfunctionpdf_1(\score_1)\reviewerfunctionpdf_2(\score_2)}{\reviewerfunctionpdf_1(\score_1)\reviewerfunctionpdf_2(\score_2) + \reviewerfunctionpdf_2(\score_1)\reviewerfunctionpdf_1(\score_2)}$, we maximize the error of the adversary by operating on the  Pareto frontier. If $\Econference([\score_1, \score_2]) \ge \frac{\reviewerfunctionpdf_1(\score_1)\reviewerfunctionpdf_2(\score_2)}{\reviewerfunctionpdf_1(\score_1)\reviewerfunctionpdf_2(\score_2) + \reviewerfunctionpdf_2(\score_1)\reviewerfunctionpdf_1(\score_2)}$, we operate at the endpoint where error of the adversary is maximum and error of the conference is no larger than the desired $\Econference([\score_1, \score_2])$. The endpoint is the point with minimum error of the conference such that error of the adversary is maximum. Therefore, it is optimal for the conference.

Similarly, if $\reviewerfunctionpdf_1(\score_1)\reviewerfunctionpdf_2(\score_2) > \reviewerfunctionpdf_2(\score_1)\reviewerfunctionpdf_1(\score_2)$, the algorithm is also optimal by maximizing error of the adversary under desired error of the conference following the Pareto frontier. Algorithm~\ref{alg:noiselessworste} follows the procedure by choosing the corresponding $\calibrateprob_1$ and $\calibrateprob_2$ for each point on the Pareto frontier and thus is optimal for the conference.

\subsection{Proof of Theorem~\ref{thm:noiselessAverage}}\label{sec:proof:thm:noiselessAverage}

Algorithm~\ref{alg:noiselessworste} with $\Econference([\score_1, \score_2]) = 1$ operates on the endpoint of the Pareto frontier when $\min \{\reviewerfunction_2(\reviewerfunction_1^{-1}(\score_2)), \reviewerfunction_1(\reviewerfunction_2^{-1}(\score_2))\} < \score_1 < \max \{\reviewerfunction_2(\reviewerfunction_1^{-1}(\score_2)), \reviewerfunction_1(\reviewerfunction_2^{-1}(\score_2))\}$. We use $\Estrategy$ to denote the error of running Algorithm~\ref{alg:noiselessworste} with $\Econference([\score_1, \score_2]) = 1$ for all $[\score_1, \score_2]$. Then we have Algorithm~\ref{alg:noiselessaveragee} that has a maximum allowable average-case error of the conference $\Econference$ as input. 

If $\Econference \ge \Estrategy$, we operate at $\Econference = \Estrategy$ by running Algorithm~\ref{alg:noiselessworste} with $\Econference([\score_1, \score_2]) = 1$. From Theorem~\ref{thm:noiselessInstance} we know that Algorithm~\ref{alg:noiselessworste} with $\Econference([\score_1, \score_2]) = 1$ is Pareto optimal for all score pairs such that the error of the adversary is maximized and no smaller error of the conference can achieve the same privacy guarantee. Increasing the error of the conference will not increase the error of the adversary. Thus, it is Pareto optimal for the allowable average-case error of the conference.

If $\Econference < \Estrategy$, the coin toss ensures that the average-case error of the conference is $\Estrategy \cdot \frac{\Econference}{\Estrategy} + 0 \cdot (1-\frac{\Econference}{\Estrategy}) = \Econference$. If we use $\eta$ to denote the average-case error of the adversary when the conference always calibrates under the true assignment, then the average-case error of the adversary when the conference runs Algorithm~\ref{alg:noiselessworste} with $\Econference([\score_1, \score_2]) = 1$ is $\Estrategy + \eta$. Because when the conference always calibrates under the true assignment, the adversary only makes error when $\score_1 \le \min \{\reviewerfunction_2(\reviewerfunction_1^{-1}(\score_2)), \reviewerfunction_1(\reviewerfunction_2^{-1}(\score_2))\}$ or $\score_1 \ge \max \{\reviewerfunction_2(\reviewerfunction_1^{-1}(\score_2)), \reviewerfunction_1(\reviewerfunction_2^{-1}(\score_2))\}$. And if the conference adopts Algorithm~\ref{alg:noiselessworste} with $\Econference([\score_1, \score_2]) = 1$, the adversary has error $\eta$ when $\score_1 \le \min \{\reviewerfunction_2(\reviewerfunction_1^{-1}(\score_2)), \reviewerfunction_1(\reviewerfunction_2^{-1}(\score_2))\}$ or $\score_1 \ge \max \{\reviewerfunction_2(\reviewerfunction_1^{-1}(\score_2)), \reviewerfunction_1(\reviewerfunction_2^{-1}(\score_2))\}$ and has error $\Estrategy$ when $\min \{\reviewerfunction_2(\reviewerfunction_1^{-1}(\score_2)), \reviewerfunction_1(\reviewerfunction_2^{-1}(\score_2))\} < \score_1 < \max \{\reviewerfunction_2(\reviewerfunction_1^{-1}(\score_2)), \reviewerfunction_1(\reviewerfunction_2^{-1}(\score_2))\}$. Therefore, the average-case error of the adversary is $(\eta + \Estrategy) \cdot \frac{\Econference}{\Estrategy} + \eta \cdot (1-\frac{\Econference}{\Estrategy}) = \Econference + \eta$. From Theorem~\ref{thm:noiselessInstance} we know that when the conference has per-instance error $\Econference$, the maximum per-instance error of the adversary is $\Econference$ if $\min \{\reviewerfunction_2(\reviewerfunction_1^{-1}(\score_2)), \reviewerfunction_1(\reviewerfunction_2^{-1}(\score_2))\} < \score_1 < \max \{\reviewerfunction_2(\reviewerfunction_1^{-1}(\score_2)), \reviewerfunction_1(\reviewerfunction_2^{-1}(\score_2))\}$. In addition, a Pareto optimal strategy when $\score_1 \le \min \{\reviewerfunction_2(\reviewerfunction_1^{-1}(\score_2)), \reviewerfunction_1(\reviewerfunction_2^{-1}(\score_2))\}$ or $\score_1 \ge \max \{\reviewerfunction_2(\reviewerfunction_1^{-1}(\score_2)), \reviewerfunction_1(\reviewerfunction_2^{-1}(\score_2))\}$ has error of the adversary being $\eta$ and error of the conference being 0. Therefore, for the average-case error of the conference being $\Econference$, the average-case error of the adversary is no larger than $\Econference + \eta$. Therefore Algorithm~\ref{alg:noiselessaveragee} is Pareto optimal.

\subsection{Proof of Theorem~\ref{thm:noisyPareto}}\label{sec:proof:thm:noisyPareto}

To find the Pareto frontier of per-instance error of the adversary against per-instance error of the conference in the noisy setting, we first find the maximum per-instance error of the adversary given per-instance error of the conference in this range.

Prior to computing the errors, we compute the posterior distribution of the quality of the papers given the assignment and scores. We have $\quality_1 | \{\scorematrix = [\score_1, \score_2], \assignmentrv = \assignment_1\} \sim N\left(\frac{a_1(\score_1-b_1)}{a_1^2+\sigma^2}, \frac{\sigma^2}{a_1^2+\sigma^2}\right)$ and $\quality_1 | \{\scorematrix = [\score_1, \score_2], \assignmentrv = \assignment_2\} \sim N\left(\frac{a_2(\score_1-b_2)}{a_2^2+\sigma^2}, \frac{\sigma^2}{a_2^2+\sigma^2}\right)$. Similarly, $\quality_2 | \{\scorematrix = [\score_1, \score_2], \assignmentrv = \assignment_1\} \sim N\left(\frac{a_2(\score_2-b_2)}{a_2^2+\sigma^2}, \frac{\sigma^2}{a_2^2+\sigma^2}\right)$ and $\quality_2 | \{\scorematrix = [\score_1, \score_2], \assignmentrv = \assignment_2\} \sim N\left(\frac{a_1(\score_2-b_1)}{a_1^2+\sigma^2}, \frac{\sigma^2}{a_1^2+\sigma^2}\right)$. We show calculation for one of the posterior distribution. Note that in continuous space, the probability is taken as the density of the corresponding distribution.

\begin{align*}
    & \Pr (\quality_1 = t | \scorematrix = [\score_1, \score_2], \assignmentrv = \assignment_1) \\
    = & \frac{\Pr(\scorematrix = [\score_1, \score_2] | \quality_1 = t, \assignmentrv = \assignment_1) \cdot \Pr(\quality_1 = t | \assignmentrv = \assignment_1)}{\Pr(\scorematrix = [\score_1, \score_2] | \assignmentrv = \assignment_1)} \\
\end{align*}
    
Then we separately compute each term in the equation above. Note that $\score_1 | \{\quality_1 = t, \assignmentrv = \assignment_1\} \sim N(a_1t + b_1, \sigma^2)$ and $\score_1 | \assignmentrv = \assignment_1 \sim N(b_1, a_1^2 + \sigma^2)$. Since $\score_2$ is independent of $\quality_1$ given that $\assignmentrv = \assignment_1$, $\score_2 | \{\quality_1 = t, \assignmentrv = \assignment_1\}$ and $\score_2 | \assignmentrv = \assignment_1$ have the same distribution. In addition, $\quality$ and $\assignment$ are independent.
\begin{align*}
    &\Pr(\scorematrix = [\score_1, \score_2] | \quality_1 = t, \assignmentrv = \assignment_1) \\
    = & \Pr(\scorematrix[1] = \score_1 | \quality_1 = t, \assignmentrv = \assignment_1) \cdot \Pr(\scorematrix[2] = \score_2 | \assignmentrv = \assignment_1) \\
    = & \frac{1}{\sqrt{2\pi} \sigma} e^{-\frac{1}{2}\left(\frac{\score_1-(a_1t+b_1)}{\sigma}\right)^2} \cdot \Pr(\scorematrix[2] = \score_2 | \assignmentrv = \assignment_1) \\
    \\
    & \Pr(\quality_2 = t | \assignmentrv = \assignment_1) \\
    = & \frac{1}{\sqrt{2\pi}} e^{-\frac{1}{2}t^2} \\
    \\
    & \Pr(\scorematrix = [\score_1, \score_2] | \assignmentrv = \assignment_1) \\
    = & \Pr(\scorematrix[1] = \score_1 | \assignmentrv = \assignment_1) \cdot \Pr(\scorematrix[2] = \score_2 | \assignmentrv = \assignment_1) \\
    = & \frac{1}{\sqrt{2\pi}\sqrt{a_1^2+\sigma^2}} e^{-\frac{1}{2}\left(\frac{\score_1-b_1}{\sqrt{a_1^2+\sigma^2}}\right)^2} \cdot \Pr(\scorematrix[2] = \score_2 | \assignmentrv = \assignment_1)
\end{align*}

Therefore, combining the terms we get
\begin{align*}
    & \Pr (\quality_1 = t | \scorematrix = [\score_1, \score_2], \assignmentrv = \assignment_1) \\
    = & \frac{\frac{1}{\sqrt{2\pi} \sigma} e^{-\frac{1}{2}\left(\frac{\score_1-(a_1t+b_1)}{\sigma}\right)^2} \cdot \Pr(\scorematrix[2] = \score_2 | \assignmentrv = \assignment_1) \cdot \frac{1}{\sqrt{2\pi}} e^{-\frac{1}{2}t^2}}{\frac{1}{\sqrt{2\pi}\sqrt{a_1^2+\sigma^2}} e^{-\frac{1}{2}\left(\frac{\score_1-b_1}{\sqrt{a_1^2+\sigma^2}}\right)^2} \cdot \Pr(\scorematrix[2] = \score_2 | \assignmentrv = \assignment_1)} \\
    = & \frac{1}{\sqrt{2\pi}} \sqrt{\frac{a_1^2+\sigma^2}{\sigma^2}} e^{-\frac{1}{2}\left(\left(\frac{\score_1-(a_1t+b_1)}{\sigma}\right)^2 + t^2 - \left(\frac{\score_1-b_1}{\sqrt{a_1^2+\sigma^2}}\right)^2 \right)} \\
    = & \frac{1}{\sqrt{2\pi}} \sqrt{\frac{a_1^2+\sigma^2}{\sigma^2}} e^{-\frac{1}{2} \left( t-\frac{a_1(\score_1-b_1)}{a_1^2+\sigma^2}\right)^2 \cdot \frac{a_1^2+\sigma^2}{\sigma^2}}.
\end{align*}
The other three posteriors are computed in a similar fashion.

Given the posterior distribution of the qualities, we can compute the posterior probability that one paper has higher quality than the other.

\begin{align*}
    & \Pr(\quality_1 > \quality_2  | \assignmentrv = \assignment_1, \scorematrix = [\score_1, \score_2]) \\
    = & \Pr \left(N\left(\frac{a_1(\score_1-b_1)}{a_1^2+\sigma^2}, \frac{\sigma^2}{a_1^2+\sigma^2}\right) > N\left(\frac{a_2(\score_2-b_2)}{a_2^2+\sigma^2}, \frac{\sigma^2}{a_2^2+\sigma^2}\right)\right) \\
    = & \Pr \left(N \left(\frac{a_1(\score_1-b_1)}{a_1^2+\sigma^2} - \frac{a_2(\score_2-b_2)}{a_2^2+\sigma^2}, \frac{\sigma^2}{a_1^2+\sigma^2} + \frac{\sigma^2}{a_2^2+\sigma^2} \right) > 0\right) \\
    = & \Pr \left(\frac{a_1(\score_1-b_1)}{a_1^2+\sigma^2} - \frac{a_2(\score_2-b_2)}{a_2^2+\sigma^2} + \sqrt{\frac{\sigma^2}{a_1^2+\sigma^2} + \frac{\sigma^2}{a_2^2+\sigma^2}} N(0,1) > 0 \right) \\
    = & \Pr \left( N(0,1) > \frac{a_2(a_1^2+\sigma^2)(\score_2-b_2)-a_1(a_2^2+\sigma^2)(\score_1-b_1)}{\sqrt{\sigma^2(a_1^2+a_2^2+2\sigma^2)(a_1^2+\sigma^2)(a_2^2+\sigma^2)}}\right) \\
    = & 1- \normalcdf\left(\frac{a_2(a_1^2+\sigma^2)(\score_2-b_2)-a_1(a_2^2+\sigma^2)(\score_1-b_1)}{\sqrt{\sigma^2(a_1^2+a_2^2+2\sigma^2)(a_1^2+\sigma^2)(a_2^2+\sigma^2)}}\right)
\end{align*}
We use $\normalcdf$ to denote the cumulative distribution function of standard Gaussian distribution. Similarly, we can compute that 
\begin{align*}
    & \Pr(\quality_1 \le \quality_2  | \assignmentrv = \assignment_1, \scorematrix = [\score_1, \score_2]) = \normalcdf\left(\frac{a_2(a_1^2+\sigma^2)(\score_2-b_2)-a_1(a_2^2+\sigma^2)(\score_1-b_1)}{\sqrt{\sigma^2(a_1^2+a_2^2+2\sigma^2)(a_1^2+\sigma^2)(a_2^2+\sigma^2)}}\right) \\
    & \Pr(\quality_1 > \quality_2  | \assignmentrv = \assignment_2, \scorematrix = [\score_1, \score_2]) = 1-\normalcdf\left(\frac{a_1(a_2^2 + \sigma^2)(\score_2-b_1)-a_2(a_1^2+\sigma^2)(\score_1-b_2)}{\sqrt{\sigma^2(a_1^2+a_2^2+2\sigma^2)(a_1^2+\sigma^2)(a_2^2+\sigma^2)}}\right) \\
    & \Pr(\quality_1 \le \quality_2  | \assignmentrv = \assignment_2, \scorematrix = [\score_1, \score_2]) = \normalcdf\left(\frac{a_1(a_2^2 + \sigma^2)(\score_2-b_1)-a_2(a_1^2+\sigma^2)(\score_1-b_2)}{\sqrt{\sigma^2(a_1^2+a_2^2+2\sigma^2)(a_1^2+\sigma^2)(a_2^2+\sigma^2)}}\right)
\end{align*}

For simplicity, let $\Phi_1 = \normalcdf\left(\frac{a_2(a_1^2+\sigma^2)(\score_2-b_2)-a_1(a_2^2+\sigma^2)(\score_1-b_1)}{\sqrt{\sigma^2(a_1^2+a_2^2+2\sigma^2)(a_1^2+\sigma^2)(a_2^2+\sigma^2)}}\right)$ and $\Phi_2 = \normalcdf\left(\frac{a_1(a_2^2 + \sigma^2)(\score_2-b_1)-a_2(a_1^2+\sigma^2)(\score_1-b_2)}{\sqrt{\sigma^2(a_1^2+a_2^2+2\sigma^2)(a_1^2+\sigma^2)(a_2^2+\sigma^2)}}\right)$. Since the conference does calibration using the posterior probabilities, the values of $\Phi_1$ and $\Phi_2$ determines the conference decision. By Proposition~\ref{proposition:calibrationfunction}, we know that the conference should accept the paper with higher estimated quality under both assignments without any calibration. Therefore, if $\Phi_1$ and $\Phi_2$ are both less than $\frac{1}{2}$, the conference should accept paper 1. Similarly, if $\Phi_1$ and $\Phi_2$ are both greater than $\frac{1}{2}$, the conference should accept paper 2. Otherwise, when $\Phi_1 -\frac{1}{2}$ and $\Phi_2 -\frac{1}{2}$ have different signs, the conference should do calibration with function $\calibratefunction$. As before, since $\scorematrix$ is a fixed realization in the analysis, we simplify the calibration strategy for the conference as 
\begin{align*}
    \calibrateprob_1 &= \calibratefunction(\scorematrix, \assignment_1)\\
    \calibrateprob_2 &= \calibratefunction(\scorematrix, \assignment_2).
\end{align*}

We first consider part $(1)$ of the theorem. If $\score_1 > \max \left\{\frac{a_2(a_1^2+\sigma^2)(\score_2-b_2)}{a_1(a_2^2+\sigma^2)} + b_1, \frac{a_1(a_2^2 + \sigma^2)(\score_2-b_1)}{a_2(a_1^2+\sigma^2)}+b_2\right\}$, which is when $\Phi_1 \le \frac{1}{2}$ and $\Phi_2 \le \frac{1}{2}$, the conference accepts paper 1 and the adversary guesses the assignment based on the scores only. Then the error of the conference is the probability that paper 2 has higher quality.

\begin{align*}
    & \Pr(\quality_1 < \quality_2  | \scorematrix = [\score_1, \score_2]) \\
    = & \Pr(\quality_1 < \quality_2  | \assignmentrv = \assignment_1, \scorematrix = [\score_1, \score_2]) \cdot \Pr(\assignmentrv = \assignment_1 | \scorematrix = [\score_1, \score_2]) \\
    & + \Pr(\quality_1 < \quality_2  | \assignmentrv = \assignment_2, \scorematrix = [\score_1, \score_2]) \cdot \Pr(\assignmentrv = \assignment_2 | \scorematrix = [\score_1, \score_2]) \\
    = & \Phi_1 \cdot \frac{\reviewerfunctionpdf_1(\score_1)\reviewerfunctionpdf_2(\score_2)}{\reviewerfunctionpdf_1(\score_1)\reviewerfunctionpdf_2(\score_2) + \reviewerfunctionpdf_2(\score_1)\reviewerfunctionpdf_1(\score_2)} + \Phi_2 \cdot \frac{\reviewerfunctionpdf_2(\score_1)\reviewerfunctionpdf_1(\score_2)}{\reviewerfunctionpdf_1(\score_1)\reviewerfunctionpdf_2(\score_2) + \reviewerfunctionpdf_2(\score_1)\reviewerfunctionpdf_1(\score_2)}
\end{align*}

Similarly, if $\score_1 < \min \left\{\frac{a_2(a_1^2+\sigma^2)(\score_2-b_2)}{a_1(a_2^2+\sigma^2)} + b_1, \frac{a_1(a_2^2 + \sigma^2)(\score_2-b_1)}{a_2(a_1^2+\sigma^2)}+b_2\right\}$, which is when $\Phi_1 \ge \frac{1}{2}$ and $\Phi_2 \ge \frac{1}{2}$, the conference accepts paper 2 and the error of the conference is the probability that paper 1 has higher quality.

\begin{align*}
    & \Pr(\quality_1 > \quality_2  | \scorematrix = [\score_1, \score_2]) \\
    = & \Pr(\quality_1 > \quality_2  | \assignmentrv = \assignment_1, \scorematrix = [\score_1, \score_2]) \cdot \Pr(\assignmentrv = \assignment_1 | \scorematrix = [\score_1, \score_2]) \\
    & + \Pr(\quality_1 > \quality_2  | \assignmentrv = \assignment_2, \scorematrix = [\score_1, \score_2]) \cdot \Pr(\assignmentrv = \assignment_2 | \scorematrix = [\score_1, \score_2]) \\
    = & (1-\Phi_1) \cdot \frac{\reviewerfunctionpdf_1(\score_1)\reviewerfunctionpdf_2(\score_2)}{\reviewerfunctionpdf_1(\score_1)\reviewerfunctionpdf_2(\score_2) + \reviewerfunctionpdf_2(\score_1)\reviewerfunctionpdf_1(\score_2)} + (1-\Phi_2) \cdot \frac{\reviewerfunctionpdf_2(\score_1)\reviewerfunctionpdf_1(\score_2)}{\reviewerfunctionpdf_1(\score_1)\reviewerfunctionpdf_2(\score_2) + \reviewerfunctionpdf_2(\score_1)\reviewerfunctionpdf_1(\score_2)}
\end{align*}

In both cases, error of the adversary is $\frac{\min \{\reviewerfunctionpdf_1(\score_1)\reviewerfunctionpdf_2(\score_2), \reviewerfunctionpdf_2(\score_1)\reviewerfunctionpdf_1(\score_2)\}}{\reviewerfunctionpdf_1(\score_1)\reviewerfunctionpdf_2(\score_2) + \reviewerfunctionpdf_2(\score_1)\reviewerfunctionpdf_1(\score_2)}$, which is the error when the adversary guesses the assignment based on scores only.

We now consider the rest scores in part $(1)$ of the theorem. If $\score_1 = \max \left\{\frac{a_2(a_1^2+\sigma^2)(\score_2-b_2)}{a_1(a_2^2+\sigma^2)} + b_1, \frac{a_1(a_2^2 + \sigma^2)(\score_2-b_1)}{a_2(a_1^2+\sigma^2)}+b_2\right\}$, without loss of generality, we assume $\max \left\{\frac{a_2(a_1^2+\sigma^2)(\score_2-b_2)}{a_1(a_2^2+\sigma^2)} + b_1, \frac{a_1(a_2^2 + \sigma^2)(\score_2-b_1)}{a_2(a_1^2+\sigma^2)}+b_2\right\} = \reviewerfunction_2(\reviewerfunction_1^{-1}(\score_2))$, then the conference accepts each paper uniform at random if calibrating under $\assignment_1$ and accepts paper 1 if calibrating under $\assignment_2$. Since paper 1 has higher or equal quality than paper 2, the conference only has error when paper 2 is accepted and $\assignmentrv = \assignment_2$.

\begin{align*}
    & \Pr(\text{conference accepts lower-quality paper} | \scorematrix = [\score_1, \score_2]) \\
    = & \Pr(\text{conference accepts lower-quality paper} | \scorematrix = [\score_1, \score_2], \decision = P_1) \Pr(\decision = P_1 | \scorematrix = [\score_1, \score_2]) \\
    & + \Pr(\text{conference accepts lower-quality paper} | \scorematrix = [\score_1, \score_2], \decision = P_2) \Pr(\decision = P_2 | \scorematrix = [\score_1, \score_2])\\
    = & \Pr(\quality_1 < \quality_2 | \scorematrix = [\score_1, \score_2]) \Pr(\decision = P_1 | \scorematrix = [\score_1, \score_2]) \\
    & + \Pr(\quality_1 > \quality_2 | \scorematrix = [\score_1, \score_2]) \Pr(\decision = P_2 | \scorematrix = [\score_1, \score_2]).
\end{align*}

Note that in this case, $\Pr(\quality_1 < \quality_2 | \scorematrix = [\score_1, \score_2]) < \Pr(\quality_1 > \quality_2 | \scorematrix = [\score_1, \score_2])$. By similar calculation as in Appendix~\ref{sec:proof:thm:noiselessInstance}, we have

\begin{align*}
    & \Pr(\decision = P_1 | \scorematrix = [\score_1, \score_2]) 
    = \frac{1}{2} (1-\calibrateprob_1) \Pr(\assignmentrv = \assignment_1 | \scorematrix = [\score_1, \score_2]) + \frac{1}{2}\calibrateprob_2 \Pr(\assignmentrv = \assignment_2 | \scorematrix = [\score_1, \score_2])\\
    & \Pr(\decision = P_2 | \scorematrix = [\score_1, \score_2]) 
    = \frac{1}{2} \calibrateprob_1 \Pr(\assignmentrv = \assignment_1 | \scorematrix = [\score_1, \score_2]) + \frac{1}{2}(1-\calibrateprob_2)\Pr(\assignmentrv = \assignment_2 | \scorematrix = [\score_1, \score_2]).
\end{align*}

Error of the conference is then a convex combination of $\Pr(\quality_1 < \quality_2 | \scorematrix = [\score_1, \score_2])$ and $\Pr(\quality_1 > \quality_2 | \scorematrix = [\score_1, \score_2])$ and is minimized when the weight of $\Pr(\quality_1 > \quality_2 | \scorematrix = [\score_1, \score_2])$ is 0.

For the adversary, if paper 1 is accepted, it gains no information on the assignment other than the scores so its error is $\frac{\min \{\reviewerfunctionpdf_1(\score_1)\reviewerfunctionpdf_2(\score_2), \reviewerfunctionpdf_2(\score_1)\reviewerfunctionpdf_1(\score_2)\}}{\reviewerfunctionpdf_1(\score_1)\reviewerfunctionpdf_2(\score_2) + \reviewerfunctionpdf_2(\score_1)\reviewerfunctionpdf_1(\score_2)}$. Otherwise, it guesses $\assignmentrv = \assignment_1$ and its error is $\Pr(\assignmentrv = \assignment_2 | \scorematrix = [\score_1, \score_2])$. Note that error of the adversary does not exceed $\frac{\min \{\reviewerfunctionpdf_1(\score_1)\reviewerfunctionpdf_2(\score_2), \reviewerfunctionpdf_2(\score_1)\reviewerfunctionpdf_1(\score_2)\}}{\reviewerfunctionpdf_1(\score_1)\reviewerfunctionpdf_2(\score_2) + \reviewerfunctionpdf_2(\score_1)\reviewerfunctionpdf_1(\score_2)}$ since in the worst case for the adversary, it guesses the assignment solely based on the scores and ignore the conference decision.

\begin{align*}
    & \Pr(\text{adversary guesses assignment wrong} | \scorematrix = [\score_1, \score_2]) \\
    = & \Pr(\text{adversary guesses assignment wrong} | \scorematrix = [\score_1, \score_2], \decision = P_1) \Pr(\decision = P_1 | \scorematrix = [\score_1, \score_2]) \\
    & + \Pr(\text{adversary guesses assignment wrong} | \scorematrix = [\score_1, \score_2], \decision = P_2) \Pr(\decision = P_2 | \scorematrix = [\score_1, \score_2]) \\
    = & \frac{\min \{\reviewerfunctionpdf_1(\score_1)\reviewerfunctionpdf_2(\score_2), \reviewerfunctionpdf_2(\score_1)\reviewerfunctionpdf_1(\score_2)\}}{\reviewerfunctionpdf_1(\score_1)\reviewerfunctionpdf_2(\score_2) + \reviewerfunctionpdf_2(\score_1)\reviewerfunctionpdf_1(\score_2)} \left( (1-\frac{1}{2}\calibrateprob_1) \Pr(\assignmentrv = \assignment_1 | \scorematrix = [\score_1, \score_2]) + (\frac{1}{2} + \frac{1}{2} \calibrateprob_2) \Pr(\assignmentrv = \assignment_2 | \scorematrix = [\score_1, \score_2]) \right) \\
    & + \Pr(\assignmentrv = \assignment_2 | \scorematrix = [\score_1, \score_2]) \left( \frac{1}{2} \calibrateprob_1 \Pr(\assignmentrv = \assignment_1 | \scorematrix = [\score_1, \score_2]) + \frac{1}{2}(1-\calibrateprob_2)\Pr(\assignmentrv = \assignment_2 | \scorematrix = [\score_1, \score_2]) \right)
\end{align*}

Therefore, we can minimize the error of the conference to 0 by choosing $\calibrateprob_1 = 0$ and $\calibrateprob_2 = 1$, which results in the conference always accepts paper 1. Then error of the adversary is $\frac{\min \{\reviewerfunctionpdf_1(\score_1)\reviewerfunctionpdf_2(\score_2), \reviewerfunctionpdf_2(\score_1)\reviewerfunctionpdf_1(\score_2)\}}{\reviewerfunctionpdf_1(\score_1)\reviewerfunctionpdf_2(\score_2) + \reviewerfunctionpdf_2(\score_1)\reviewerfunctionpdf_1(\score_2)}$, which is maximized. Further increase of error of the conference cannot increase error of the adversary. So the Pareto optimal point is $(\Phi_1 \cdot \frac{\reviewerfunctionpdf_1(\score_1)\reviewerfunctionpdf_2(\score_2)}{\reviewerfunctionpdf_1(\score_1)\reviewerfunctionpdf_2(\score_2) + \reviewerfunctionpdf_2(\score_1)\reviewerfunctionpdf_1(\score_2)} + \Phi_2 \cdot \frac{\reviewerfunctionpdf_2(\score_1)\reviewerfunctionpdf_1(\score_2)}{\reviewerfunctionpdf_1(\score_1)\reviewerfunctionpdf_2(\score_2) + \reviewerfunctionpdf_2(\score_1)\reviewerfunctionpdf_1(\score_2)}, \frac{\min \{\reviewerfunctionpdf_1(\score_1)\reviewerfunctionpdf_2(\score_2), \reviewerfunctionpdf_2(\score_1)\reviewerfunctionpdf_1(\score_2)\}}{\reviewerfunctionpdf_1(\score_1)\reviewerfunctionpdf_2(\score_2) + \reviewerfunctionpdf_2(\score_1)\reviewerfunctionpdf_1(\score_2)})$. The same argument follows when $\score_1 = \min \left\{\frac{a_2(a_1^2+\sigma^2)(\score_2-b_2)}{a_1(a_2^2+\sigma^2)} + b_1, \frac{a_1(a_2^2 + \sigma^2)(\score_2-b_1)}{a_2(a_1^2+\sigma^2)}+b_2\right\}$.

We then look at part $(2)$ of the theorem where the scores lie in the region $\min \left\{\frac{a_2(a_1^2+\sigma^2)(\score_2-b_2)}{a_1(a_2^2+\sigma^2)} + b_1, \frac{a_1(a_2^2 + \sigma^2)(\score_2-b_1)}{a_2(a_1^2+\sigma^2)}+b_2\right\} < \score_1 < \max \left\{\frac{a_2(a_1^2+\sigma^2)(\score_2-b_2)}{a_1(a_2^2+\sigma^2)} + b_1, \frac{a_1(a_2^2 + \sigma^2)(\score_2-b_1)}{a_2(a_1^2+\sigma^2)}+b_2\right\}$.
We will then show the proof with the assumptions that  $\reviewerfunctionpdf_1(\score_1)\reviewerfunctionpdf_2(\score_2) < \reviewerfunctionpdf_2(\score_1)\reviewerfunctionpdf_1(\score_2)$ and $\Phi_1 = \frac{1}{2} - \varphi_1$ and $\Phi_2 = \frac{1}{2} + \varphi_2$ with $0 < \varphi_2 < \varphi_1$. The analysis is of the same procedure for different assumptions on the values of $\reviewerfunctionpdf_1(\score_1)\reviewerfunctionpdf_2(\score_2)$, $\reviewerfunctionpdf_2(\score_1)\reviewerfunctionpdf_1(\score_2)$, $\Phi_1$ and $\Phi_2$ with $\Phi_1 -\frac{1}{2}$ and $\Phi_2 - \frac{1}{2}$ having different signs. The notations are of the same meaning as in Section~\ref{sec:proof:thm:noiselessInstance}. In the noisy setting, even if the conference calibrates under the true assignment, there is still possibility to accept the lower-quality paper due to the noise in the scores given by the reviewers. Note that with the assumptions and when $\min \left\{\frac{a_2(a_1^2+\sigma^2)(\score_2-b_2)}{a_1(a_2^2+\sigma^2)} + b_1, \frac{a_1(a_2^2 + \sigma^2)(\score_2-b_1)}{a_2(a_1^2+\sigma^2)}+b_2\right\} < \score_1 < \max \left\{\frac{a_2(a_1^2+\sigma^2)(\score_2-b_2)}{a_1(a_2^2+\sigma^2)} + b_1, \frac{a_1(a_2^2 + \sigma^2)(\score_2-b_1)}{a_2(a_1^2+\sigma^2)}+b_2\right\}$, the conference accepts paper 1 if calibrates under $\assignment_1$ and accepts paper 2 if calibrates under $\assignment_2$ by the assumptions on $\Phi_1$ and $\Phi_2$. So we have

\begin{align*}
    & \Pr(\text{conference accepts lower-quality paper} | \scorematrix = [\score_1, \score_2]) \\
    = & \Pr(\text{conference accepts }P_1, \quality_1 < \quality_2 | \scorematrix = [\score_1, \score_2]) + \Pr(\text{conference accepts }P_2, \quality_1 > \quality_2 | \scorematrix = [\score_1, \score_2]) \\
    = & \Pr(\text{conference accepts }P_1 | \quality_1 < \quality_2, \scorematrix = [\score_1, \score_2]) \cdot \Pr(\quality_1 < \quality_2 | \scorematrix = [\score_1, \score_2]) \\
    & + \Pr(\text{conference accepts }P_2 | \quality_1 > \quality_2, \scorematrix = [\score_1, \score_2]) \cdot \Pr(\quality_1 > \quality_2 | \scorematrix = [\score_1, \score_2]).
\end{align*}
We then expand each of the two terms.
\begin{align*}
    & \Pr(\text{conference accepts }P_1 | \quality_1 < \quality_2, \scorematrix = [\score_1, \score_2]) \\
    = & \Pr(\text{conference accepts }P_1, \assignmentrv = \assignment_1 | \quality_1 < \quality_2, \scorematrix = [\score_1, \score_2]) + \Pr(\text{conference accepts }P_1, \assignmentrv = \assignment_2 | \quality_1 < \quality_2, \scorematrix = [\score_1, \score_2]) \\
    = & \Pr(\text{conference accepts }P_1 | \assignmentrv = \assignment_1, \quality_1 < \quality_2, \scorematrix = [\score_1, \score_2]) \cdot P(\assignmentrv = \assignment_1 | \quality_1 < \quality_2, \scorematrix = [\score_1, \score_2]) \\
    & + \Pr(\text{conference accepts }P_1 | \assignmentrv = \assignment_2, \quality_1 < \quality_2, \scorematrix = [\score_1, \score_2]) \cdot \Pr(\assignmentrv = \assignment_2 | \quality_1 < \quality_2, \scorematrix = [\score_1, \score_2]) \\
    = & \Pr(\calibrationrv=T | \assignmentrv = \assignment_1, \quality_1 < \quality_2, \scorematrix = [\score_1, \score_2]) \Pr(\assignmentrv = \assignment_1 | \quality_1 < \quality_2, \scorematrix = [\score_1, \score_2]) \\
    & + \Pr(\calibrationrv=F | \assignmentrv = \assignment_2, \quality_1 < \quality_2, \scorematrix = [\score_1, \score_2]) \Pr(\assignmentrv = \assignment_2 | \quality_1 < \quality_2, \scorematrix = [\score_1, \score_2]) \\
    = & \calibrateprob_1 \Pr(\assignmentrv = \assignment_1 | \quality_1 < \quality_2, \scorematrix = [\score_1, \score_2]) + (1-\calibrateprob_2) \Pr(\assignmentrv = \assignment_2 | \quality_1 < \quality_2, \scorematrix = [\score_1, \score_2]) \\
    = & \calibrateprob_1 \frac{\Pr(\assignmentrv = \assignment_1, \quality_1 < \quality_2 | \scorematrix = [\score_1, \score_2])}{\Pr(\quality_1 < \quality_2 | \scorematrix = [\score_1, \score_2])}  + (1-q_2) \frac{\Pr(\assignmentrv = \assignment_2, \quality_1 < \quality_2 | \scorematrix = [\score_1, \score_2])}{\Pr(\quality_1 < \quality_2 | \scorematrix = [\score_1, \score_2])}\\
    = & \calibrateprob_1 \frac{\Pr(\quality_1 < \quality_2 | \assignmentrv = \assignment_1, \scorematrix = [\score_1, \score_2]) \cdot \Pr(\assignmentrv = \assignment_1 | \scorematrix = [\score_1, \score_2])}{\Pr(\quality_1 < \quality_2 | \scorematrix = [\score_1, \score_2])} \\
    & + (1-\calibrateprob_2) \frac{\Pr(\quality_1 < \quality_2 | \assignmentrv = \assignment_2, \scorematrix = [\score_1, \score_2]) \cdot \Pr(\assignmentrv = \assignment_2 | \scorematrix = [\score_1, \score_2])}{\Pr(\quality_1 < \quality_2 | \scorematrix = [\score_1, \score_2])}.
\end{align*}
Similarly, 
\begin{align*}
    & \Pr(\text{conference accepts }P_2 | \quality_1 > \quality_2, \scorematrix = [\score_1, \score_2]) \\
    = & (1-\calibrateprob_1) \frac{\Pr(\quality_1 > \quality_2 | \assignmentrv = \assignment_1, \scorematrix = [\score_1, \score_2]) \cdot \Pr(\assignmentrv = \assignment_1 | \scorematrix = [\score_1, \score_2])}{\Pr(\quality_1 > \quality_2 | \scorematrix = [\score_1, \score_2])} \\
    & + \calibrateprob_2 \frac{\Pr(\quality_1 > \quality_2 | \assignmentrv = \assignment_2, \scorematrix = [\score_1, \score_2]) \cdot \Pr(\assignmentrv = \assignment_2 | \scorematrix = [\score_1, \score_2])}{\Pr(\quality_1 > \quality_2 | \scorematrix = [\score_1, \score_2])}
\end{align*}

Therefore, we have
\begin{align*}
    & \Pr(\text{conference accepts lower-quality paper} | \scorematrix = [\score_1, \score_2]) \\
    = & \calibrateprob_1 \Pr(\quality_1 < \quality_2 | \assignmentrv = \assignment_1, \scorematrix = [\score_1, \score_2]) \cdot \Pr(\assignmentrv = \assignment_1 | \scorematrix = [\score_1, \score_2]) \\
    & + (1-\calibrateprob_2) \Pr(\quality_1 < \quality_2 | \assignmentrv = \assignment_2, \scorematrix = [\score_1, \score_2]) \cdot \Pr(\assignmentrv = \assignment_2 | \scorematrix = [\score_1, \score_2]) \\
    & + (1-\calibrateprob_1) \Pr(\quality_1 > \quality_2 | \assignmentrv = \assignment_1, \scorematrix = [\score_1, \score_2]) \cdot \Pr(\assignmentrv = \assignment_1 | \scorematrix = [\score_1, \score_2]) \\
    & + \calibrateprob_2 \Pr(\quality_1 > \quality_2 | \assignmentrv = \assignment_2, \scorematrix = [\score_1, \score_2]) \cdot \Pr(\assignmentrv = \assignment_2 | \scorematrix = [\score_1, \score_2]) \\
    = & \frac{\reviewerfunctionpdf_1(\score_1)\reviewerfunctionpdf_2(\score_2)}{\reviewerfunctionpdf_1(\score_1)\reviewerfunctionpdf_2(\score_2) + \reviewerfunctionpdf_2(\score_1)\reviewerfunctionpdf_1(\score_2)} \cdot (\calibrateprob_1 \Phi_1 + (1-\calibrateprob_1)(1-\Phi_1)) \\
    & + \frac{\reviewerfunctionpdf_2(\score_1)\reviewerfunctionpdf_1(\score_2)}{\reviewerfunctionpdf_1(\score_1)\reviewerfunctionpdf_2(\score_2) + \reviewerfunctionpdf_2(\score_1)\reviewerfunctionpdf_1(\score_2)} \cdot ((1-\calibrateprob_2) \Phi_2 + \calibrateprob_2 (1-\Phi_2)).
\end{align*}

Under the assumptions that $\Phi_1 = \frac{1}{2} - \varphi_1$ and $\Phi_2 = \frac{1}{2} + \varphi_2$ where $0 < \varphi_2 < \varphi_1$ and $\reviewerfunctionpdf_1(\score_1)\reviewerfunctionpdf_2(\score_2) < \reviewerfunctionpdf_2(\score_1)\reviewerfunctionpdf_1(\score_2)$, we analyze the per-instance error of the adversary similar to the procedure in Section~\ref{sec:proof:thm:noiselessPareto}. There are 4 scenarios combining the decision and the true assignment.

\begin{enumerate}
    \item Scenario 1: $\assignmentrv = \assignment_1$ and $\decisionrv = P_1$ 
    
    This scenario happens with probability $\Pr(\assignmentrv = \assignment_1, \decisionrv = P_1 | \scorematrix = [\score_1, \score_2]) = \frac{\reviewerfunctionpdf_1(\score_1)\reviewerfunctionpdf_2(\score_2)\calibrateprob_1}{\reviewerfunctionpdf_1(\score_1)\reviewerfunctionpdf_2(\score_2) + \reviewerfunctionpdf_2(\score_1)\reviewerfunctionpdf_1(\score_2)}$. 
    In this scenario, the adversary guesses wrong if $\calibrateprob_1 \reviewerfunctionpdf_1(\score_1)\reviewerfunctionpdf_2(\score_2) < (1-\calibrateprob_2)\reviewerfunctionpdf_2(\score_1)\reviewerfunctionpdf_1(\score_2)$. 
    
    \item Scenario 2: $\assignmentrv = \assignment_1$ and $\decisionrv = P_2$
    
    This scenario happens with probability $\Pr(\assignmentrv = \assignment_1, \decisionrv = P_1 | \scorematrix = [\score_1, \score_2]) = \frac{\reviewerfunctionpdf_1(\score_1)\reviewerfunctionpdf_2(\score_2)(1-\calibrateprob_1)}{\reviewerfunctionpdf_1(\score_1)\reviewerfunctionpdf_2(\score_2) + \reviewerfunctionpdf_2(\score_1)\reviewerfunctionpdf_1(\score_2)}$.
    In this scenario, the adversary guesses wrong if $(1-\calibrateprob_1) \reviewerfunctionpdf_1(\score_1)\reviewerfunctionpdf_2(\score_2) < \calibrateprob_2\reviewerfunctionpdf_2(\score_1)\reviewerfunctionpdf_1(\score_2)$. 
    
    \item Scenario 3: $\assignmentrv = \assignment_2$ and $\decisionrv = P_1$
    
    This scenario happens with probability $\Pr(\assignmentrv = \assignment_1, \decisionrv = P_1 | \scorematrix = [\score_1, \score_2]) = \frac{\reviewerfunctionpdf_2(\score_1)\reviewerfunctionpdf_1(\score_2)(1-\calibrateprob_2)}{\reviewerfunctionpdf_1(\score_1)\reviewerfunctionpdf_2(\score_2) + \reviewerfunctionpdf_2(\score_1)\reviewerfunctionpdf_1(\score_2)}$.
    In this scenario, the adversary guesses wrong if $\calibrateprob_1 \reviewerfunctionpdf_1(\score_1)\reviewerfunctionpdf_2(\score_2) > (1-\calibrateprob_2)\reviewerfunctionpdf_2(\score_1)\reviewerfunctionpdf_1(\score_2)$. 
    
     \item Scenario 4: $\assignmentrv = \assignment_2$ and $\decisionrv = P_2$
     
     This scenario happens with probability $\Pr(\assignmentrv = \assignment_1, \decisionrv = P_1 | \scorematrix = [\score_1, \score_2]) = \frac{\reviewerfunctionpdf_2(\score_1)\reviewerfunctionpdf_1(\score_2)\calibrateprob_2}{\reviewerfunctionpdf_1(\score_1)\reviewerfunctionpdf_2(\score_2) + \reviewerfunctionpdf_2(\score_1)\reviewerfunctionpdf_1(\score_2)}$.
     In this scenario, the adversary guesses wrong if $(1-\calibrateprob_1) \reviewerfunctionpdf_1(\score_1)\reviewerfunctionpdf_2(\score_2) > \calibrateprob_2\reviewerfunctionpdf_2(\score_1)\reviewerfunctionpdf_1(\score_2)$. 
\end{enumerate}

To compute the error of the adversary, we need to compare $\reviewerfunctionpdf_1(\score_1)\reviewerfunctionpdf_2(\score_2)$ and $\reviewerfunctionpdf_2(\score_1)\reviewerfunctionpdf_1(\score_2)$. So we suppose $\reviewerfunctionpdf_1(\score_1)\reviewerfunctionpdf_2(\score_2) < \reviewerfunctionpdf_2(\score_1)\reviewerfunctionpdf_1(\score_2)$. From the above 4 scenarios, 2 of them compare $\reviewerfunctionpdf_1(\score_1)\reviewerfunctionpdf_2(\score_2)\calibrateprob_1$ with $\reviewerfunctionpdf_2(\score_1)\reviewerfunctionpdf_1(\score_2)(1-\calibrateprob_2)$ and 2 of them compare $\reviewerfunctionpdf_1(\score_1)\reviewerfunctionpdf_2(\score_2)\calibrateprob_1$ with $\reviewerfunctionpdf_1(\score_1)\reviewerfunctionpdf_2(\score_2) - \reviewerfunctionpdf_2(\score_1)\reviewerfunctionpdf_1(\score_2)\calibrateprob_2$. To analyze the error of the adversary, we consider 5 cases of the value of $\reviewerfunctionpdf_1(\score_1)\reviewerfunctionpdf_2(\score_2)\calibrateprob_1$ separated by $\reviewerfunctionpdf_2(\score_1)\reviewerfunctionpdf_1(\score_2)(1-\calibrateprob_2)$ and $\reviewerfunctionpdf_1(\score_1)\reviewerfunctionpdf_2(\score_2) - \reviewerfunctionpdf_2(\score_1)\reviewerfunctionpdf_1(\score_2)\calibrateprob_2$. We refer to the 4 scenarios of $(\assignmentrv, \decisionrv)$ above. 

\begin{itemize}
    \item If $\calibrateprob_1 \reviewerfunctionpdf_1(\score_1)\reviewerfunctionpdf_2(\score_2) < \reviewerfunctionpdf_1(\score_1)\reviewerfunctionpdf_2(\score_2) - \calibrateprob_2\reviewerfunctionpdf_2(\score_1)\reviewerfunctionpdf_1(\score_2)$, the adversary guesses wrong in scenarios 1 and 4. Error of the adversary $\Eadversary([\score_1, \score_2])$ is $\frac{\calibrateprob_1\reviewerfunctionpdf_1(\score_1)\reviewerfunctionpdf_2(\score_2) + \calibrateprob_2\reviewerfunctionpdf_2(\score_1)\reviewerfunctionpdf_1(\score_2)}{\reviewerfunctionpdf_1(\score_1)\reviewerfunctionpdf_2(\score_2) + \reviewerfunctionpdf_2(\score_1)\reviewerfunctionpdf_1(\score_2)}$.
    
    \item If $\calibrateprob_1 \reviewerfunctionpdf_1(\score_1)\reviewerfunctionpdf_2(\score_2) = \reviewerfunctionpdf_1(\score_1)\reviewerfunctionpdf_2(\score_2) - \calibrateprob_2\reviewerfunctionpdf_2(\score_1)\reviewerfunctionpdf_1(\score_2)$, the adversary makes random guess in scenarios 2 and 4 and guesses wrong in scenario 1. Error of the adversary $\Eadversary([\score_1, \score_2])$ is $\frac{\calibrateprob_1\reviewerfunctionpdf_1(\score_1)\reviewerfunctionpdf_2(\score_2)}{\reviewerfunctionpdf_1(\score_1)\reviewerfunctionpdf_2(\score_2) + \reviewerfunctionpdf_2(\score_1)\reviewerfunctionpdf_1(\score_2)} + \frac{1}{2}(\frac{(1-\calibrateprob_1)\reviewerfunctionpdf_1(\score_1)\reviewerfunctionpdf_2(\score_2)}{\reviewerfunctionpdf_1(\score_1)\reviewerfunctionpdf_2(\score_2) + \reviewerfunctionpdf_2(\score_1)\reviewerfunctionpdf_1(\score_2)} + \frac{\calibrateprob_2\reviewerfunctionpdf_2(\score_1)\reviewerfunctionpdf_1(\score_2)}{\reviewerfunctionpdf_1(\score_1)\reviewerfunctionpdf_2(\score_2) + \reviewerfunctionpdf_2(\score_1)\reviewerfunctionpdf_1(\score_2)}) = \frac{\reviewerfunctionpdf_1(\score_1)\reviewerfunctionpdf_2(\score_2)}{\reviewerfunctionpdf_1(\score_1)\reviewerfunctionpdf_2(\score_2) + \reviewerfunctionpdf_2(\score_1)\reviewerfunctionpdf_1(\score_2)}$.
    
    \item If $\reviewerfunctionpdf_1(\score_1)\reviewerfunctionpdf_2(\score_2) - \calibrateprob_2\reviewerfunctionpdf_2(\score_1)\reviewerfunctionpdf_1(\score_2) < \calibrateprob_1 \reviewerfunctionpdf_1(\score_1)\reviewerfunctionpdf_2(\score_2) < \reviewerfunctionpdf_2(\score_1)\reviewerfunctionpdf_1(\score_2) - \calibrateprob_2\reviewerfunctionpdf_2(\score_1)\reviewerfunctionpdf_1(\score_2)$, the adversary guesses wrong in scenarios 1 and 2. Error of the adversary $\Eadversary([\score_1, \score_2])$ is $\frac{\reviewerfunctionpdf_1(\score_1)\reviewerfunctionpdf_2(\score_2)}{\reviewerfunctionpdf_1(\score_1)\reviewerfunctionpdf_2(\score_2) + \reviewerfunctionpdf_2(\score_1)\reviewerfunctionpdf_1(\score_2)}$.
    
    \item If $\calibrateprob_1 \reviewerfunctionpdf_1(\score_1)\reviewerfunctionpdf_2(\score_2) = \reviewerfunctionpdf_2(\score_1)\reviewerfunctionpdf_1(\score_2) - \calibrateprob_2\reviewerfunctionpdf_2(\score_1)\reviewerfunctionpdf_1(\score_2)$, the adversary makes random guess in scenarios 1 and 3 and guesses wrong in scenario 2. Error of the adversary $\Eadversary([\score_1, \score_2])$ is $\frac{(1-\calibrateprob_1)\reviewerfunctionpdf_1(\score_1)\reviewerfunctionpdf_2(\score_2)}{\reviewerfunctionpdf_1(\score_1)\reviewerfunctionpdf_2(\score_2) + \reviewerfunctionpdf_2(\score_1)\reviewerfunctionpdf_1(\score_2)} + \frac{1}{2} (\frac{\calibrateprob_1\reviewerfunctionpdf_1(\score_1)\reviewerfunctionpdf_2(\score_2)}{\reviewerfunctionpdf_1(\score_1)\reviewerfunctionpdf_2(\score_2) + \reviewerfunctionpdf_2(\score_1)\reviewerfunctionpdf_1(\score_2)} + \frac{(1-\calibrateprob_2)\reviewerfunctionpdf_2(\score_1)\reviewerfunctionpdf_1(\score_2)}{\reviewerfunctionpdf_1(\score_1)\reviewerfunctionpdf_2(\score_2) + \reviewerfunctionpdf_2(\score_1)\reviewerfunctionpdf_1(\score_2)}) = \frac{\reviewerfunctionpdf_1(\score_1)\reviewerfunctionpdf_2(\score_2)}{\reviewerfunctionpdf_1(\score_1)\reviewerfunctionpdf_2(\score_2) + \reviewerfunctionpdf_2(\score_1)\reviewerfunctionpdf_1(\score_2)}$.
    
    \item If $\calibrateprob_1 \reviewerfunctionpdf_1(\score_1)\reviewerfunctionpdf_2(\score_2) > \reviewerfunctionpdf_2(\score_1)\reviewerfunctionpdf_1(\score_2) - \calibrateprob_2\reviewerfunctionpdf_2(\score_1)\reviewerfunctionpdf_1(\score_2)$, the adversary guesses wrong in scenarios 2 and 3. Error of the adversary $\Eadversary([\score_1, \score_2])$ is $1 - \frac{\calibrateprob_1\reviewerfunctionpdf_1(\score_1)\reviewerfunctionpdf_2(\score_2) + \calibrateprob_2\reviewerfunctionpdf_2(\score_1)\reviewerfunctionpdf_1(\score_2)}{\reviewerfunctionpdf_1(\score_1)\reviewerfunctionpdf_2(\score_2) + \reviewerfunctionpdf_2(\score_1)\reviewerfunctionpdf_1(\score_2)}$.
\end{itemize}

To find the maximum error of the adversary given error of the conference, we solve an optimization problem. In order to formulate the optimization problem, we can combine the 5 cases above into 3 cases for simplicity.

\begin{itemize}
    \item If $\calibrateprob_1 \reviewerfunctionpdf_1(\score_1)\reviewerfunctionpdf_2(\score_2) \le \reviewerfunctionpdf_1(\score_1)\reviewerfunctionpdf_2(\score_2) - \calibrateprob_2\reviewerfunctionpdf_2(\score_1)\reviewerfunctionpdf_1(\score_2)$, error of the adversary $\Eadversary([\score_1, \score_2])$ is $\frac{\calibrateprob_1\reviewerfunctionpdf_1(\score_1)\reviewerfunctionpdf_2(\score_2) + \calibrateprob_2\reviewerfunctionpdf_2(\score_1)\reviewerfunctionpdf_1(\score_2)}{\reviewerfunctionpdf_1(\score_1)\reviewerfunctionpdf_2(\score_2) + \reviewerfunctionpdf_2(\score_1)\reviewerfunctionpdf_1(\score_2)}$.
    
    \item If $\reviewerfunctionpdf_1(\score_1)\reviewerfunctionpdf_2(\score_2) - \calibrateprob_2\reviewerfunctionpdf_2(\score_1)\reviewerfunctionpdf_1(\score_2) \le \calibrateprob_1 \reviewerfunctionpdf_1(\score_1)\reviewerfunctionpdf_2(\score_2) \le \reviewerfunctionpdf_2(\score_1)\reviewerfunctionpdf_1(\score_2) - \calibrateprob_2\reviewerfunctionpdf_2(\score_1)\reviewerfunctionpdf_1(\score_2)$, error of the adversary $\Eadversary([\score_1, \score_2])$ is $\frac{\reviewerfunctionpdf_1(\score_1)\reviewerfunctionpdf_2(\score_2)}{\reviewerfunctionpdf_1(\score_1)\reviewerfunctionpdf_2(\score_2) + \reviewerfunctionpdf_2(\score_1)\reviewerfunctionpdf_1(\score_2)}$.
    
    \item If $\calibrateprob_1 \reviewerfunctionpdf_1(\score_1)\reviewerfunctionpdf_2(\score_2) \ge \reviewerfunctionpdf_2(\score_1)\reviewerfunctionpdf_1(\score_2) - \calibrateprob_2\reviewerfunctionpdf_2(\score_1)\reviewerfunctionpdf_1(\score_2)$, error of the adversary $\Eadversary([\score_1, \score_2])$ is $1 - \frac{\calibrateprob_1\reviewerfunctionpdf_1(\score_1)\reviewerfunctionpdf_2(\score_2) + \calibrateprob_2\reviewerfunctionpdf_2(\score_1)\reviewerfunctionpdf_1(\score_2)}{\reviewerfunctionpdf_1(\score_1)\reviewerfunctionpdf_2(\score_2) + \reviewerfunctionpdf_2(\score_1)\reviewerfunctionpdf_1(\score_2)}$.
\end{itemize}

We let $T(\Econference) = \Econference (u + v) - u \cdot (1-\Phi_1) - v \cdot \Phi_2$ to be a function that takes the error of the conference as input.

\begin{itemize}
    \item Maximize $\frac{\calibrateprob_1\reviewerfunctionpdf_1(\score_1)\reviewerfunctionpdf_2(\score_2) + \calibrateprob_2\reviewerfunctionpdf_2(\score_1)\reviewerfunctionpdf_1(\score_2)}{\reviewerfunctionpdf_1(\score_1)\reviewerfunctionpdf_2(\score_2) + \reviewerfunctionpdf_2(\score_1)\reviewerfunctionpdf_1(\score_2)}$ subject to $\Econference([\score_1, \score_2]) (\reviewerfunctionpdf_1(\score_1)\reviewerfunctionpdf_2(\score_2) + \reviewerfunctionpdf_2(\score_1)\reviewerfunctionpdf_1(\score_2)) - \reviewerfunctionpdf_1(\score_1)\reviewerfunctionpdf_2(\score_2) \cdot (1-\Phi_1) - \reviewerfunctionpdf_2(\score_1)\reviewerfunctionpdf_1(\score_2) \cdot \Phi_2 = \reviewerfunctionpdf_1(\score_1)\reviewerfunctionpdf_2(\score_2) (2\Phi_1-1) \calibrateprob_1 + \reviewerfunctionpdf_2(\score_1)\reviewerfunctionpdf_1(\score_2) \cdot (1-2\Phi_2) \calibrateprob_2$ and $\calibrateprob_1 \reviewerfunctionpdf_1(\score_1)\reviewerfunctionpdf_2(\score_2) \le \reviewerfunctionpdf_1(\score_1)\reviewerfunctionpdf_2(\score_2) - \calibrateprob_2\reviewerfunctionpdf_2(\score_1)\reviewerfunctionpdf_1(\score_2)$. 
    
    The maximum occurs at $\calibrateprob_1 \reviewerfunctionpdf_1(\score_1)\reviewerfunctionpdf_2(\score_2) = \reviewerfunctionpdf_1(\score_1)\reviewerfunctionpdf_2(\score_2) - \calibrateprob_2\reviewerfunctionpdf_2(\score_1)\reviewerfunctionpdf_1(\score_2)$. Then the intersection of the two lines is $\calibrateprob_1 = 1 - \frac{(2\Phi_1 - 1)u-T(\Econference([\score_1, \score_2]))}{(2\Phi_1+2\Phi_2-2)u}$ and $\calibrateprob_2 = \frac{(2\Phi_1 - 1)u-T(\Econference([\score_1, \score_2]))}{(2\Phi_1+2\Phi_2-2)v}$.

    \begin{itemize}
        \item If the intersection point can be reached, $\calibrateprob_1, \calibrateprob_2 \in [0,1]$, $(2\Phi_1 - 1)u \le T(\Econference([\score_1, \score_2])) \le (1 - 2\Phi_2)u$, then error of the conference $\Econference([\score_1, \score_2])$ ranges from 
        $\frac{\reviewerfunctionpdf_1(\score_1)\reviewerfunctionpdf_2(\score_2) \Phi_1}{\reviewerfunctionpdf_1(\score_1)\reviewerfunctionpdf_2(\score_2) + \reviewerfunctionpdf_2(\score_1)\reviewerfunctionpdf_1(\score_2)}  + \frac{ \reviewerfunctionpdf_2(\score_1)\reviewerfunctionpdf_1(\score_2) \Phi_2}{\reviewerfunctionpdf_1(\score_1)\reviewerfunctionpdf_2(\score_2) + \reviewerfunctionpdf_2(\score_1)\reviewerfunctionpdf_1(\score_2)}$ to 
        $\frac{\reviewerfunctionpdf_1(\score_1)\reviewerfunctionpdf_2(\score_2) (2-\Phi_1 - 2\Phi_2)}{\reviewerfunctionpdf_1(\score_1)\reviewerfunctionpdf_2(\score_2) + \reviewerfunctionpdf_2(\score_1)\reviewerfunctionpdf_1(\score_2)} + \frac{ \reviewerfunctionpdf_2(\score_1)\reviewerfunctionpdf_1(\score_2) \Phi_2}{\reviewerfunctionpdf_1(\score_1)\reviewerfunctionpdf_2(\score_2) + \reviewerfunctionpdf_2(\score_1)\reviewerfunctionpdf_1(\score_2)}$.
        
        Error of the adversary $\Eadversary([\score_1, \score_2])$ is $\frac{\reviewerfunctionpdf_1(\score_1)\reviewerfunctionpdf_2(\score_2)}{\reviewerfunctionpdf_1(\score_1)\reviewerfunctionpdf_2(\score_2) + \reviewerfunctionpdf_2(\score_1)\reviewerfunctionpdf_1(\score_2)}$.
        
        \item If the intersection point can not be reached and $T(\Econference([\score_1, \score_2])) < (2\Phi_1 - 1)u$, then no $\calibrateprob_1, \calibrateprob_2$ are qualified for the constraints.
        
        \item If the intersection point can not be reached and  $T(\Econference([\score_1, \score_2])) > (1 - 2\Phi_2)u$.
        \begin{itemize}
            \item If $(1 - 2\Phi_2)u < T(\Econference([\score_1, \score_2])) \le 0$ then the maximum is reached when $\calibrateprob_1 = 0$ and $\calibrateprob_2 = \frac{T(\Econference([\score_1, \score_2]))}{(1 - 2\Phi_2)v}$. 
        
            Error of the conference $\Econference([\score_1, \score_2])$ ranges from $\frac{(2-\Phi_1-2\Phi_2)u+\Phi_2v}{u+v}$ (when $T(\Econference([\score_1, \score_2])) = (1 - 2\Phi_2)u$) to $\frac{(1-\Phi_1)u+\Phi_2v}{u+v}$ (when $T(\Econference([\score_1, \score_2])) = 0$). 
        
            Error of the adversary $\Eadversary([\score_1, \score_2])$ is $\frac{T(\Econference([\score_1, \score_2]))}{(1-2\Phi_2)(u+v)}$, ranges from $\frac{u}{u+v}$ (when $T(\Econference([\score_1, \score_2])) = (1 - 2\Phi_2)u$) to 0 (when $T(\Econference([\score_1, \score_2])) = 0$).
            
            \item If $ T(\Econference([\score_1, \score_2])) > 0$ then no $\calibrateprob_1, \calibrateprob_2$ are qualified for the constraints.
        \end{itemize}
    \end{itemize}

    \item Error of the adversary $\Eadversary([\score_1, \score_2])$ is $\frac{\reviewerfunctionpdf_1(\score_1)\reviewerfunctionpdf_2(\score_2)}{\reviewerfunctionpdf_1(\score_1)\reviewerfunctionpdf_2(\score_2) + \reviewerfunctionpdf_2(\score_1)\reviewerfunctionpdf_1(\score_2)}$ subject to $\reviewerfunctionpdf_1(\score_1)\reviewerfunctionpdf_2(\score_2) - \calibrateprob_2\reviewerfunctionpdf_2(\score_1)\reviewerfunctionpdf_1(\score_2) \le \calibrateprob_1 \reviewerfunctionpdf_1(\score_1)\reviewerfunctionpdf_2(\score_2) \le \reviewerfunctionpdf_2(\score_1)\reviewerfunctionpdf_1(\score_2) - \calibrateprob_2\reviewerfunctionpdf_2(\score_1)\reviewerfunctionpdf_1(\score_2)$.
    
    From Figure~\ref{fig:noisycase2} we can see that error of the conference $\Econference([\score_1, \score_2])$ has its extremes at $\calibrateprob_1 = 0, \calibrateprob_2 = \frac{u}{v}$ and $\calibrateprob_1 = 1, \calibrateprob_2 = 1-\frac{u}{v}$. Therefore, error of the conference ranges from $\frac{(2-\Phi_1-2\Phi_2)u+\Phi_2v}{u+v}$ to $\frac{(\Phi_1+2\Phi_2-1)u+(1-\Phi_2)v}{u+v}$.
    
\begin{figure}[t!]
\centering
  \includegraphics[scale=0.3]{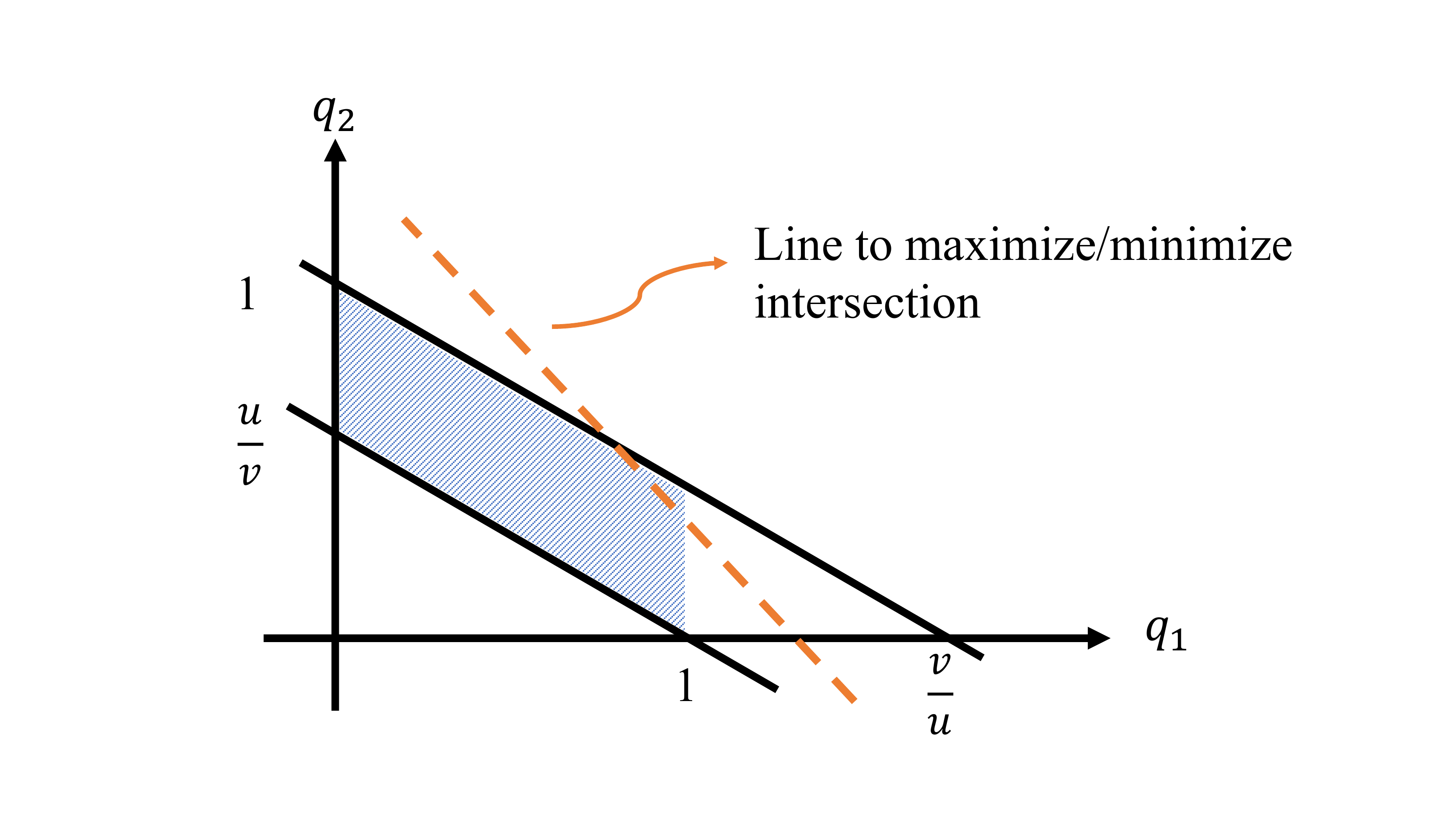}
  \caption{\label{fig:noisycase2}A diagram illustrates the optimization problem in this case.}
\end{figure}

     \item Maximize $1 - \frac{\calibrateprob_1\reviewerfunctionpdf_1(\score_1)\reviewerfunctionpdf_2(\score_2) + \calibrateprob_2\reviewerfunctionpdf_2(\score_1)\reviewerfunctionpdf_1(\score_2)}{\reviewerfunctionpdf_1(\score_1)\reviewerfunctionpdf_2(\score_2) + \reviewerfunctionpdf_2(\score_1)\reviewerfunctionpdf_1(\score_2)}$ subject to $\Econference([\score_1, \score_2]) (\reviewerfunctionpdf_1(\score_1)\reviewerfunctionpdf_2(\score_2) + \reviewerfunctionpdf_2(\score_1)\reviewerfunctionpdf_1(\score_2)) - \reviewerfunctionpdf_1(\score_1)\reviewerfunctionpdf_2(\score_2) \cdot (1-\Phi_1) - \reviewerfunctionpdf_2(\score_1)\reviewerfunctionpdf_1(\score_2) \cdot \Phi_2 = \reviewerfunctionpdf_1(\score_1)\reviewerfunctionpdf_2(\score_2) (2\Phi_1-1) \calibrateprob_1 + \reviewerfunctionpdf_2(\score_1)\reviewerfunctionpdf_1(\score_2) \cdot (1-2\Phi_2) \calibrateprob_2$ and $\calibrateprob_1 \reviewerfunctionpdf_1(\score_1)\reviewerfunctionpdf_2(\score_2) \ge \reviewerfunctionpdf_2(\score_1)\reviewerfunctionpdf_1(\score_2) - \calibrateprob_2\reviewerfunctionpdf_2(\score_1)\reviewerfunctionpdf_1(\score_2)$.
     
     The maximum occurs at $\calibrateprob_1 \reviewerfunctionpdf_1(\score_1)\reviewerfunctionpdf_2(\score_2) = \reviewerfunctionpdf_2(\score_1)\reviewerfunctionpdf_1(\score_2) - \calibrateprob_2\reviewerfunctionpdf_2(\score_1)\reviewerfunctionpdf_1(\score_2)$. Then the intersection of the two lines is $\calibrateprob_1 = \frac{(1-2\Phi_2)v-T(\Econference([\score_1, \score_2]))}{(2-2\Phi_1-2\Phi_2)u}$ and $\calibrateprob_2 = \frac{T(\Econference([\score_1, \score_2]))-(2\Phi_1 - 1)v}{(2 - 2\Phi_1-2\Phi_2)v}$.
     
    \begin{itemize}
        \item If the intersection point can be reached, $\calibrateprob_1, \calibrateprob_2 \in [0,1]$, $(1-2\Phi_2)v - (2-2\Phi_1 - 2\Phi_2)u \le T(\Econference([\score_1, \score_2])) \le (1 - 2\Phi_2)v$, then error of the conference $\Econference([\score_1, \score_2])$ ranges from
        $\frac{\reviewerfunctionpdf_1(\score_1)\reviewerfunctionpdf_2(\score_2) (1-\Phi_1)}{\reviewerfunctionpdf_1(\score_1)\reviewerfunctionpdf_2(\score_2) + \reviewerfunctionpdf_2(\score_1)\reviewerfunctionpdf_1(\score_2)} + \frac{\reviewerfunctionpdf_2(\score_1)\reviewerfunctionpdf_1(\score_2) (1-\Phi_2)}{\reviewerfunctionpdf_1(\score_1)\reviewerfunctionpdf_2(\score_2) + \reviewerfunctionpdf_2(\score_1)\reviewerfunctionpdf_1(\score_2)}$ (when $T(\Econference([\score_1, \score_2])) = (1 - 2\Phi_2)v$) to 
        $\frac{\reviewerfunctionpdf_1(\score_1)\reviewerfunctionpdf_2(\score_2) (\Phi_1 + 2\Phi_2 -1)}{\reviewerfunctionpdf_1(\score_1)\reviewerfunctionpdf_2(\score_2) + \reviewerfunctionpdf_2(\score_1)\reviewerfunctionpdf_1(\score_2)} + \frac{\reviewerfunctionpdf_2(\score_1)\reviewerfunctionpdf_1(\score_2) (1-\Phi_2)}{\reviewerfunctionpdf_1(\score_1)\reviewerfunctionpdf_2(\score_2) + \reviewerfunctionpdf_2(\score_1)\reviewerfunctionpdf_1(\score_2)}$ (when $T(\Econference([\score_1, \score_2])) = (1-2\Phi_2)v - (2-2\Phi_1 - 2\Phi_2)u$).
        
        Error of the adversary $\Eadversary([\score_1, \score_2])$ is $\frac{\reviewerfunctionpdf_1(\score_1)\reviewerfunctionpdf_2(\score_2)}{\reviewerfunctionpdf_1(\score_1)\reviewerfunctionpdf_2(\score_2) + \reviewerfunctionpdf_2(\score_1)\reviewerfunctionpdf_1(\score_2)}$.
        
        \item If the intersection point can not be reached and $T(\Econference([\score_1, \score_2])) > (1 - 2\Phi_2)v$, then no $\calibrateprob_1, \calibrateprob_2$ are qualified for the constraints.
        
         \item If the intersection point can not be reached and $T(\Econference([\score_1, \score_2])) < (1-2\Phi_2)v - (2-2\Phi_1 - 2\Phi_2)u$.
         \begin{itemize}
            \item If $(2\Phi_1 - 1)u + (1-2\Phi_2)v \le T(\Econference([\score_1, \score_2])) < (1-2\Phi_2)v - (2-2\Phi_1 - 2\Phi_2)u$ then the maximum is reached when $\calibrateprob_1 = 1$ and $\calibrateprob_2 = \frac{T(\Econference([\score_1, \score_2]))-(2\Phi_1-1)u}{(1 - 2\Phi_2)v}$. 
        
            Error of the conference $\Econference([\score_1, \score_2])$ ranges from $\frac{(\Phi_1+2\Phi_2-1)u+(1-\Phi_2)v}{u+v}$ (when $T(\Econference([\score_1, \score_2])) = (1-2\Phi_2)v - (2-2\Phi_1 - 2\Phi_2)u$) to $\frac{\Phi_1u+(1-\Phi_2)v}{u+v}$ (when $T(\Econference([\score_1, \score_2])) = (2\Phi_1 - 1)u + (1-2\Phi_2)v$). 
        
            Error of the adversary $\Eadversary([\score_1, \score_2])$ is $1 - \frac{T(\Econference([\score_1, \score_2]))+(2-2\Phi_1-2\Phi_2)u}{(1-2\Phi_2)(u+v)}$, ranges from $\frac{u}{u+v}$ (when $T(\Econference([\score_1, \score_2])) = (1-2\Phi_2)v - (2-2\Phi_1 - 2\Phi_2)u$) to 0 (when $T(\Econference([\score_1, \score_2])) = (2\Phi_1 - 1)u + (1-2\Phi_2)v$).
            
            \item If $ T(\Econference([\score_1, \score_2])) < (2\Phi_1 - 1)u + (1-2\Phi_2)v$ then no $\calibrateprob_1, \calibrateprob_2$ are qualified for the constraints.
        \end{itemize}
    \end{itemize}
\end{itemize}

\begin{figure}[t!]
\centering
\includegraphics[scale=0.22]{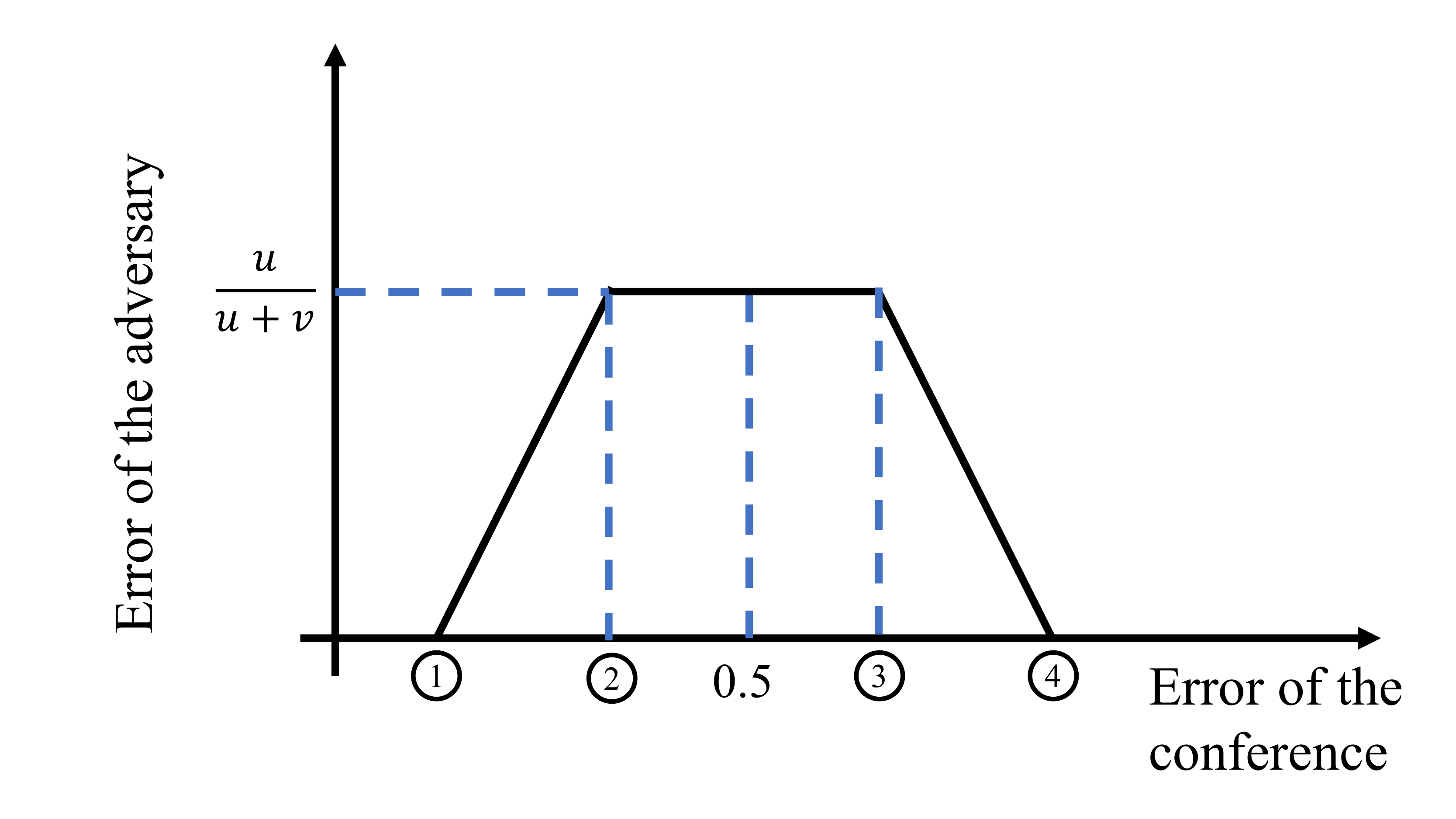}
  \caption{\label{fig:EAECnoisy1}Maximum per-instance error of the adversary given per-instance error of the conference when $u < v$, $\Phi_1 = \frac{1}{2} - \varphi_1$ and $\Phi_2 = \frac{1}{2} + \varphi_2$ with $0 < \varphi_2 < \varphi_1$. The coordinates in the plot are: \textcircled{1} = $\frac{u\Phi_1+v(1-\Phi_2)}{u+v}$, \textcircled{2} = $\frac{u(\Phi_1+2\Phi_2-1)+v(1-\Phi_2)}{u+v}$,
  \textcircled{3} = $\frac{u(2-\Phi_1-2\Phi_2)+v\Phi_2}{u+v}$, \textcircled{4} = $\frac{u(1-\Phi_1)+v\Phi_2}{u+v}$.}

\end{figure}

Therefore, the relation between error of the adversary and error of the conference when $\Phi_1 = \frac{1}{2} - \varphi_1$ and $\Phi_2 = \frac{1}{2} + \varphi_2$ where $0 < \varphi_2 < \varphi_1$ and $\reviewerfunctionpdf_1(\score_1)\reviewerfunctionpdf_2(\score_2) < \reviewerfunctionpdf_2(\score_1)\reviewerfunctionpdf_1(\score_2)$ is of the shape of a trapezoid in $[0,1]$ as in Figure~\ref{fig:EAECnoisy1}. Note that the relation between the per-instance errors does not change with the relation between values of $\reviewerfunctionpdf_1(\score_1)\reviewerfunctionpdf_2(\score_2)$ and $\reviewerfunctionpdf_2(\score_1)\reviewerfunctionpdf_1(\score_2)$ or with the values of $\Phi_1$ and $\Phi_2$. 

From Figure~\ref{fig:EAECnoisy1} we see that the conference should keep its per-instance error between $\frac{u\Phi_1+v(1-\Phi_2)}{u+v}$ and $\frac{u(\Phi_1+2\Phi_2-1)+v(1-\Phi_2)}{u+v}$ to stay optimal. The conference cannot have its error less than $\frac{u\Phi_1+v(1-\Phi_2)}{u+v}$ due to the reviewers' noise. If error of the conference is greater than $\frac{u(\Phi_1+2\Phi_2-1)+v(1-\Phi_2)}{u+v}$, increasing its error does not increase error the adversary and thus is not optimal. Thus, the Pareto frontier of per-instance error of the adversary against error of the conference is the first line segment with positive slope in Figure~\ref{fig:EAECnoisy1} when $\min \left\{\frac{a_2(a_1^2+\sigma^2)(\score_2-b_2)}{a_1(a_2^2+\sigma^2)} + b_1, \frac{a_1(a_2^2 + \sigma^2)(\score_2-b_1)}{a_2(a_1^2+\sigma^2)}+b_2\right\} < \score_1 < \max \left\{\frac{a_2(a_1^2+\sigma^2)(\score_2-b_2)}{a_1(a_2^2+\sigma^2)} + b_1, \frac{a_1(a_2^2 + \sigma^2)(\score_2-b_1)}{a_2(a_1^2+\sigma^2)}+b_2\right\}$.

\subsection{Proof of Theorem~\ref{thm:noisyInstance}}\label{sec:proof:thm:noisyInstance}

We prove that Algorithm~\ref{alg:noisyworste} is optimal for each instance of scores $\scorematrix = [\score_1, \score_2]$ with desired error of the conference $\Econference([\score_1, \score_2])$ in the noisy setting. We carry the assumptions from Section~\ref{sec:proof:thm:noisyPareto} that $\Phi_1 = \frac{1}{2} - \varphi_1$ and $\Phi_2 = \frac{1}{2} + \varphi_2$ where $0 < \varphi_2 < \varphi_1$ and $\reviewerfunctionpdf_1(\score_1)\reviewerfunctionpdf_2(\score_2) < \reviewerfunctionpdf_2(\score_1)\reviewerfunctionpdf_1(\score_2)$.

From Proposition~\ref{proposition:calibrationfunction} we know that if a paper has higher estimated quality under both assignments, the conference should accept the paper. This is the optimal calibration strategy for the conference.

Otherwise when the scores are in the region $\min \left\{\frac{a_2(a_1^2+\sigma^2)(\score_2-b_2)}{a_1(a_2^2+\sigma^2)} + b_1, \frac{a_1(a_2^2 + \sigma^2)(\score_2-b_1)}{a_2(a_1^2+\sigma^2)}+b_2\right\} < \score_1 < \max \left\{\frac{a_2(a_1^2+\sigma^2)(\score_2-b_2)}{a_1(a_2^2+\sigma^2)} + b_1, \frac{a_1(a_2^2 + \sigma^2)(\score_2-b_1)}{a_2(a_1^2+\sigma^2)}+b_2\right\}$, we use the Pareto frontiers analyze the optimality of our algorithm. Theorem~\ref{thm:noisyPareto} shows that the Pareto frontier in the noiseless setting within this region. The analysis is similar to the one in the noiseless setting in Section~\ref{sec:proof:thm:noiselessInstance}.

Suppose $\reviewerfunctionpdf_1(\score_1)\reviewerfunctionpdf_2(\score_2) < \reviewerfunctionpdf_2(\score_1)\reviewerfunctionpdf_1(\score_2)$, then the endpoint on the Pareto frontier has error of the adversary being $\frac{\reviewerfunctionpdf_1(\score_1)\reviewerfunctionpdf_2(\score_2)}{\reviewerfunctionpdf_1(\score_1)\reviewerfunctionpdf_2(\score_2) + \reviewerfunctionpdf_2(\score_1)\reviewerfunctionpdf_1(\score_2)}$ and error of the conference being $\frac{\reviewerfunctionpdf_1(\score_1)\reviewerfunctionpdf_2(\Phi_1+2\Phi_2-1)+\reviewerfunctionpdf_2(\score_1)\reviewerfunctionpdf_1(\score_2)(1-\Phi_2)}{\reviewerfunctionpdf_1(\score_1)\reviewerfunctionpdf_2+\reviewerfunctionpdf_2(\score_1)\reviewerfunctionpdf_1(\score_2)}$. 
If $\frac{\reviewerfunctionpdf_1(\score_1)\reviewerfunctionpdf_2(\score_2)\Phi_1+\reviewerfunctionpdf_2(\score_1)\reviewerfunctionpdf_1(\score_2)(1-\Phi_2)}{\reviewerfunctionpdf_1(\score_1)\reviewerfunctionpdf_2(\score_2)+\reviewerfunctionpdf_2(\score_1)\reviewerfunctionpdf_1(\score_2)} \le \Econference([\score_1, \score_2]) < \frac{\reviewerfunctionpdf_1(\score_1)\reviewerfunctionpdf_2(\score_2)(\Phi_1+2\Phi_2-1)+\reviewerfunctionpdf_2(\score_1)\reviewerfunctionpdf_1(\score_2)(1-\Phi_2)}{\reviewerfunctionpdf_1(\score_1)\reviewerfunctionpdf_2(\score_2)+\reviewerfunctionpdf_2(\score_1)\reviewerfunctionpdf_1(\score_2)}$, we maximize the error of the adversary by operating on the Pareto frontier. If $\Econference([\score_1, \score_2]) \ge \frac{\reviewerfunctionpdf_1(\score_1)\reviewerfunctionpdf_2(\score_2)(\Phi_1+2\Phi_2-1)+\reviewerfunctionpdf_2(\score_1)\reviewerfunctionpdf_1(\score_2)(1-\Phi_2)}{\reviewerfunctionpdf_1(\score_1)\reviewerfunctionpdf_2(\score_2)+\reviewerfunctionpdf_2(\score_1)\reviewerfunctionpdf_1(\score_2)}$, we operate at the endpoint where error of the adversary is maximum and error of the conference is no larger than the desired $\Econference([\score_1, \score_2])$. The endpoint is the point with minimum error of the conference such that error of the adversary is maximum. Therefore, it is optimal for the conference.

Algorithm~\ref{alg:noisyworste} follows the procedure by choosing the corresponding $\calibrateprob_1$ and $\calibrateprob_2$ for each point on the Pareto frontier and thus is optimal for the conference.

\end{document}